\documentclass[a4paper,fleqn,usenatbib]{mnras}

\usepackage{bm,mathptmx}

\usepackage[T1]{fontenc}
\usepackage{ae,aecompl}
\usepackage[brazil]{babel}
\usepackage[utf8]{inputenc}

\usepackage{graphicx}	
\usepackage{amsmath}	
\usepackage{amssymb}	
\usepackage{makecell}

\usepackage{longtable}
\usepackage{lipsum}
\usepackage{txfonts}
\usepackage{threeparttable}
\usepackage[bottom]{footmisc}

\usepackage{soulutf8} 

\title[VIVA catalogue from the New Insights into Time Series Analysis project]{\centering The VVV Infrared Variability Catalog (VIVA-I)}

\author[C. E. Ferreira Lopes et al.]{
\centering C. E. Ferreira Lopes$^{1}$\thanks{E-mail: ferreiralopes1011@gmail.com}, N. J. G.~Cross$^2$, M. Catelan$^{3,4}$, D. Minniti$^{4,5,6}$, M. Hempel$^{5}$,
\newauthor
 P. W. Lucas$^{7}$, R. Angeloni$^{8,9}$, F. Jablonsky$^{1}$,  V. F. Braga$^{4,5}$, I. C. Le\~ao$^{10}$, F. R. Herpich$^{15}$,
\newauthor
 J. Alonso-Garc\'ia$^{11,4}$, A.   Papageorgiou$^{3,4}$, K. Pichara$^{12,13,4}$, R. K. Saito$^{14}$,  A. Bradley$^{2}$,
\newauthor
 J. C. Beamin$^{16}$, C. Cort\'es$^{17}$, J. R. De Medeiros$^{10}$, Christopher. M. P. Russell$^{3}$
\\
\textit{(Affiliations can be found after the references)}}

\date{Accepted XXX. Received YYY; in original form ZZZ}

\pubyear{2018}

\begin{document}
\label{firstpage}
\pagerange{\pageref{firstpage}--\pageref{lastpage}}
\maketitle

\begin{abstract}
High extinction and crowding create a natural limitation for optical surveys towards the central regions of the Milky Way where the gas and dust are mainly confined. Large scale near-IR surveys of the Galactic Plane and Bulge are a good opportunity to explore open scientific questions as well as to test our capability to explore future datasets efficiently. Thanks to the \textit{VISTA Variables in the Vía Láctea} (VVV) ESO Public Survey it is now possible to explore a large number of objects in those regions. This paper addresses the variability analysis of all VVV point sources having more than $10$ observations in VVVDR4 using a novel approach. In total, the near-IR light curves of $288,378,769$ sources were analysed using methods developed in the New Insight Into Time Series Analysis project. 
As a result, we present a complete sample having  $44,998,752$ variable star candidates (VVV-CVSC), which include accurate individual coordinates, near-IR magnitudes ($ZYJH_s$), extinctions $A(K_s)$, variability indices, periods, amplitudes, among other parameters to assess the science. Unfortunately, a side effect of having a highly complete sample, is also having a high level of contamination by non-variable (contamination ratio of non-variables to variables is slightly over 10:1). To deal with this, we also provide some flags and parameters that can be used by the community to decrease the number of variable candidates without heavily decreasing the completeness of the sample. In particular, we icross-dentified $339,601$ of our sources with Simbad and AAVSO databases, which provide us with information for these objects at other wavelegths. This subsample constitutes a unique resource to study the corresponding near-IR variability of known sources as well as to assess the IR variability related with X-ray and Gamma-Ray sources. On the other hand, the other $\sim99.5\%$ sources in our sample constitutes a number of potentially new objects with variability information for the heavily crowded and reddened regions of the Galactic Plane and Bulge. The present results also provide an important queryable resource to perform variability analysis and to characterize ongoing and future surveys like TESS and LSST.
\end{abstract}



\section{Introduction}\label{sec_introduction}	

The first infrared (IR) light curve was probably that obtained for the Cepheid Zeta Geminorum ($\zeta$ Gem) by John S. Hall using a caesium oxide photoelectric cell \citep[][]{Hall-1932,Hall-1934}. The author found that the infrared maximum (at $7400\AA$) of the light curve occurs at $\sim0.024$ periods later than that observed in optical light curves \citep[][]{Hoffleit-1987}. Indeed, the characterization of different physical processes is better enabled when photometry across the whole electromagnetic spectrum is available. On the other hand, the interstellar environment is noticeably more transparent in IR and Near-IR (\textit{NIR}) light than at visible light. Thus, photometric surveys at infrared wavelengths can reveal different physical processes and explore unknown Milky Way (MW) regions at low Galactic latitudes that are usually obscured at visible wavelengths by the absorption of light by the interstellar medium. Atmospheric transparency is a strong function of wavelength, and many parts of the electromagnetic spectrum are not visible from the ground, and the technical capabilities of instruments tend to be poorer outside of the visible because the technologies are newer and have fewer commercial applications: hence the scientific discoveries have been limited by technology. The MW inner structure and details of its formation and evolution have been poorly understood due to the lack of variability datasets in these regions. Gas and dust in MW are mostly confined to the disk, where high extinction and crowding limit the usefulness of optical wavelengths. According to this natural limitation, most current optical surveys avoid the innermost MW plane. The detailed shapes of disk galaxies can hold clues to understanding the role that dynamical instabilities, hierarchical merging, and dissipative collapse played in the assembly history of the entire host galaxy \citep[][]{Athanassoula-2005}. In particular, the resolved stellar populations of the bulge, in connection with those of the disc and halo, provide us with a unique laboratory to investigate the fossil records of such fundamental processes \citep[see][]{Gonzalez-2016}. 

There are many large new studies of stellar variability due to improved telescopes/instruments with large entente and in particular the better access to publicly available datasets from large variability surveys.
For instance, at optical wavelengths this has led to improvements in the understanding of the stellar astrophysics of rotational modulation of stellar activity \citep[e.g.][]{McQuillan-2014,FerreiraLopes-2015cycles,Cortes-2015,Suarez-2016,Balona-2019}, stellar pulsation \citep[e.g.][]{Andersson-1996,Garcia-2014,Angeloni-2014ogle,FerreiraLopes-2015mgiant,Catelan-2015book,Braga-2019}, exoplanets \citep[e.g.][]{Fernandez-2006,Minniti-2007exo,Pietrukowicz-2010,Paz-Chinchon-2015,Gillon-2017,Almeida-2019,CortesDM-2019}, young stellar objects \citep[e.g.][]{ContrerasPenaKU-2017,ContrerasPenaDM-2017,LucasSM-2017,Guo-2019}, novae \citep[e.g.][]{SaitoTel-2012ATel,Banerjee-2018}, gravitational microlensing events \citep[e.g.][]{MinnitiCR-2015,Navarro-2017,Navarro-2018,Navarro-2019}, and eclipsing binaries
 \citep[e.g.][]{Torres-2010,Angeloni-2012,Helminiak-2013,Deleuil-2018}. On the other hand, new studies based on IR variability data at low Galactic latitudes may now become more accessible. 

For the past 10 years the ESO Public Survey VVV\footnote{\url{https://vvvsurvey.org/}} Survey and its extension VVVX (VISTA Variables in the Via Lactea, VVV eXtended, respectively) have been mapping the \textit{NIR} variability ($K_{s}$-band), of the Milky Way Bulge and the adjacent southern Disk, complemented by multi-colour observations. The VVV included the ZYJH$K_s$ bands \citep[][]{Minniti-2010}, whereas VVVX was restricted to the JHKs bands. The variability campaign in the $K_s$-waveband observed about $100$ $K_s$ epochs per field over the period 2010-2016 (for more details see Sect. \ref{sec_pawprint}). 



The VVV complements other public optical and mid-IR variability surveys of the Milky Way such as the Optical Gravitational Lensing Experiment \citep[OGLE -][]{Soszynski-2009}, Gaia \citep[][]{Perryman-2005}, the Transiting Exoplanet Survey Satellite \citep[TESS - ][]{Ricker-2015}, the Panoramic Survey Telescope and Rapid Response System \citep[Pan-STARRS - ][]{Kaiser-2002}, A High-cadence All-sky Survey System \citep[ATLAS - ][]{Tonry-2018}, Zwicky Transient Facility \citep[ZTF - ][]{Bellm-2019} as well as the next generation of surveys like PLAnetary Transits and Oscillation of stars \citep[PLATO - ][]{Rauer-2014}, \textbf{the Large Synoptic Survey Telescope \citep[LSST - ][]{Ivezic-2019}} and the Wide-field Infrared Survey Explorer \citep{NEOWISE} by covering the dust-encompassed central bulge regions and far-side of the disk at higher spatial resolution than is possible at longer wavelengths and adding additional important spectral information to all objects observed.

Large volumes of data containing potential scientific results are still unexplored or delayed due to our current inventory of tools that are unable to select clean samples. Despite great efforts having been undertaken, we run the risk of underusing a large part of these data. In the last decade, much effort has been made in automating, for example, the classification of variable stars \citep[e.g.][]{Debosscher-2007,Ivezic-2008,Richards-2011,Kim-2011,Bloom-2012,Pichara-2013,Nun-2014,Angeloni-2014vvv,Pichara-2016,Cabrera-Vives-2017,Benavente-2017,Graham-2017,Valenzuela-2018}. Usually these methods invest lots of efforts to extract features able to represent the peculiarities of different signals. These features can vary in number from a few to many tens of parameters \citep[e.g.][]{kim2014epoch,nun2015fats}. On the other hand, approaches where the  light curves are transformed into a two-dimensional array to perform classification with a convolutional neural network \citep[][]{mahabal2017deep} and unsupervised feature learning algorithms \citep[][]{mackenzie2016clustering} can find most of the underlying patterns that represent every light curve. Moreover, approaches using automatic learning of features are also being tested \citep[e.g.][]{mackenzie2016clustering}. Indeed, the light curves of the same source observed by different surveys would normally have different values for their features. However, if we use noise and periodicities to match distributions of features we avoid having to re-train from scratch for each new classification problem \citep[][]{long2012optimizing}.

The classification procedure presupposes that all parameters are accurately measured. For instance, a few percent of observed stars have non-stochastic variability and $75\%$ of the parameters used to characterize light curves are derived from variability periods \citep[][]{Richards-2011}. Inaccurate parameters may lead to a considerable increase of machine processing time and greater misclassification rates \citep[e.g.][]{Dubath-2011,FerreiraLopes-2015wfcam}. On the other hand, the New Insight into Time Series Analysis (\textit{NITSA}) project took a step back in order to review and improve all time-varying procedures \citep[][]{FerreiraLopes-2016papI,FerreiraLopes-2017papII,FerreiraLopes-2018papIII}. As a result, the \textit{NITSA} project provides optimized constraints to select a clean sample, i.e. a sample having only variable stars, on which the classification methods can be applied properly.

Unlike many variability surveys, the VVV survey is carried out in the near-IR. Despite several fundamental advantages, mostly due to the ability to probe deeper into the heavily reddened regions, the use of near-IR also presents important challenges. In particular, high-quality templates that are needed for training the automated variable star classification algorithms are not available \citep[e.g.][]{Debosscher-2007,Richards-2011,Dubath-2012,Bloom-2012,Pichara-2016}. Many variable-star classes have not yet been observed extensively in the near-IR, so that proper light curves are entirely lacking for these classes. The VVV Templates Project \footnote{\url{http://www2.astro.puc.cl/VVVTemplates/}} \citep[][]{Angeloni-2014vvv} has turned out to be a large observational effort in its own right, aimed at creating the first database on stellar variability in the near-IR, i.e. producing a large database of well-defined, high-quality, near-IR light curves. This project is in working progress and the variability analysis of the entire VVV database will be a very important step for such achievements. In order to reduce misclassification and mislabelling, accurate detections of true stellar variations are required. Moreover, the algorithms of classification need phased data to extract the main light curve features. 

\textit{NITSA} results were used to analyze the largest NIR survey of the MW bulge and disk. The text is organized as follows. Section \ref{sec_pawprint} describes the VVV processing and in particular the multi-epoch pawprint data. The variability analysis is described in Sect. \ref{sec_selection}, where the discrimination of sources into correlated and non correlated data is presented (see Sects. \ref{sec_noncorrelated} and \ref{sec_correlated}). In particular, all constraints used to perform this step are tested on real data (see Sect. \ref{sec_cutoffs}). Section \ref{sec_cross} discusses the variable stars previously identified in the literature. These sources were used to check the reliability of the variability periods determined by us in Sect. \ref{sec_periods}. Next, we discuss using the height of the periodogram peaks (related to the likelihood that the frequency is periodic), for the different methods, to produce more reliable samples in order to reduce the misselection in Sect. \ref{sec_reliablesel}. A new approach that improves the VVV data quality was proposed recently and hence we present the major implications in the current work \ref{sec_newcalibration}. Discussions and final remarks are presented in Sects. \ref{sec_results} and \ref{sec_conclusion}. All parameters released in this work are described in Appendix \ref{sec_columndescription}.

\section{Data}\label{sec_pawprint}	

The \textit{VVV} is an ESO public survey that uses the Visible and Infrared Survey Telescope for Astronomy (VISTA) to map the bulge ($-10.0^{\circ} \lesssim l \lesssim +10.5^{\circ}$ and  $-10.3^{\circ} \lesssim b \lesssim +5.1^{\circ}$) and the inner southern part of the Galactic disk ($294.7^{\circ} \lesssim l \lesssim +350.0^{\circ}$ and  $-2.25^{\circ} \lesssim b \lesssim +2.25^{\circ}$) of our Galaxy using five near-IR wavebands (Z, Y, J, H and $K_s$) plus a variability campaign in $K_s$ waveband over the period 2010-2017 \citep[][]{Minniti-2010}. 

We select our data from the VISTA Science Archive\citep[VSA\footnote{\url{http://surveys.roe.ac.uk/vsa/}}][]{Cross-2012}, and in particular from the VVVDR4 release, which contains all VVV data up to the end of ESO period P91 (30/09/2013). The VISTA data comes as two types of image product with derived catalogues: {\it pawprint} and {\it tile}. We use the {\it pawprint} data throughout our analysis, since these measurements are observed in a way which allows us to use correlation indices. However, the standard products, and tables used for light-curves in the VVVDR4 release contain {\it tile} data, so some additional linking, as described below, is necessary to create light-curves from pawprint data. 

The VIRCAM instrument on the \textit{VISTA} telescope has 16 detectors, arranged in a $4\times4$ pattern, with $90\%$ of a detector separation between each detector in the x-direction and $42.5\%$ in the y-direction. An individual observation labelled as a {\it normal} in the VSA is a multi-extension FITS file containing 16 image extensions, one for each detector. Several of these frames are jittered and co-averaged to form {\it pawprint} stacks. We use the catalogues from these in our analysis. 6 pawprint stacks are mosaiced together to form a 1.5 sq. deg. {\it tile}. These pawprints are arranged in a 2 by 3 grid, with a shift of almost one-detector in the x-direction and almost a half-detector in the y-direction, so that a typical part of the tile has twice the integration time\footnote{\url{http://casu.ast.cam.ac.uk/surveys-projects/vista/technical/tiles}}. The \textit{VVV} pointings are divided into different disk and bulge tile pointings which are labelled from d001 to d152 and from b201 to b396, respectively. 

We have decided to use stacked pawprint photometry for the following reasons:

\begin{itemize}
    \item Our analysis relies heavily on correlation indices and the overlapping pawprints within a tile provide between 2 to 6 independent measurements on short timescales (i.e. timescales much shorter than the epoch to epoch timescales, and therefore much shorter than the timescales of variability that we can measure), and can be considered to be correlated. 
    \item Tile photometry extraction is a complex process and corrections for saturation, scattered light, aperture loss and distortion are more difficult to model in tiles. These problems arise because both the sky and point-spread-function (PSF) is highly variable in the near-infrared on time-scales shorter than observation length of the tile, so the individual pawprints have different values.
    \item VVVDR4, on which this version of VIVA is based, is on CASU version 1.3, and the newer version 1.5 includes many improvements to tile photometry, but the pawprint photometry remains the same apart from some zeropoint changes.
    \item While tiles have twice the exposure times of the pawprint stacks this does not always give the much increased depth in the crowded regions of the VVV bulge where source confusion is significant.
    \item There are typically twice as many pawprint measurements as tile measurements.
\end{itemize}

The raw data is processed by the Cambridge Astronomy Survey Unit \citep[CASU][]{Irwin-2004} to produce the science quality stacked pawprint frames and standard $1.5$ sq. deg. tile frames and the catalogues from both image types. Up to date details about the nightly image and catalogue processing and calibration can be found at CASU\footnote{\url{http://casu.ast.cam.ac.uk/surveys-projects/vista/technical}}. These images and catalogues are stored in FITS format and are transferred to the VSA, where further processing is done to create deeper images and catalogues, band-merged products, light-curves and simple variability statistics and crossmatches to multi-wavelength surveys, which are stored as tables in a SQLServer relational database management system (\textit{RBDMS}). This allows scientists to rapidly select data, and only download what is relevant to their science case. In addition, these \textit{VDFS} products are linked to other products developed by the \textit{VVV} team, such as proper-motion catalogues \cite{VIRAC}, or PSF photometry catalogues  \citep[e.g.][]{Alonso-Garcia-2018}. The VIVA catalog provided in the present paper will also be linked into the VSA, so it can be searched along with all the other VVV data and be used as part of complex queries that can select out particular samples of variable stars. 

Light-curves can be extracted from the VSA VVVDR4 database using the \verb+vvvSourceXDetectionBestMatch+ table. However, this is based on tile detections, so to get the pawprint light-curves, we must join to the \verb+vvvTilePawprint+ table\footnote{VVVDR5 links to the pawprints on the request of the Principle Investigators, so this second step is no longer necessary.}. An example SQL selection is shown in App~\ref{app:sql}. 

Light-curves in the VSA do not just link all frames in a tile pointing, but also find all matches in overlapping pointings \citep[see][]{Cross-2009}. If a star is in a region overlapping two tiles, where there have been 49 observations in the first and 53 in the second, and it is in a region of the first where it has measurements on 2 \textit{pawprints} and of the second where it has measurements on 4 \textit{pawprints}, we  have 310 \textit{pawprint} measurements of the star altogether. 

The overlaps and short time between the \textit{pawprint} measurements  return data that match the necessary conditions to analyse variability using correlated indices \citep[][]{FerreiraLopes-2016papI,FerreiraLopes-2017papII}, i.e. two or more measurements close in time, where the interval between the measurements used in a correlation are much less than the variability period. The correlated indices only provided trustful information about variability under this condition. The conditions for correlation are discussed in detail in \cite{FerreiraLopes-2016papI}, where the case of VISTA observations is also considered.

We have used the standard aperture-corrected aperture photometry in our analysis and in particular the default aperture of 1.0 arcsec radius (aper3, named as \textit{A3}) for the  photometry as it usually gives the best signal-to-noise for the typical seeing of \textit{VVV} data \citep[see][for more details]{FerreiraLopes-2017papII}. This has a radius of 3 pixels and contains $\sim75\%$ of the total flux in stellar images, and most of the seeing dependency is removed by the aperture-correction. However, we must keep in mind that, mainly in crowded regions, nearby stars can affect the observations by adding an additional noise component from deblending images that relies on some imperfect modelling \citep[e.g.][]{Cross-2009,ContrerasRamos-2017,Alonso-Garcia-2018,Medina-2018}. For such regions, the PSF photometry is being performed by VVV teams, e.g. \citep{Alonso-Garcia-2018B,Surot-2019}. 

\section{Selection of Targets}\label{sec_selection}	

The selection of variable stars using variability indices is mandatory because the later steps on variability analysis, like the detection of variability periods, are more time-consuming, so an early reduction in the number of possible targets leads to significantly less processing overall. The detection of reliable variations is intrinsically related to the number of observations since the statistical significance of the parameters used to discriminate variable stars from noise increases with the number of measurements. Fewer correlated measurements are required to compute correlated variability indices than the number of measurements needed to calculate  non-correlated indices (statistical parameters) to the same accuracy. The number of observations required to compute reliable statistical parameters is not analytically defined. On the other hand, five is the minimum number of correlated measurements required to use correlated flux independent indices \citep[for more details see][]{FerreiraLopes-2016papI}. Indeed, this limit can be extrapolated for all correlated indices. The efficiency rate of correlated indices is higher than non-correlated indices and hence correlated indices will be adopted in preference when they are available.

Photometric surveys can be divided into two main groups from the viewpoint of the number of observations: databases where the variability signal can be viewed in time, i.e. very well-sampled light-curves like CoRoT and Kepler light curves, and those ones which the variability signal can only be observed in the folded phase diagram like the large majority of sources observed by the VVV survey. For the latter ones, the variability indices will not be enough to determine the reliability of signals. Therefore, the variability periods are required to create phase diagrams for forthcoming analysis. To determine the period accurately we need enough measurements to cover all the main variability phases. For instance, some eclipsing binaries have eclipses that only cover a small fraction of the phase diagram and hence the signal can be lost if this region is not covered or only very sparsely covered, for example Algol type stars (see the OGLEII DIA BUL-SC35 V1058 in Fig.   \ref{fig_lcscross} and OGLEII DIA BUL-SC19 V4104 in Fig. \ref{fig_lcscrossbest}). The lack of coverage of specific phases is less of a problem if the variability signature is a more smoothly varying signal along the whole phase diagram, like pulsating variable stars. Therefore, a reasonable number of measurements ($N$) is required to determine correctly the period and variability signature, but this is dependent on the type of variable star.
 
Photometric time series can be divided in four main groups in terms of variability indices and variability periods, as following: 
\begin{itemize}
    \item  Noise (noise) - non-variable stars with random variations due to noise, which have variability indices that are consistent with the a non-variable source with noise or variations below the detection limit;  
    \item Misclassified sources (\textit{MIS}) - variable stars having variability indices around the noise level or noisy data having variability indices larger than that expected for the noise. As a result we will miss some real variable stars as well as including some noisy data in the target list; 
    \item Variable stars with a non-detected variability period (\textit{VSNP}) - variable stars where no variability period was detected either because they are aperiodic or the measurements were not sufficient to recover the period. This class also includes those sources having enough variation to be detected by variability indices but the data quality are not good enough to determine the light curve morphology, like saturated LPVs. 
    \item Periodic variable stars (\textit{VSP}) - variable stars where the variability period detected returns a smooth phase diagrams.
\end{itemize}

Indeed, statistical fluctuations, a small number of good measurements ($N$), outliers, correlated-noise, and seasonal variations are factors that are usually present in the data and hence a fraction of \textit{MIS} are expected. The \textit{MIS} rate varies for a particular dataset when using different techniques \citep[][]{FerreiraLopes-2016papI,FerreiraLopes-2017papII}. On the other hand, the \textit{MIS} rate also depends on the signal-to-noise distribution of the reliable signals as well as the data quality. The present work concerns the selection of \textit{VSNP} and \textit{VSP} targets observed by the VVV survey.

\subsection{VVV Data Analysis}\label{sec_vvvdataanalyse}	

The New Insights into Time-Series Analysis (\textit{NITSA}) project reviewed and improved the variability indices and the selection criteria for variable star candidates \citep[][]{FerreiraLopes-2016papI,FerreiraLopes-2017papII}. The authors defined the criteria to determine which sources that can be analyzed with variability indices based on correlation measurements. Therefore, the data must be separated into two subsets: Correlated-Data (\textit{CD}) and Non-Correlated Data (\textit{NCD}), i.e those sources that should be analysed using correlated indices and non-correlated (statistical parameters) variability indices, respectively. The \textit{CD} set includes those sources having more than 4 correlated measurements. The remaining data must be labelled as \textit{NCD}. This identification is crucial to ensure the correct use of the variability indices. Non-correlated indices are not dependent on the arrangement of the observations and hence they can be computed for all sources. Therefore, both correlated and non-correlated variability indices can be combined to analyze \textit{CD} sources while the \textit{NCD} can only be analyzed using non-correlated variability indices. The correlated indices are more efficient than non-correlated indices \citep[see left panel Fig. 8 of][]{FerreiraLopes-2017papII}, giving much better discrimination if available, so should be used if possible. 

\begin{figure}
  \centering
  \includegraphics[width=0.5\textwidth,height=0.45\textwidth]{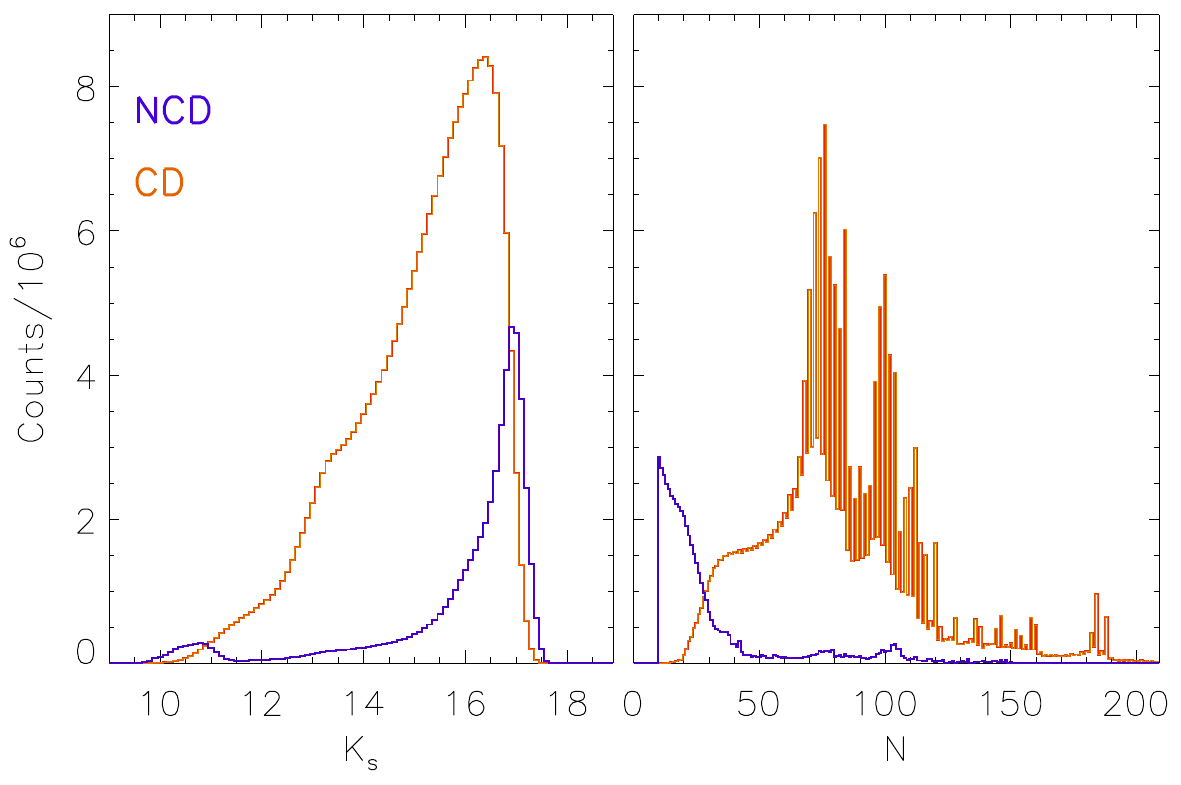} 
  \caption{Histograms of $K_{s}$ magnitude and number of measurements ($N$) for the VVV initial sample. The results for both NCD (black lines) and CD (orange lines) are shown. }
  \label{fig_histinitial}
\end{figure}

The observations of VVV pawprints necessary for the creation of tiles (see Sect \ref{sec_pawprint}) provide correlated data as a standard VISTA product, so we can optimize the search for variable stars since the correlated indices are freely available. Typically the observations necessary to make all 6 pawprint stacks in a tile are taken within $400s$, including the readout time that allows accurate correlated indices for variable stars having periods less than $\sim130$min. The released table contains the values of non-correlated indices for \textit{NCD} and \textit{CD} data while the correlated indices only for the latter (for more details see Sects. \ref{sec_noncorrelated} and \ref{sec_correlated}).

All VVV sources having more than $10$ measurements were considered in the current work. An initial sample of $288,378,769$ VVV sources found in the DR4 release were analyzed in the present work. The interval time between consecutive measurements of  $0.01$days was used to select close observations. These measurements were used to compute the correlated indices and determine the number of correlations \citep[for more details see][]{FerreiraLopes-2016papI}. VVV data having more than four correlated measurements were labelled as \textit{CD} otherwise \textit{NCD}. About $82\%$ of the initial sample corresponds to \textit{CD} type while the remaining sources are \textit{NCD}. The NCD sources are mostly those which are in the single exposure "ears" of each tile, and a small number of faint sources which were not detected on many frames. Indeed, those measurements having quality bit flags corresponding to more serious conditions  were removed. These were measurements with flags with values larger than 256\footnote{See ppErrBits at \url{http://horus.roe.ac.uk/vsa/www/gloss\_p.html}}.

\begin{figure*}
  \centering
  \includegraphics[width=0.49\textwidth,height=0.5\textwidth]{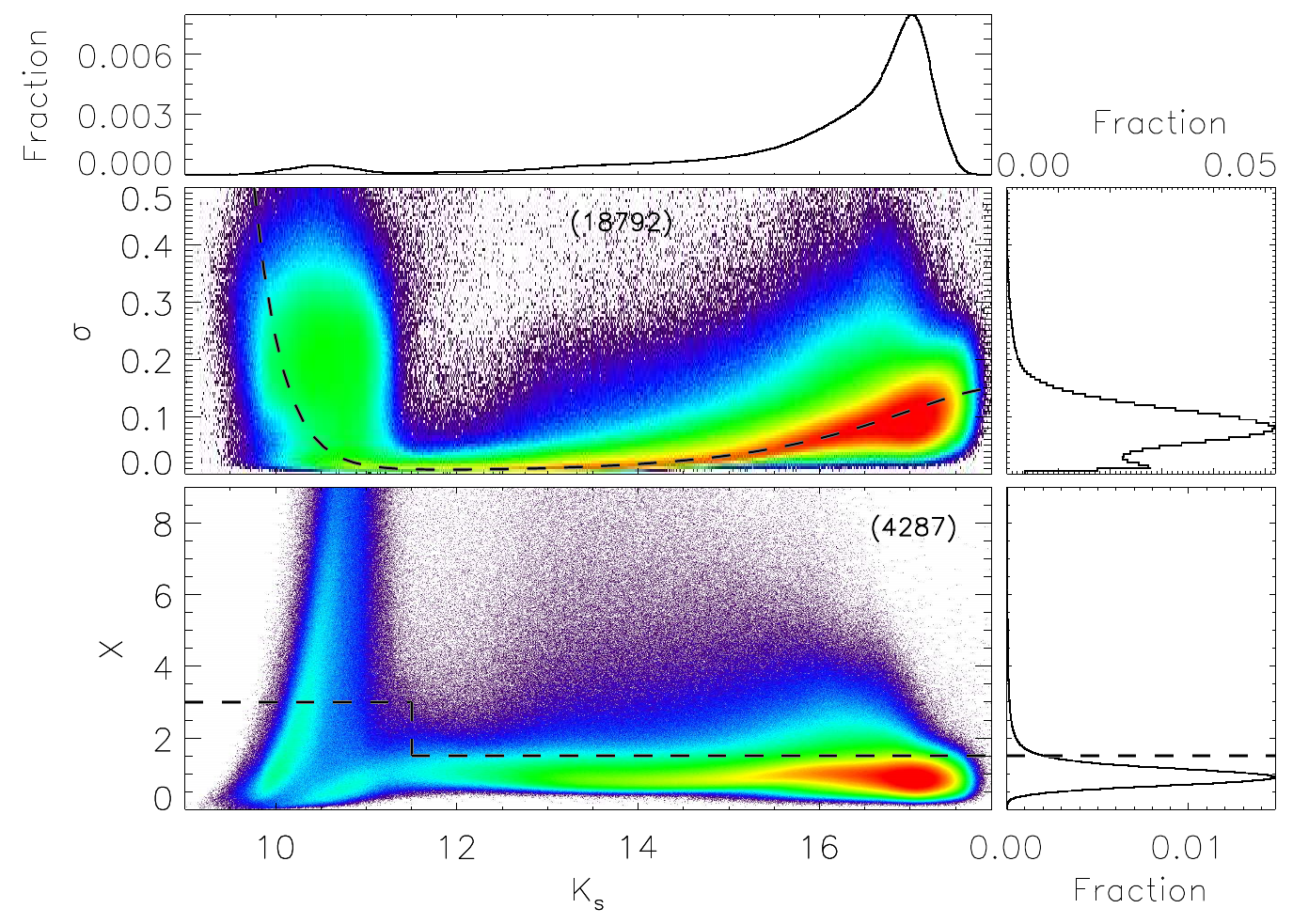} 
  \includegraphics[width=0.49\textwidth,height=0.5\textwidth]{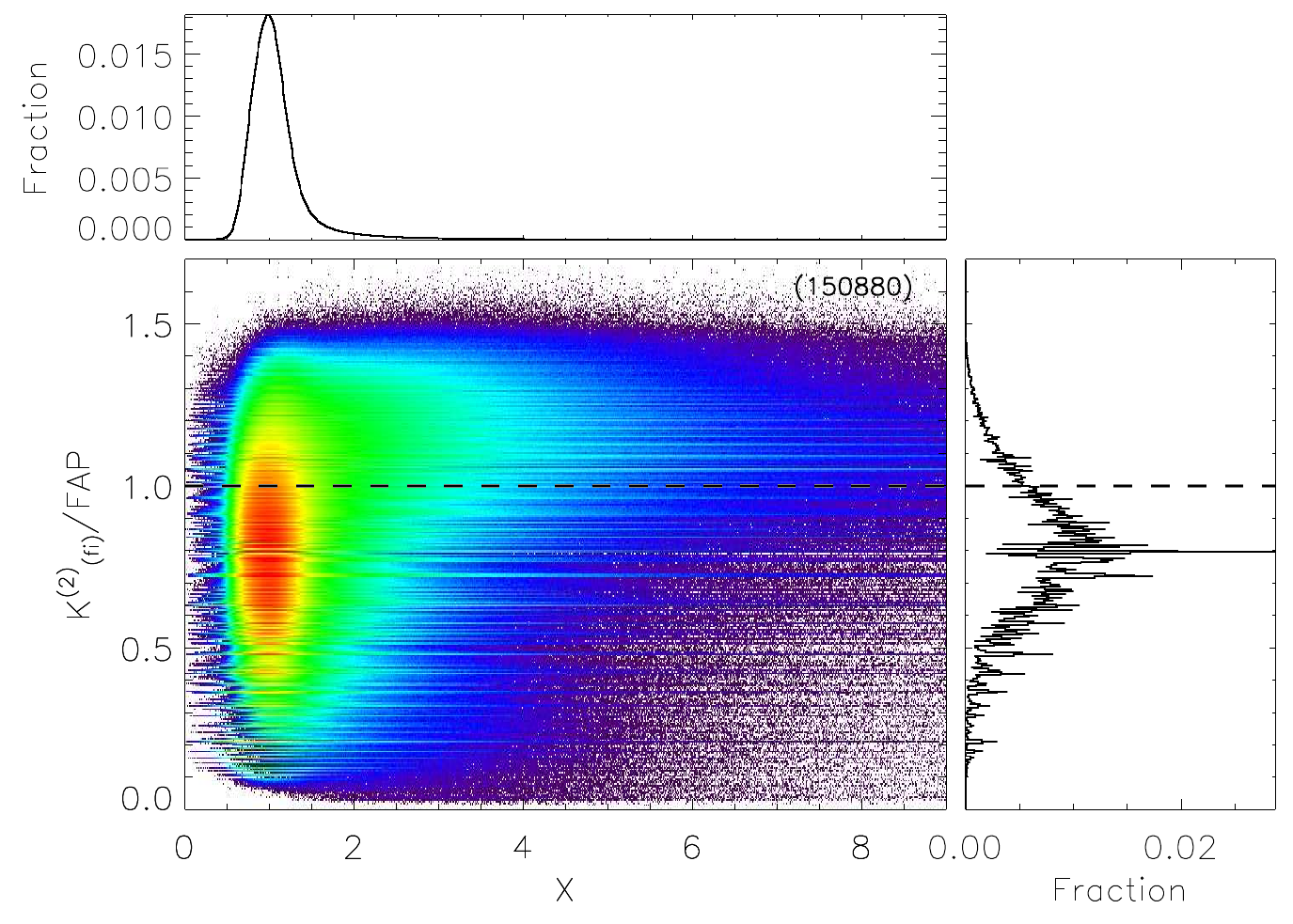} 
 
  \caption{Relative density plots  of $\sigma$ (upper left panel) and $X$ parameters (lower left panel) as function of magnitude as well as the $K_{(fi)}^{(2)}/Fap$ versus $X$ (right panel) index for our initial sample. The noise model is set by dark dashed line  (upper left panels) while this line in the lower left panel and right panel mark the cut-off above which the variable star candidates of non-correlated and correlated data were selected, respectively. The histograms at the top and right side show the normalized distribution of x and y axis respectively.}
  \label{fig_initialdata}
\end{figure*}

Figure  \ref{fig_histinitial} shows the histograms of $K_s$ magnitude and number of measurements ($N$) for the VVV initial sample where the \textit{NCD} and \textit{CD} samples are set by colours. The faint and bright stars contribute about $84\%$ and $5\%$ of \textit{NCD} (see upper panel blue line), respectively. The pronounced relative frequency of fainter sources found as \textit{NCD} is related with the reduction in the number of detections for these sources since a particular observation can drop below the detection threshold if the sky background is higher, the seeing is worse, their intrinsic flux dims, or even random photon statistics. Indeed, $73\%$ of \textit{NCD} have fewer than $30$ good measurements. Therefore, statistical fluctuations and systematics related to the faint and bright stars together with a small number of data will increase the misclassification rate for \textit{NCD}. On the other hand, only $2\%$ of \textit{CD} have $N$ smaller than $30$. Moreover, the centre of the histogram of $K_s$ magnitude is no longer concentrated on the region of faint stars. The reliability of analyses performed on \textit{CD} will be better than \textit{NCD}. The following subsections summarize the variability indices and describe the selection of \textit{NCD} variable stars candidates (\textit{NCD-CVSC}) and of \textit{CD} variable stars candidates (\textit{CD-CVSC}).

\subsection{Non-correlated Data (NCD)}\label{sec_noncorrelated}	
 
The recommendations provided by \citet[][]{FerreiraLopes-2017papII} to analyse \textit{NCD} sources were adopted. The main steps can be summarized as follows;

\begin{itemize}
  \item Photometric observations using a standard photometric aperture (aper3), see Sect.~\ref{sec_pawprint}.   
  \item  Compute the even-dispersion ($ED$) using only those measurements within twice of $ED_{\sigma\mu}$ about the even-median (BAS approach), i.e. $\sim95\%$ of data about the even-median. Removing outliers this way improves the performance by about $30\%$ according to \citet[][]{FerreiraLopes-2017papII}.
  \item Estimate the sample size correction factor for $ED$ in order to reduce the statistical fluctuations related to the number of measurements. As result, the adjusted $\sigma = ED\times w_{ED}$ values are obtained, where,  $w_{ED}$ is an weight related to the number of measurements.
  \item  Determine the noise model from the Strateva-modified function ($\zeta(K_s)$) \citep[$\zeta$ - ][]{FerreiraLopes-2017papII}. This model is obtained from the diagram of $K_s$ magnitudes as function of $ED$ (see black line of up panel of Fig. \ref{fig_initialdata}). This function fits the locus of non-variable point-sources and determines the expected noise value as a function of magnitude. 
  \item Finally, the non correlated indices are computed as the ratio of $\sigma$ by its expected noise value, given by \citet[see ][ for more detail]{FerreiraLopes-2017papII}

  \begin{equation}
    X = \frac{ \sigma }{\zeta}.
    \label{eq_varind1}  
    \nonumber
  \end{equation}

\noindent As result, the sources having $X\lesssim1$ should be related to the noise while larger values should indicate variable stars, i.e. this approach assumes that for the same magnitude stochastic (noisy data) and non-stochastic variation (variable stars) have different statistical properties. 
  \item The \textit{NCD-CVSC} stars were selected as those having $X > 1.5$ and $K_s > 11.5$mag or $X > 3.0$ and $K_s < 11.5$mag (for more details see Sect. \ref{sec_cutoffs}).
  \item ALL the above steps were performed on each VVV tile. 
\end{itemize}

Figure \ref{fig_initialdata} left hand side shows $\sigma$ (middle panel) and $X$ variability index (lower panel) as a function of $K_{s}$ magnitude for \textit{NCD}. The dark detached line indicates the Strateva-modified function (or noise model - middle panel) and the cut-off value used to select \textit{NCD-CVSC} stars (lower panel). The noise model was obtained using \textit{NCD} and \textit{CD} data in order to increase the statistical significance of the coefficients to the model. However, the left-hand plots only show the \textit{NCD}. The maximum number of \textit{NCD} sources per pixel is shown in brackets in the top right of the panel. The modified-Strateva function provides an improved fit to bright sources where an exponential increase is found for saturated stars. However, the dispersion about $\sigma$ is so high for bright sources implying a large dispersion for $K_{s}\lesssim11$ mag. The saturation level varies with the sky level, i.e a brighter background saturates the detector quicker. Therefore, a single noise model for entire VVV dataset is not recommended. Indeed, this behaviour also can be found in a single VVV pointing. As result, the number of \textit{MIS} increases for very bright sources. Indeed, $5\%$ of \textit{NCD} has a $K_s$  magnitude less than $11$ mag. 

\subsection{Correlated Data}\label{sec_correlated}	

The flux independent correlated index of order two \citep[$K_{(fi)}^{(2)}$ - ][]{FerreiraLopes-2016papI} was adopted to  analyse the VVV \textit{CD}. An order equal to two calculates the correlation between pairs of measurements close together in time ($\Delta\,T<0.01$ days). This index is defined as

\begin{equation}
  K_{(fi)}^{(2)} = \frac{ N^{+}_{\rm co} } {N_{\rm co}} 
  \label{eq_varind2}  
  \nonumber
\end{equation}

\noindent where $N_{\rm co}$ and $N^{+}_{\rm co}$  mean the total number of correlations and the number of positive correlations, respectively \citep[see ][ for more detail]{FerreiraLopes-2016papI}. The quantities ($N_{\rm co}$ and $N^{+}_{\rm co}$) used to compute the index are not dependent on the amplitude and hence $K_{(fi)}^{(2)}$ is weakly dependent on outliers and instrumental properties allowing a straightforward comparison between data observed in different telescopes at different or equal wavelengths (see Sect. \ref{sec_cutoffs}). Moreover, it has the highest efficiency for selecting variable stars among the correlated variability indices according to the authors. The following main steps were taken to analyze the \textit{CD} data:

\begin{itemize}
  \item Photometric measurements using the standard photometric aperture (A3) as for non-correlated data.   
  \item Use clipping of $ED_{\sigma\mu}$ about the even-median like that performed in Sect. \ref{sec_noncorrelated} to remove outlier measurements. The $K_{(fi)}^{(2)}$ is not dependent on the signal amplitude but it depends on the average value. This approach reduces the misselection rate true by the $K_{(fi)}^{(2)}$ index according to the authors.
  \item Measurements observed within $0.01$ days of each other were set as correlated measurements. The observations within each correlation box were then combined in each possible permutation of pairs, i.e. if there were 2 measurements there would be 1 correlation pair, if there were 3 measurements, 3 correlation pairs, if there were 4 measurements, 6 correlation pairs and so on. These correlations come mainly from the multiple pawprint measurements within a single tile (2-6), but may occasionally come from overlapping pawprints in the adjacent tiles if they were observed in quick succession.  
 \item Light curves having more than 4 correlated measurements were assigned as \textit{CD} and the $K_{(fi)}^{(2)}$ was computed. Indeed, the minimum number of correlated measurements necessary to use correlated indices is four according to the authors \citep[for more details see][]{FerreiraLopes-2016papI}.
 \item The $X$ index was computed as for the \textit{NCD} data.
 \item The false alarm probability for $K_{(fi)}^{(2)}$ as proposed by \citet[][]{FerreiraLopes-2016papI} was calculated as follows, 
 \begin{equation}
   FAP = 1 - \alpha\times\left(1 - \sqrt{\frac{ 4 }{ N_{\rm co}}} \right)
   \label{eq_fap}
 \end{equation}
 where $\alpha$ is a real positive number and $N_{\rm co}$ is the number of correlations. The theoretical value for the minimum number of correlations (four correlated measurements) and $\alpha = 0.45$ were adopted (for more details see Sect. \ref{sec_cutoffs}). $10^6$ Monte Carlo simulations of white noise considering $N_{\rm co}$ ranging from 10 to $1000$ correlated measurements were performed to verify how many spurious noisy data sources we expect to find above the cutoff of the FAP. As result, $\sim99\%$ of white noise dominated sources were found below this cut-off. Indeed, we could select a smaller fraction of spurious sources using a higher cutoff but, as result, a higher fraction of low signal to noise variables would be missed according to our tests (see Fig. \ref{fig_probe}). 
  \item The \textit{CD-CVSC} stars were selected as those having $K_{(fi)}^{(2)}/FAP > 1.0$ (for more details see Sect. \ref{sec_cutoffs}). The $X$ index was not used to select the \textit{CD-CVSC} sample but this information is available in the tables. The sources in the region limited by  $K_{(fi)}^{(2)}/FAP > 1.0$ and  $X < 1.0$ can be related with the correlated noise. On the other hand, the same region also can include those sources having overestimated noise values (for more details see Sect. \ref{sec_cutoffs}).
  \item ALL the above steps were performed in each VVV pointing.
\end{itemize}

Indeed, $K_{(fi)}^{(2)}$ is not dependent on the noise model and hence the sky background, unlike the X index. However, correlated noise must increase the number of \textit{MIS} since the FAP limits were estimated using white noise. The minimum number of correlations necessary to discriminate variable stars from noise is five according to \citet[][]{FerreiraLopes-2015wfcam}. However, the  $K_{(fi)}^{(2)}$ index assumes discrete values and hence small fluctuations in the correlation numbers can remove variable stars or increase the number of \textit{MIS}. Four correlated measurements were adopted as a minimum but a larger value increases the statistical significance of this correlated index.

\subsection{Cut-off and variable stars candidates}\label{sec_cutoffs}


Ideally only true variables should be included in the data analysis. Spurious contributions, e.g. related to seasonal variations or statistical fluctuations do in fact hamper the analysis of light curves. Therefore, the cut-off criteria are used to get complete samples ($\sim100\%$ of variable stars and a large number of \textit{MIS}), reliable samples ($\sim70\%$ of variable stars and a reduced number of \textit{MIS}), or ''genuine'' sample (only a small number of true detections). From the viewpoint of variability indices, genuine samples are only achievable for those variable stars having a high signal-to-noise and a reasonable number of observations. 
 For instance, the sample selected to contain about $95\%$ of \textit{WFSC1} variable stars (almost complete) is thrice as big as that selected to contain $72\%$ where the latter sample has, on average, higher amplitudes. Indeed, considering the WFSC1 catalogue, for each ''genuine'' source, there are at least three \textit{MIS} sources that will be misselected using correlated indices. This ratio of misselected to true sources increases to fourteen if non-correlated indices are used \citep[see ][ for more details]{FerreiraLopes-2016papI,FerreiraLopes-2017papII}. We point out that these ratios between genuine variables and MIS are only valid for data sets similar in S/N, since the efficiency rate decreases near the noise level. In this work, we create a complete sample in order to widen the utility of this catalog. The released data has parameters that allow users to select reliable or genuine samples (for more details see \ref{sec_reliablesel}).

A complete sample includes a small fraction of the entire database and hence it is a starting point to apply slower procedures. Indeed, reliable and genuine samples can be selected from the complete sample. Empirical cut-offs using different methods have been adopted to select targets in different surveys \citep[e.g.][]{Akerlof-2000,Damerdji-2007,Bhatti-2010,Shappee-2011,DeMedeiros-2013,Drake-2014,Rice-2015,Wang-2017,Ita-2018}. A comparative performance of selected variability detection techniques in photometric time series have been made by \citet[][]{Sokolovsky-2017} where the authors show that the $\eta$ correlated variability index provides the best performance. However, this is not a general result according to \citet[][]{FerreiraLopes-2017papII}, i.e. it is only valid for the sample analyzed by the authors. The best recommendations for analysing variability in photometric surveys can be found in the \textit{NITSA} project since these studies address how to set a common cut-off for a generic survey. Indeed, the cut-off is not unique for correlated indices based on amplitude or non-correlated indices since the noise properties and variability amplitudes can change from one survey to another. On the other hand, the panchromatic flux independent indices ($K_{(fi)}^{(s)}$) allow us to achieve this goal since they are only weakly dependent on the amplitude and instrument properties. Therefore, this cut-off must be valid for any survey.

\begin{figure}
  \centering
  \includegraphics[width=0.45\textwidth,height=0.5\textwidth]{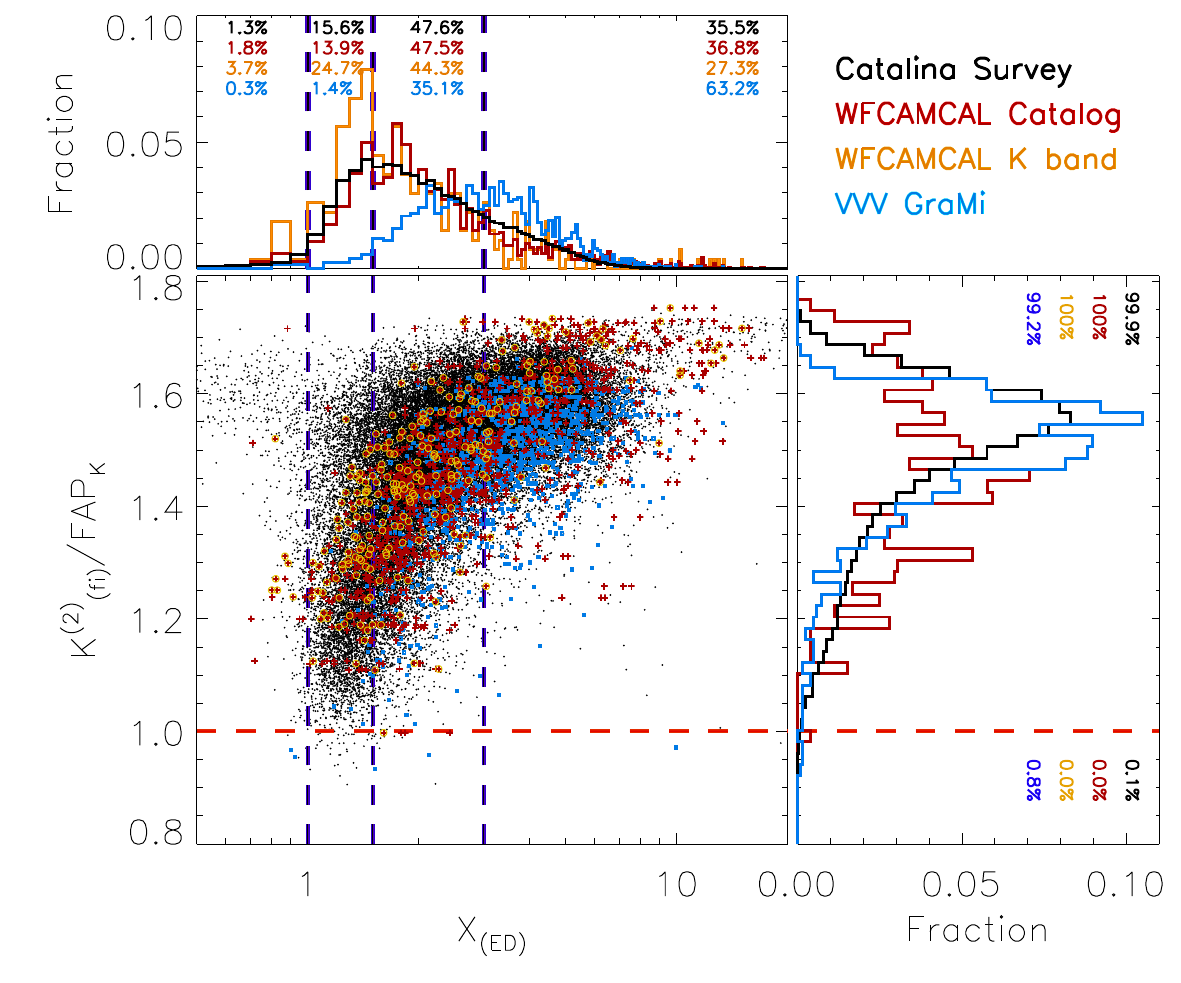} 
  \caption{$X$ non-correlated variability index versus the ratio of the  $K_{(fi)}^{(2)}$ correlated index to the $FAP$. The Catalina (black dots - V waveband), RR Lyrae found in the VVV survey (blue squares - $K_s$ waveband), and WFCAM variable stars (red and yellow crosses - ZYJHK wavebands and K-band respectively) are shown. The WFCAM results obtained in $K$ waveband are indicated by yellow circles. The lines set the cut-off values regarding those used to select \textit{NCD} (blue detached line)  and \textit{CD} (orange detached line) variable star candidates. The percentage of data enclosed by these lines are displayed in the upper and right panels along with their respective histograms.}
  \label{fig_probe}
\end{figure}

Moreover, three datasets were used to verify how many variable stars are being missed using our cutoffs for \textit{NCD} and \textit{CD} data: the WFCAM variable star catalogue (\textit{WFSC1}) having $275$ clearly periodic variable stars and $44$  other variable sources showing reasonably coherent light curves in ZYJHK wavebands; the Catalina Survey Periodic Variable  star catalogue (\textit{CVSC1}) having $\sim47000$ variable stars in the V waveband; \citep[][]{Drake-2014}; the catalogue of RRLyr stars found by \citet[][]{Gran-2016} and \citet[][]{Minniti-2017} selected from the VVV Survey (\textit{GraMi}). No special considerations are required to compute the $X$ index. On the other hand, the $K_{(fi)}^{(2)}$ index needs more than four correlated measurements to be computed. The \textit{CVSC1} and \textit{GraMi} have enough correlated measurements in a single filter to calculate the $K_{(fi)}^{(2)}$ index, in contrast to the \textit{WFSC1} sample. Therefore, all wavebands were used to compute $K_{(fi)}^{(2)}$ for \textit{WFSC1} sample as demonstrated in \citet[][]{FerreiraLopes-2015wfcam}. As a result, a single $X$ index value is computed for each waveband while $K_{(fi)}^{(2)}$ is estimated using all wavebands together \cite[for more details see][]{FerreiraLopes-2016papI}. Figure \ref{fig_probe} shows the ratio of $K_{(fi)}^{(2)}$ to FAP as function of $X$ index for the \textit{WFSC1-ZYJHK},  \textit{WFSC1-K}, \textit{CVSC1}, and \textit{GraMi} catalogues. The main results about that can be summarized as following;

\begin{itemize}
\item The \textit{WFSC1-ZYJHK}, \textit{WFSC1-K}, and \textit{CVSC1} show similar distributions of $X$ values (see top panel). On the other hand, the \textit{GraMi} shows a large number of sources having $X$ index bigger than $3$. This means that the \textit{WFSC1-ZYJHK}, \textit{WFSC1-K}, and \textit{CVSC1} samples have quite similar signal to noise distribution \citep[][]{FerreiraLopes-2018papIII} and they are more representative than the \textit{GraMi} sample, i.e. those samples are more mixed, and include a larger variety of variable stars. In fact, the \textit{GraMi} is a sample of RR Lyrae stars which have amplitudes that are, on average, larger than in the others samples.

\item The amplitude found in optical light curves is usually larger than those found in the near-infrared light curves for the majority of variable stars \citep[e.g.][]{FerreiraLopes-2015wfcam,Huang-2018}.  Therefore, on average, the number of sources having $X$ index close to the  noise limit will be bigger.
 Indeed, $28.4\%$ of \textit{WFSC1-K} have $X < 1.5$ while the proportion of \textit{CVSC1} is $16.9\%$ and \textit{WFSC1-ZYJHK} is $15.7\%$ at the same cut-off. On the other hand, only $1.7\%$ of \textit{GraMi} data are found in this range as expected, given the nature of the sample discussed in the previous paragraph. This indicates that a fraction of RR Lyr stars having lower amplitudes in the fields analysed by \citet[][]{Gran-2016} and \citet[][]{Minniti-2017} were missed. 

\item The \textit{CVSC1} and \textit{GraMi} show a peak at  $K_{(fi)}^{(2)}/FAP \simeq 1.55$. However, the \textit{WFSC1-ZYJHK} has more stars for high or lower  $K_{(fi)}^{(2)}/FAP$ values than the other distributions. It indicates that  \textit{CVSC1} and \textit{GraMi} missed some variable stars or it is only a sampling effect. Indeed, the \textit{CVSC1} and \textit{GraMi} were not investigated using the $K_{(fi)}^{(2)}$, a new variability analysis using NITSA recommendations will resolve this question.

\item About $0.1\%$ of \textit{GraMi} sources do not have enough correlated measurements and so they only can be analysed using the $X$ index. Therefore the efficiency rate using $K_{(fi)}^{(2)}/FAP > 1$ is nearly $100\%$. On the other hand, all of the sources in the \textit{WFSC1-ZYJHK}, \textit{WFSC1-K}, and \textit{CVSC1} samples are above this limit.

\item The cut-off used to create the \textit{CD-CVSC} implies that $\sim99\%$ of variable stars are included in the VVV database based on the analysis of the \textit{WFSC1-K}, \textit{CVSC1}, \textit{WFSC1-ZYJHK}, and \textit{GraMi} samples. The variability indices should detect all correlated signal types, including ones not present in the already analysed catalogs, since these indices were not designed to detect any particular signal. On the other hand, the \textit{NCD-CVSC} selects $\sim71.6\%$ of the true variable sources and  $\sim27.3\%$ for $K_s>11.5$ and $K_s<11.5$, respectively. Indeed, this statistic is biased by the signal-to-noise distribution (see discussion above).
\end{itemize}

The current analysis validates the cut-offs used to create \textit{CD-CVSC} and \textit{NCD-CVSC}. Indeed, this diagram can be extended for past, ongoing, and forthcoming projects since the $K_{(fi)}^{(2)}/FAP$ is weakly dependent on the wavelength observed or instrumental properties. This means a real improvement on variability analysis since a single and universal parameter is enough to select complete samples.

\begin{figure*}
  \centering
  \includegraphics[width=1.0\textwidth,height=0.6\textwidth]{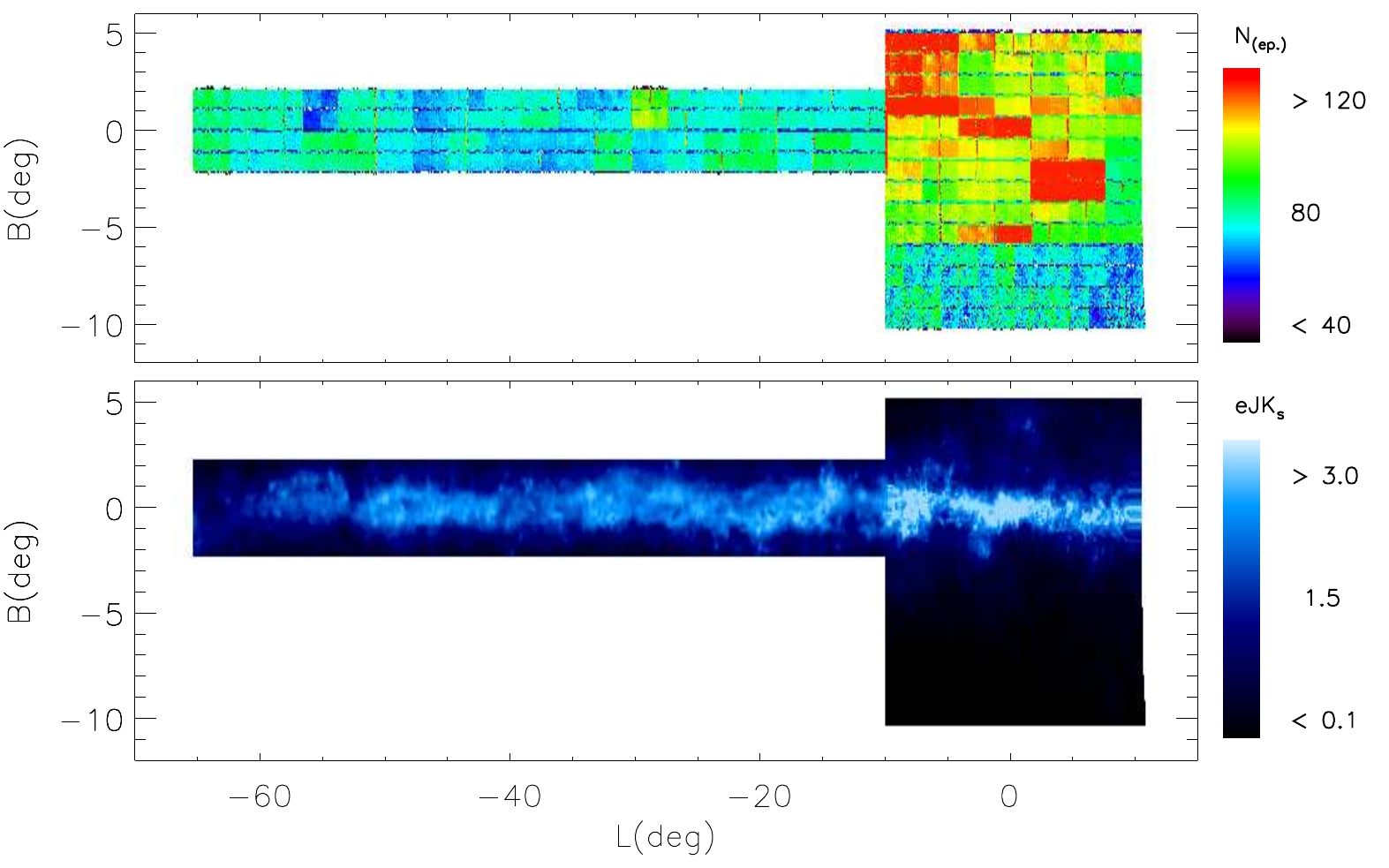} 
  \caption{Spatial distribution for \textit{CD-CVSC} and \textit{NCD-CVSC} VVV stars with the data-points colour coded according to the number of detections (upper panel) and extinction $A_{K_s}$ (lower panel). The tile edges are seen in the upper panel. }
  \label{fig_galdist}
\end{figure*}

\section{\textit{CD-CVSC} and \textit{NCD-CVSC} VVV stars}\label{sec_vsc}

Using \textit{CD-CVSC} and \textit{NCD-CVSC}, we have selected a sample containing $44,998,752$ sources (\textit{VVV-CVSC}). About $99\%$ of variable stars detectable by the VVV survey are included in our catalogue according to our analysis (for more details see \ref{sec_cutoffs}). Indeed, for each true detection there are at least $10$ \textit{MIS} sources according to \citet[][]{FerreiraLopes-2017papII}. A smaller number of \textit{MIS} sources can be achieved using higher cut-off values available in the released tables (see Sect. \ref{sec_columndescription}). Additionally, the $ZYJHK_s$ VVV photometry and the total extinction in the $K_s$-band ($A_{K_{s}}$) provided by the VVV extinction maps presented in \citet[][]{Minniti-2018ebv} are available in the released tables. The mean $A_{K_{s}}$ over an area of $10\times10$ arcmin$^2$ around the target position was used for the disc area. On the other hand, the total extinction $A_{K_{s}}$ was taken directly from the Bulge Extinction And Metallicity (BEAM) Calculator  \citet[][]{Gonzalez-2012}. The \citet[][]{Cardelli-1989} extinction law was assumed in both estimations.

Figure \ref{fig_galdist} shows the spatial distribution of \textit{CD-CVSC} and \textit{NCD-CVSC} VVV stars. The number of detections taken in the bulge is greater than in the disc. The highest number of measurements are found is b293, b294, b295, b296, b307, b308, b309, b310, as well as b333 (the tile containing the Galactic centre), where $N_{\rm ep}>120$. The specification of each tile can be found in the released table. Indeed, VVVX may improve the period detection or increase the variable candidate list as observations will be taken for all of VVV fields. There are often similar numbers of observations in groups of 4 tiles (arranged 2x2). The observing tool allows combining them in a so-called concatenation, i.e. these tiles are observed back to back together, without any other observations interloping. This is done to calculate the sky background, which in the Ks-waveband changes rather quickly. Indeed, the difference in the number of measurements within a concatenation will arise because some observations were declared failed, and deprecated: maybe the seeing degraded or there are some other concerns (like very bright stars).

\begin{table}
 \centering     
 \caption[]{Variability types and counts for the crossmatched sources. The meaning of the acronyms can be found at the  AAVSO\footnote{\url{https://www.aavso.org/vsx/index.php?view=about.vartypes}} and SIMBAD\footnote{\url{http://simbad.u-strasbg.fr/simbad/sim-display?data=otypes}} repositories.) }\label{tab_dep} 
 \begin{tabular}{c c c}        
   \hline\hline                 
    Type & Other types & Counts  \\   
   \hline    
ACV  &  ACVO, $*$alf2CVn, RotV$*$alf2CVn, & $ 24 $ \\ 
APER  &   & $ 27 $ \\ 
BE  &  GCAS, Be$*$, Ae$*$, .... & $ 145 $ \\ 
BY  &  BY$*$, & $ 21 $ \\ 
CEP  &  CEP(B), Cepheid, Ce$*$, ... & $ 64 $ \\ 
CV  &  CataclyV$*$, IBWD, V838MON, ... & $ 54 $ \\ 
CW  &  CWA, CWB, CW-FU, CW-FO, & $ 814 $ \\ 
DCEP  &  DCEP(B), DCEPS, DCEP-FU, ... & $ 393 $ \\ 
DSCT  &  DSCTC, DSCTr, dS$*$, DS & $ 101 $ \\ 
E  &  AR, D, DM, ECL, SD, $*$in$**$, SB$*$, & $ 90687 $ \\ 
EA  &  EA-BLEND, ED, EB$*$Algol, Al$*$, & $ 1867 $ \\ 
EB  &  ESD, EB$*$WUMa, EB$*$betLyr, ... & $ 3167 $ \\ 
EW  &  DW, K, KE, WU$*$, KW & $ 24722 $ \\ 
EC  &  EC & $ 47498 $ \\ 
FKCOM  &  RS, RSCVn, SXARI, ... & $ 1556 $ \\ 
GRB  &  gam, gB, SNR, SNR?, ... & $ 27 $ \\ 
HADS  &  HADS(B), SXPHE, SXPHE(B), & $ 40 $ \\ 
HMXB  &  HXB, HX?, ... & $ 13 $ \\ 
I  &  IA, IB, $*$iA, & $ 20 $ \\ 
IN  &  IT, INA, INB, INT, ... & $ 33 $ \\ 
IR  &  IR$<10\mu m$, IR$>30\mu m$, OH/IR, NIR, & $ 2274 $ \\ 
ISM  &  PoC, CGb, bub, EmO, ... & $ 1166$ \\ 
L  &  LB, LC, ... & $ 215 $ \\ 
LMXB  &  LXB, & $ 13 $ \\ 
LPV  &  LP$*$, LPV$*$, ... & $ 745 $ \\ 
M  &  Mira, Mi?, Mi$*$, ... & $ 1689 $ \\ 
Microlens  &  LensingEv, Lev & $ 231 $ \\ 
N  &  NA, NB, NC, NL, NR, Nova-like, ... & $ 830$ \\ 
NSIN  &  EllipVar, ELL, & $ 14564 $ \\ 
Others  &  PoC, CGb, bub, EmO,  ... & $ 72355 $ \\ 
PER  &   & $ 261 $ \\ 
PUL  &  PULS, Pu$*$, Psr, ... & $ 308 $ \\ 
Planet  &  PN, Pl, ... & $ 290 $ \\ 
RCB  &  DYPer, FF, DPV, DIP, ... & $ 18 $ \\ 
RGB  &  RGB$*$, RG$*$, ... & $ 702 $ \\ 
ROT  &  R, RotV$*$, RotV, CTTS & $ 343 $ \\ 
RR  &  RR(B), RRD, RRAB, RRC, RRLyr, RR$*$ & $ 30923 $ \\ 
RV  &  RVA, RVB, ... & $ 143 $ \\ 
Radio  &  mm, cm, smm, FIR, Mas, ... & $ 1084 $ \\ 
SR  &  SRA, SRB, SRC, SRD, SRS, ... & $ 71297 $ \\ 
TTS  &  WTTS & $ 193 $ \\ 
TTau  &  TTau$*$, TT$*$ & $ 31 $ \\ 
UG  &   & $ 41 $ \\ 
V$*$  &  V$*$?, & $ 1299 $ \\ 
WR  &  WR$*$ & $ 88 $ \\ 
X  &  XB, XF, XI, XJ, XND, ... & $ 2073 $ \\ 
YSO  &  Y$*$O, Y$*$, Y$*$? & $ 7123 $ \\ 
ZAND  &   & $ 26 $ \\ 
iC  &  $*$iC, $*$iN, AGB$*$, ... & $ 6167 $ \\ 
 \hline                                   
 \end{tabular}
\label{tab_crosstable}
\end{table}
 
Within the VVV tiles, we found a tiny region having a smaller number of detections, the blue stripes in contrast with the green and red region in the upper panel of Figure \ref{fig_galdist}. This can be related with a smaller efficiency of the detector in its boundaries. On the other side, the region that links the disk and bulge VVV areas shows an increase in the number of detections (see a red line in the crossed region between bulge and disk tiles). This happens because the intersection region between the disk and bulge VVV areas has a higher number of measurements. The spatial distribution of $eJK_{s}$ values varies from  $ < 0.1$ mag in the outer bulge up to  $eJK_{s} \simeq 3$ mag for objects near the Galactic Centre. A note of caution: the total extinction as calculated by the VVV maps is certainly overestimated according to \citet[][]{Gonzalez-2018}. 

\subsection{Cross-identification}\label{sec_cross}

$339,601$ \textit{VVV-CVSC} sources were previously recorded by the \textit{AAVSO} International Variable Star Index \citep[VSX;][]{Watson-2014} or \textit{SIMBAD} database\footnote{\url{http://simbad.u-strasbg.fr/simbad/}}. This subsample was named as \textit{VVV-CVSC-CROS}. SIMBAD contains about $9,795,519$ objects across the sky while VSX contains $1,432,959$ sources to date. These repositories contain the widest compilations of variable stars known so far that can contain names, positions, photometric information, period, variability types, and astronomical parameters such as constellation and the passband used to measure the variability.  The Tool for OPerations on Catalogs And Tables \citep[\textit{TOPCAT} - ][]{Taylor-2005}\footnote{\url{http://www.star.bris.ac.uk/~mbt/topcat/}} was used to crossmatch our catalogue with the SIMBAD database. The allowed tolerance of the crossmatch was $1''$ in the sky coordinates for VVV where the nearest source was assumed as the crossmatched source.  

The data found in these repositories does not contain all available information in the literature. For instance, the main table of \textit{SIMBAD} has variability types but does not include the variability periods. On the other hand, the \textit{VSX} table contains both information. Moreover, multiple classifications or different nomenclature can be found in these tables. The acronyms identifying the variability types\footnote{https://www.aavso.org/vsx/index.php?view=about.vartypes} were used to group the sources in different branches. We took the first classification for those objects having multiple classification. Therefore we have added two columns to our table giving information about the variability type: the notation adopted by us (column $cfl.mainVarType$) and the one that comes from literature (column $cfl.literatureVarType$). The full description of available tables is given in the Sect. \ref{sec_columndescription}.

The main information about \textit{VVV-CVSC-CROS} are released in a secondary table having the following pieces of information; VVV identifiers, literature names, variability periods, and variability types when available. The VVV identifiers can be crossmatched with the \textit{VVV-CVSC} table (for more details see Sect. \ref{sec_columndescription}) to access full VVV information about these sources. Besides, further information about them can be accessed using the literature names or coordinates in web services (for more details see Sect. \ref{sec_columndescription}). Table \ref{tab_crosstable} shows a summary of \textit{VVV-CVSC-CROS} having more than 10 object per variability type. The main results from this crosscorrelated database are summarized below; 

\begin{itemize}
\item (E) About $27\%$ of the crossmatched sources are classified as eclipsing binaries, matching the $49\%$ of stars being found in double or multiple systems. Hence a larger number of eclipsing binaries is to be expected. If we include E, EA, EB, EW, EC, NSIN, and X the final rate rises to $54\%$.

\item (RR) The variability type having the second largest number of crossmatched sources are the RR Lyrae. These types of stars have quite a high amplitude and short periods \citep[e.g.][]{FerreiraLopes-2015wfcam,Huang-2018}. These properties increase the identification rate of these sources.

\item (SR) Semiregular variable stars are giants or supergiants of intermediate and late spectral type showing considerable periodicity in their light changes, accompanied or sometimes interrupted by various irregularities. Their amplitudes may be from hundredths of a magnitude to several magnitudes. On the other hand, the variability periods are quite long (the range from $20$ to $>2000$ days) compared with the RR Lyrae. Therefore a smaller detection rate for these sources are expected. Indeed, the long period variables (LPVs) and Miras (M) can be included in this class.

\item (FKCOM) FK Comae Berenices-type variables are rapidly rotating giants with non-uniform surface brightnesses with a wide range of variability periods and amplitudes about several tenths of a magnitude. Their detection rate is not so different from that found for X-ray type stars.

\item There are many \textit{VVV-CVSC-CROS} sources which have not been assigned a variability type. The identification can be related to their localization like a star in a cluster (iC), young stellar object (YSO), or part of cloud (Poc) for example. On the other hand, they also can be classified as peculiar emitters like metric/centimetric/milimetric/sub-millimetric radio sources, far/near infrared sources, or objects having emission lines.
\end{itemize}

The \textit{VVV-CVSC-CROS} is a unique catalogue which can be used to study many open stellar astrophysics questions about the IR variability of a wide range of variable stars. In fact, stellar populations or a deeper analysis about the IR variability are beyond the scope of this paper. However, the light curve shapes and some comments about these objects are explored in Sect. \ref{sec_varproperties}.

\subsection{Variability periods}\label{sec_periods}

The variability period of \textit{VVV-CVSC} were estimated using five methods; Generalized Lomb-Scargle \citep[LSG: ][]{Lomb-1976,Scargle-1982,Zechmeister-2009}, String Length Minimization \citep[STR: ][]{Dworetsky-1983}, Phase Dispersion Minimization method \citep[PDM: ][]{Stellingwerf-1978, Dupuy-1985}, and Flux Independent and L Panchromatic Period method \citep[PK and PL: ][]{FerreiraLopes-2018papV}. We combined these five different period estimations with our statistics to reduce the number of \textit{MIS} sources as well as to set the reliability of signal detection. A range of frequencies between $f_{min} = 2/T_{tot}$d$^{-1}$ to $f_{max} = 30$d$^{-1}$ and a frequency sampling of  $N_{freq.} = 20\times f_{max} \times T_{tot}$ were used. This frequency sampling has higher resolution than that commonly used in surveys like OGLE, Catalina, WFCAM, Gaia, as well as previous works using VVV data. However signals like EA can still be missed using this frequency grid accordingly to \citet[][]{FerreiraLopes-2018papIII}.  Indeed, a procedure adopting a lower resolution grid that then steps up to higher resolutions if a sufficiently good quality period is not found may improve processing time. However, how to set the criteria to define a good quality period is an open question. For all the above, the choice of frequency sampling is a compromise between efficiency rate, signal type, and processing time.

Moreover, the best period estimation is determined by the signal-to-noise ratio. We created the phase diagram using each period estimation and with Fourier harmonic the fit was obtained. The signal-to-noise ratio was calculated by dividing the peak to peak amplitude by the standard deviation of the residue. The period with the highest signal-to-noise was determined to be the best one. Two columns related with the best period (FreqSNR) ant its signal to noise (SNRfit) are available in the table.

\begin{figure*}
  \centering
  \includegraphics[width=0.98\textwidth,height=0.35\textwidth]{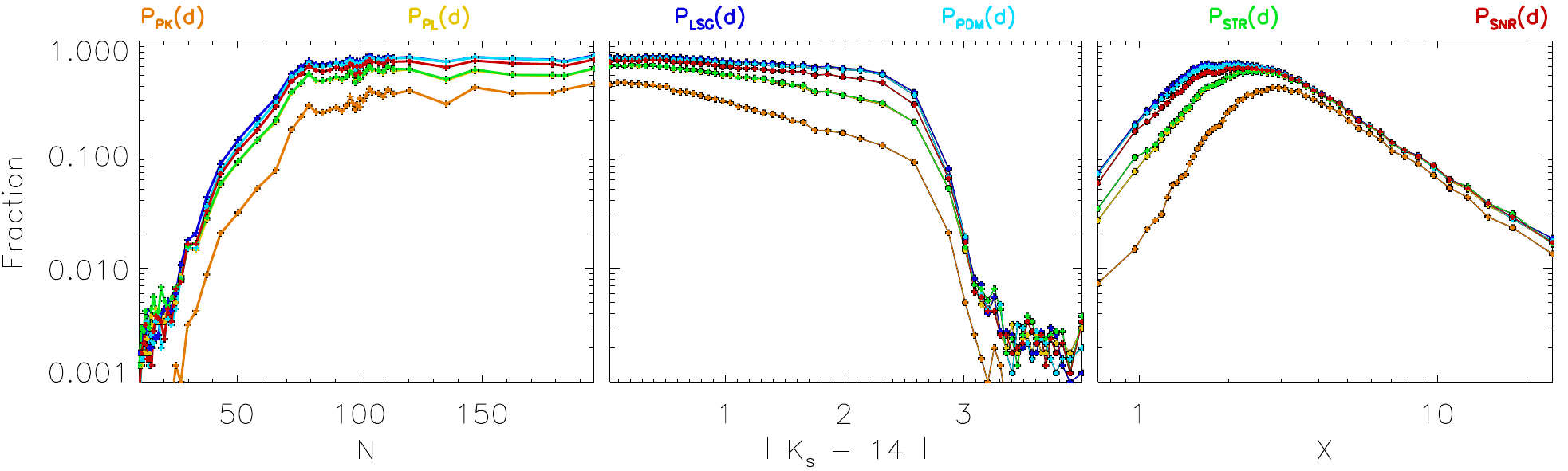} 
  \caption{The rate of agreement between our period estimations in comparison with the literature period ($P_{cr.}$) as a function of the number of observations (left panel), $K_s$ magnitude (middle panel), and X variability index (right panel). The results considering each period estimation method are shown by different colours identified by the key at the top.}  
  \label{fig_periodcross}
\end{figure*}

Crossmatched sources having previous estimations of variability periods from independent groups, and usually with independent data, were used to check our results. Three considerations must be kept in mind when performing an accurate analysis of the crossmatched periods: \textit{i)} typos or incorrect variability periods found in the literature; \textit{ii)} the signal to noise also depends on telescope and observing strategy, whereas amplitude is mainly dependent on wavelength usually varies for different wavelengths and hence the detection of a signal can be difficult if the signal to noise in the $K_s$ waveband is very small; \textit{iii)} the data quality, number of measurements, and arrangement of observations can hinder the signal detection. Figure \ref{fig_periodcross} shows the the rate of agreement between the periods determined in this work with the literature as function of number of observations, $K_s$ magnitude, and the X-index. Each data point was computed using five thousand sources of the VIVA catalog. The main remarks are summarized below;

\begin{figure*}
  \centering

  \includegraphics[width=0.33\textwidth,height=0.3\textwidth]{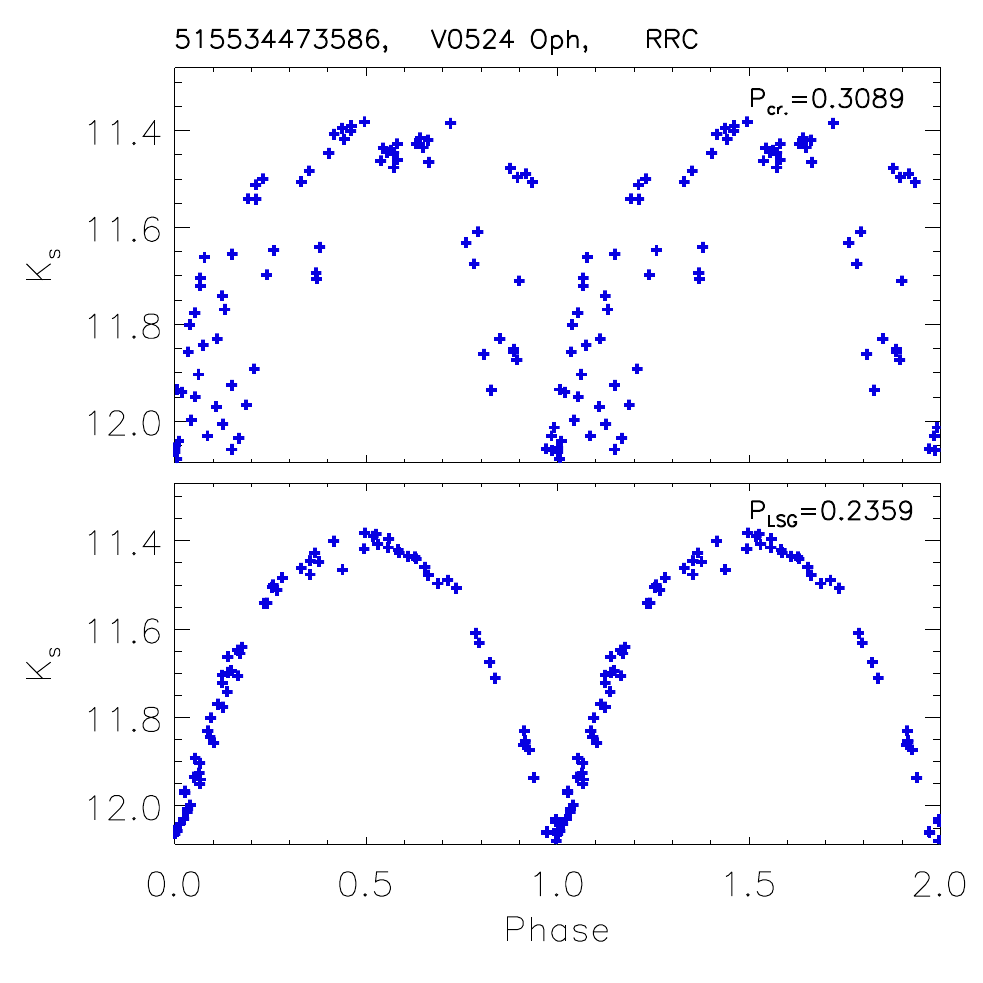} 
  \includegraphics[width=0.33\textwidth,height=0.3\textwidth]{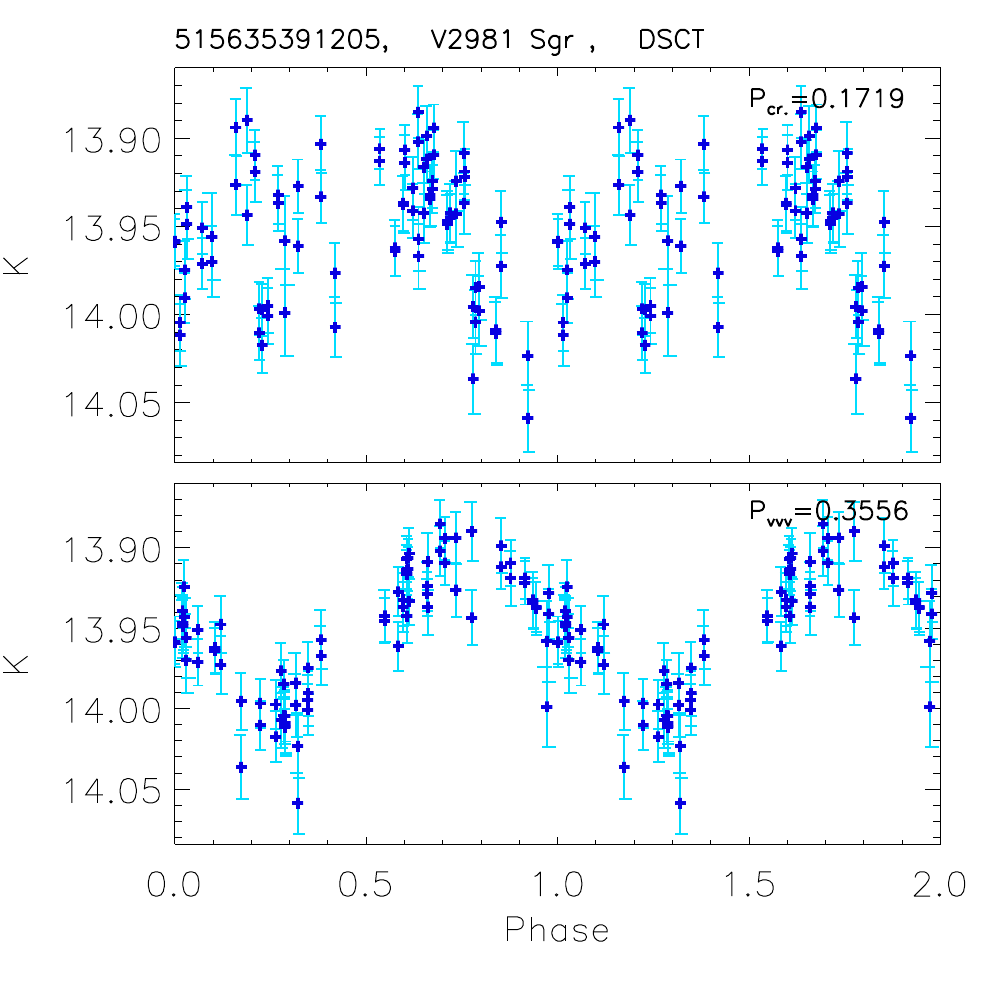} 
  \includegraphics[width=0.33\textwidth,height=0.3\textwidth]{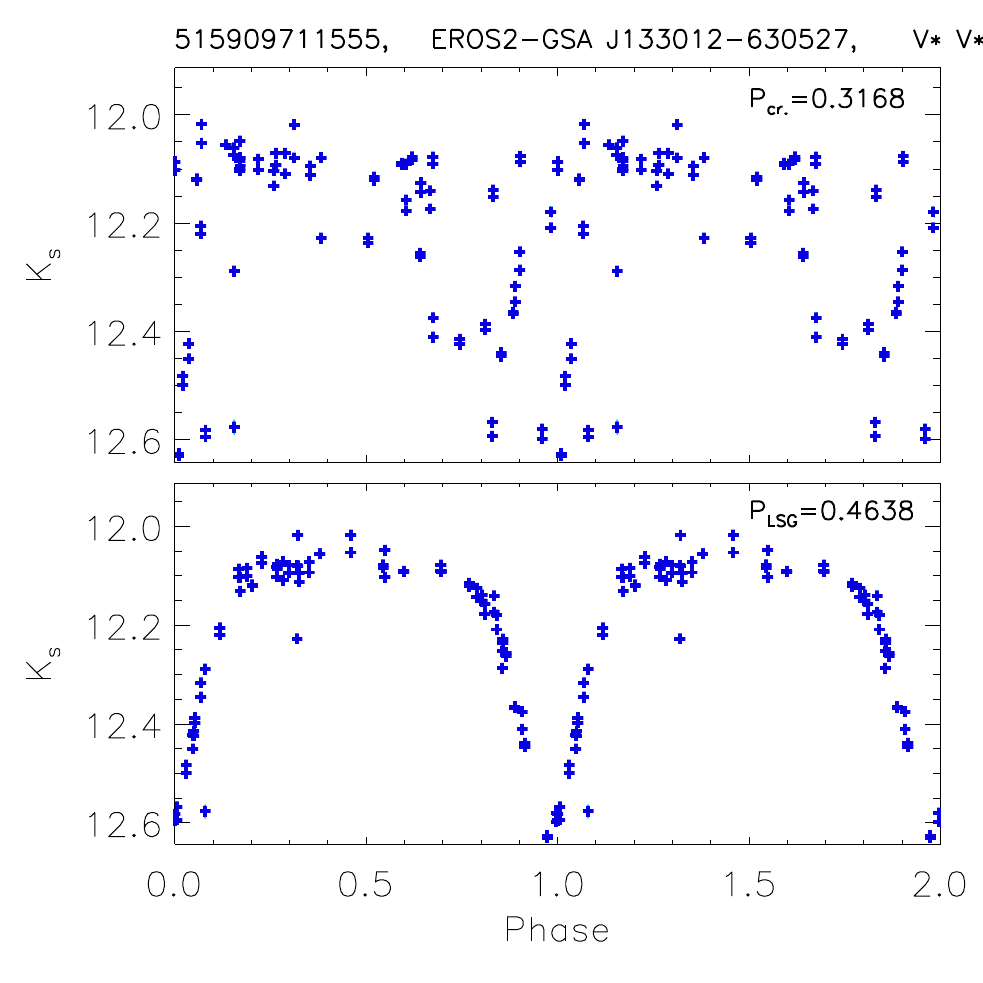} 

  \includegraphics[width=0.33\textwidth,height=0.3\textwidth]{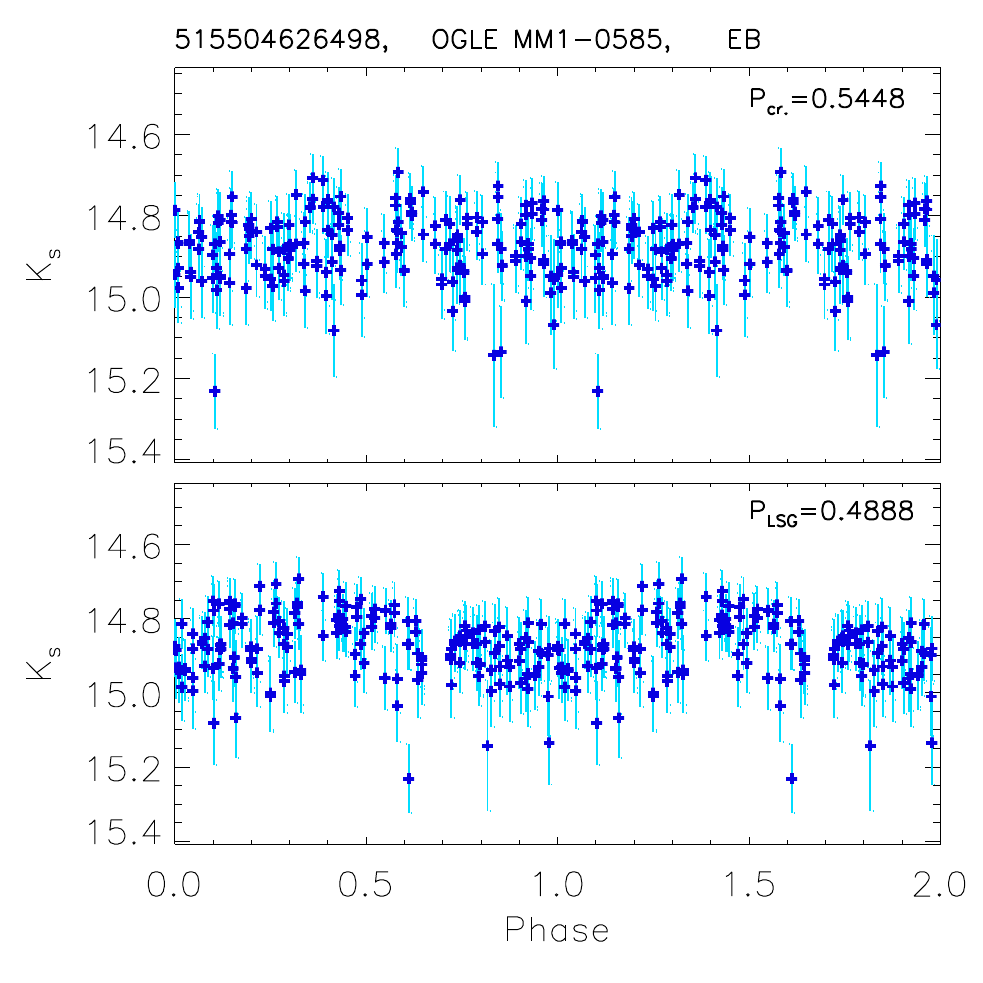} 
  \includegraphics[width=0.33\textwidth,height=0.3\textwidth]{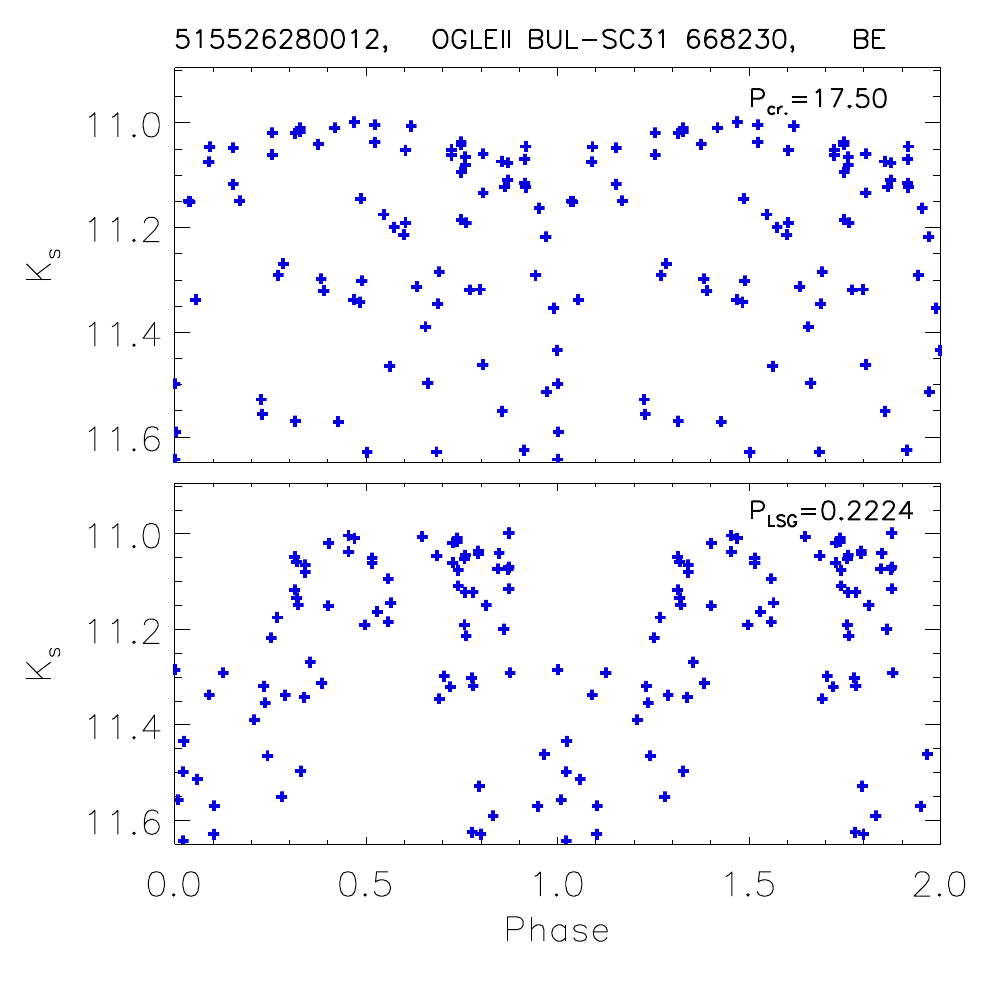} 
  \includegraphics[width=0.33\textwidth,height=0.3\textwidth]{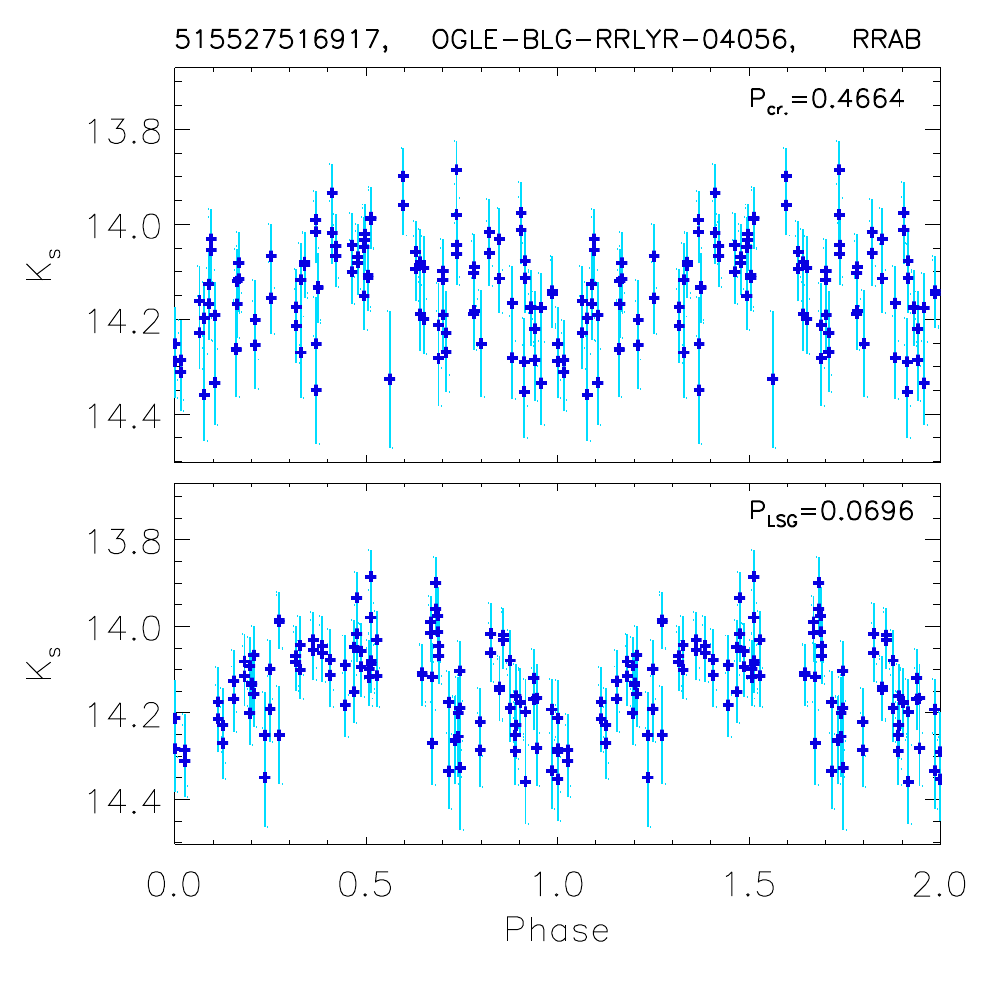} 

  \includegraphics[width=0.33\textwidth,height=0.3\textwidth]{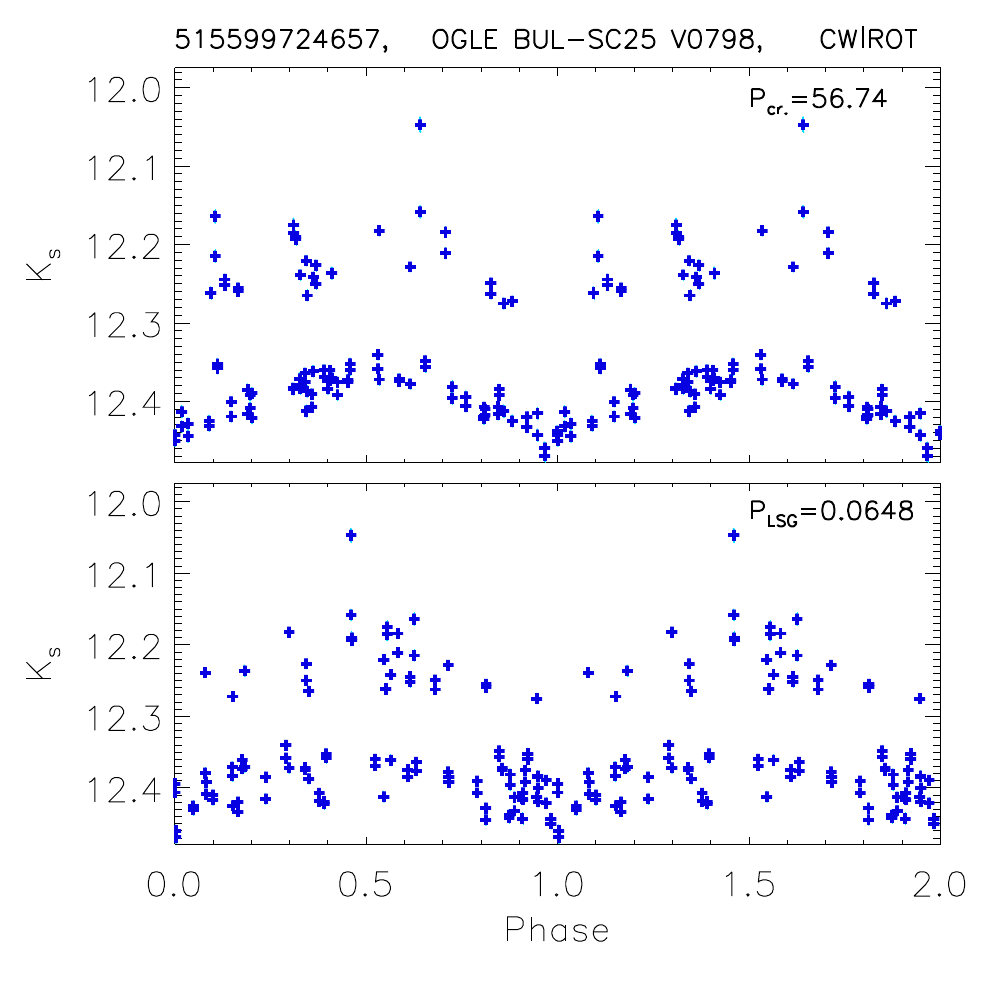} 
  \includegraphics[width=0.33\textwidth,height=0.3\textwidth]{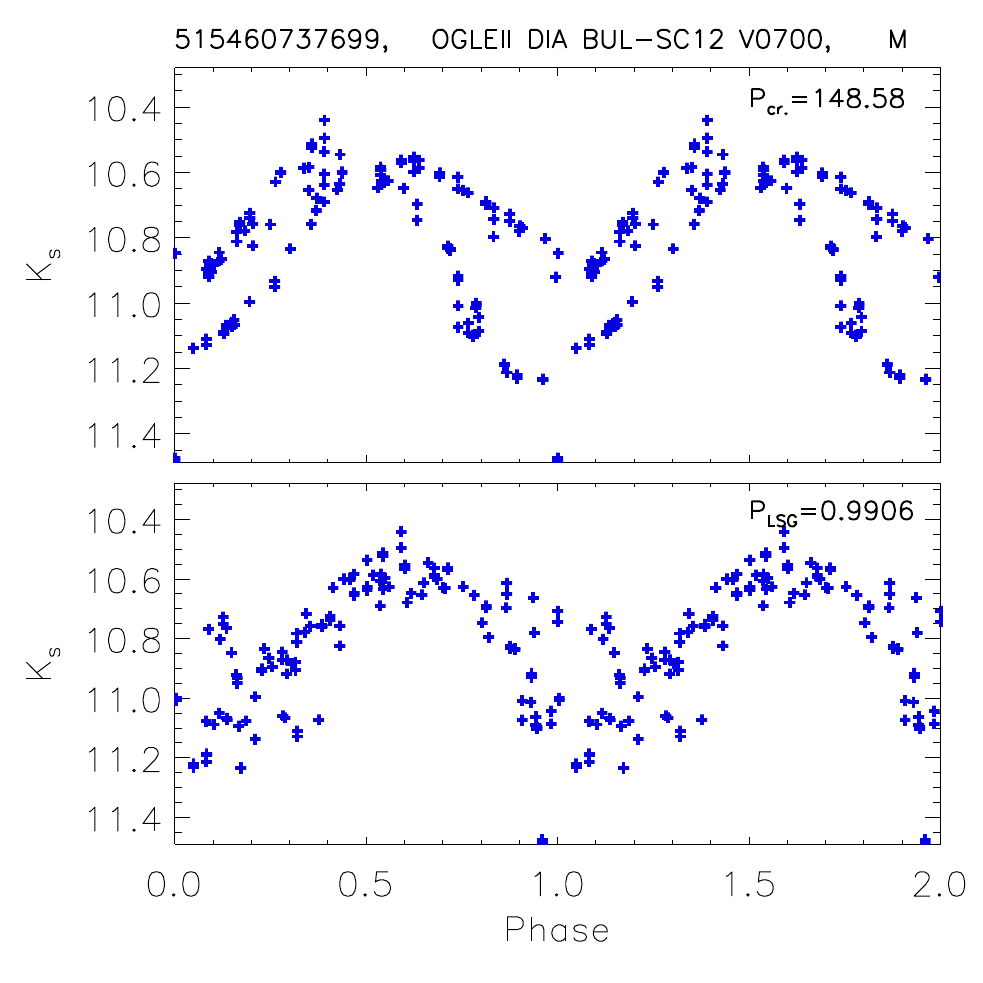} 
  \includegraphics[width=0.33\textwidth,height=0.3\textwidth]{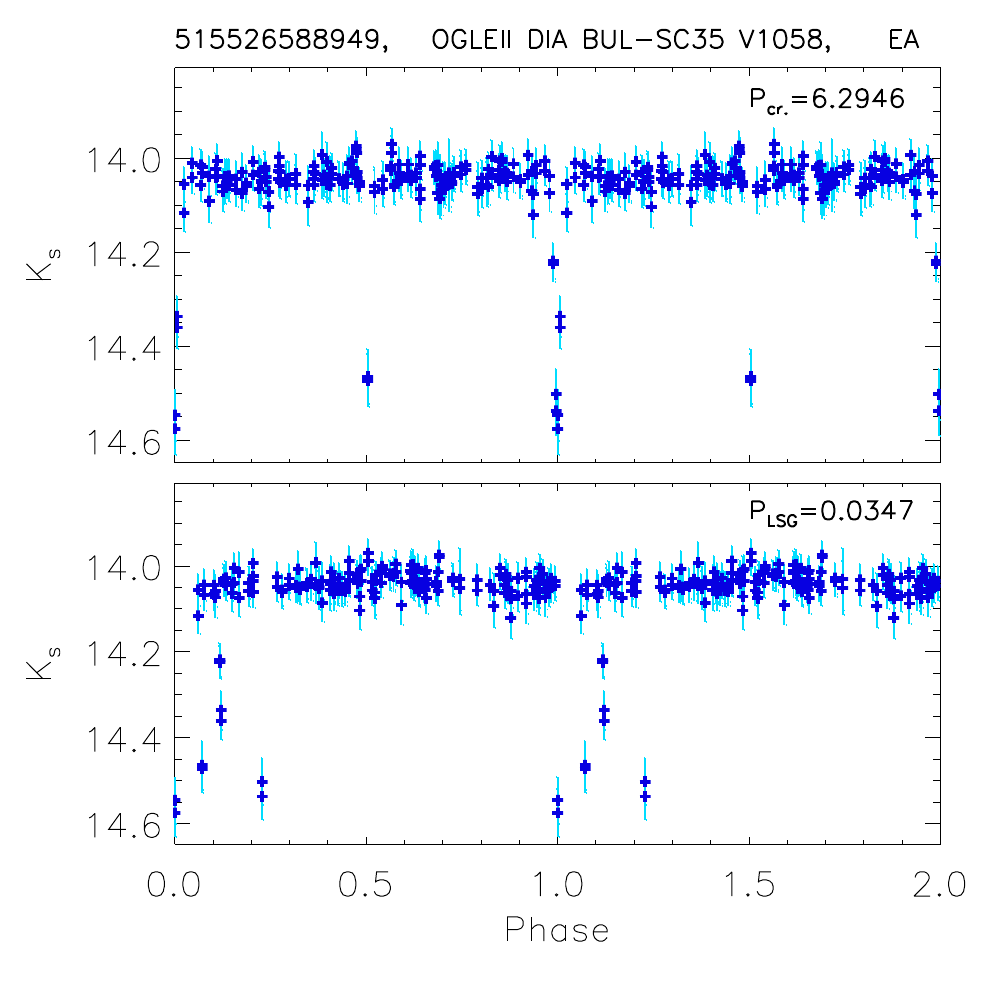} 
  
  \caption{Phase diagrams of crossmatched sources. In each set of panels we show the phase diagrams created using $P_{cr.}$ (upper panel) and with our period (lower panel). The periods used are displayed in the upper right corner of each panel while the VVV-ID, name, and variability types are in the title. }
  \label{fig_lcscross}
\end{figure*}

\begin{itemize}
  \item The yield rates for $P_{LSG}$ and $P_{PDM}$ are the highest and similar to each other. $P_{PL}$ and $P_{STR}$ are slightly lower, but not too dissimilar. On the other hand, a lower yield rate is found for $P_{PK}$. The $P_{SNR}$ has a rate of agreement slightly lower than that found by $P_{LSG}$ and $P_{PDM}$. 
  
  \item  The $P_{cr.}$ found in the crossmatched tables are often truncated, only providing a smaller number of decimal places than those presented in this work. The large majority of these periods were from the VSX table and hence the original works of these results can have a better estimation.

  \item None of the parameters used as a reference leads to a yield rate of $100\%$, The highest yield rate is found when a magnitude selection is considered ($72\%$). This means that a clean sample cannot be achieved using any single parameter alone. 
  
  \item The  literature periods in disagreement with those computed  in this work are mainly those related with semi-regular variables and eclipsing binaries. Eclipsing binaries and semi-regular variables are strongly dependent on the number of measurements and signal-to-noise 
  ratio since these sources can have low amplitudes and  the statistical significance of all variability sources  depends on these parameters. In particular, eclipsing binaries having a small phase range in eclipse are easily missed with a few measurements (see bottom right panel of Fig. \ref{fig_lcscross}). On the other hand, the rate of agreement for the RR stars can achieve $\sim92\%$ if the X index is take in account.

  \item  About $39\%$ of detected periods are harmonics or aliases of $P_{cr.}$. These peculiarities must be taken into account when classifying the variables.
    
  \item Seasonal periods are more likely to be selected using the  LSG, PDM, and STR methods. On the other hand, the $P_{PK}$ and $P_{PL}$ do not show strong lines related with seasonal variations but they show more sources related with higher harmonics of $P_{cr.}$. Moreover, some parallel lines that do not correspond to harmonics also appear when the periods are compared.   
\end{itemize}

The rate of agreement depends of the number of observations, magnitude, variability indices, among other factors. Therefore, we visually inspected the phase diagrams folded with $P_{cr.}$ as well as those periods estimated by us in order to understand the differences.  Our conclusions are based on a quick visualisation of sources having more than 30 measurements. Three main groups can be found when the estimations of variability periods are different (see Fig. \ref{fig_lcscross}), such as;

\begin{itemize}
  \item (Upper panels of Fig. \ref{fig_lcscross}) - $P_{cr.}$ is not accurately estimated or the corresponding variation is not found in the VVV-$K_s$ data.  Indeed, sources that change their period over time can provide different results for different epochs. However, if these sources are not changing their periods, this result indicates that $P_{cr.}$ is wrong since the period estimated by us provides a smooth phase diagram. On the other hand, a second possibility although unlikely, is that the variations observed in the $K_s$ band may be different to those ones observed in other bands. The third possibility is that the available $P_{cr.}$ is not accurate enough to return smooth phase diagrams. In this case, both estimations may be correct or they may be harmonics of the main period. 

  \item (Middle panels of Fig. \ref{fig_lcscross}) - Neither the folded phase diagram with $P_{cr.}$ nor that using our period estimate are smooth. The phase diagram folded with our periods seems smoother than those found by $P_{cr.}$ for a large number of sources. These types of objects are the vast majority of not-matching crossmatched periods. Indeed, we are using aperture photometry and hence nearby stars, diffraction spikes and other biases related with crowded regions may affect the measurements.  
  
  \item (Lower panels of Fig. \ref{fig_lcscross}) - The period estimated by us is wrong or it is not in agreement with $P_{cr.}$. The arrangement of measurements found for the periods estimated resemble a smooth phase diagram but they are related with seasonal variations.  Variations on zero point calibration also can cause such variations. Indeed, such cases correspond to a small fraction of crossmatched sources. This highlights the importance to check other information besides the folded phase diagrams to determine the true variability periods in order to return a reliable classification.
\end{itemize}

\begin{figure*}
  \centering

  \includegraphics[width=0.33\textwidth,height=0.3\textwidth]{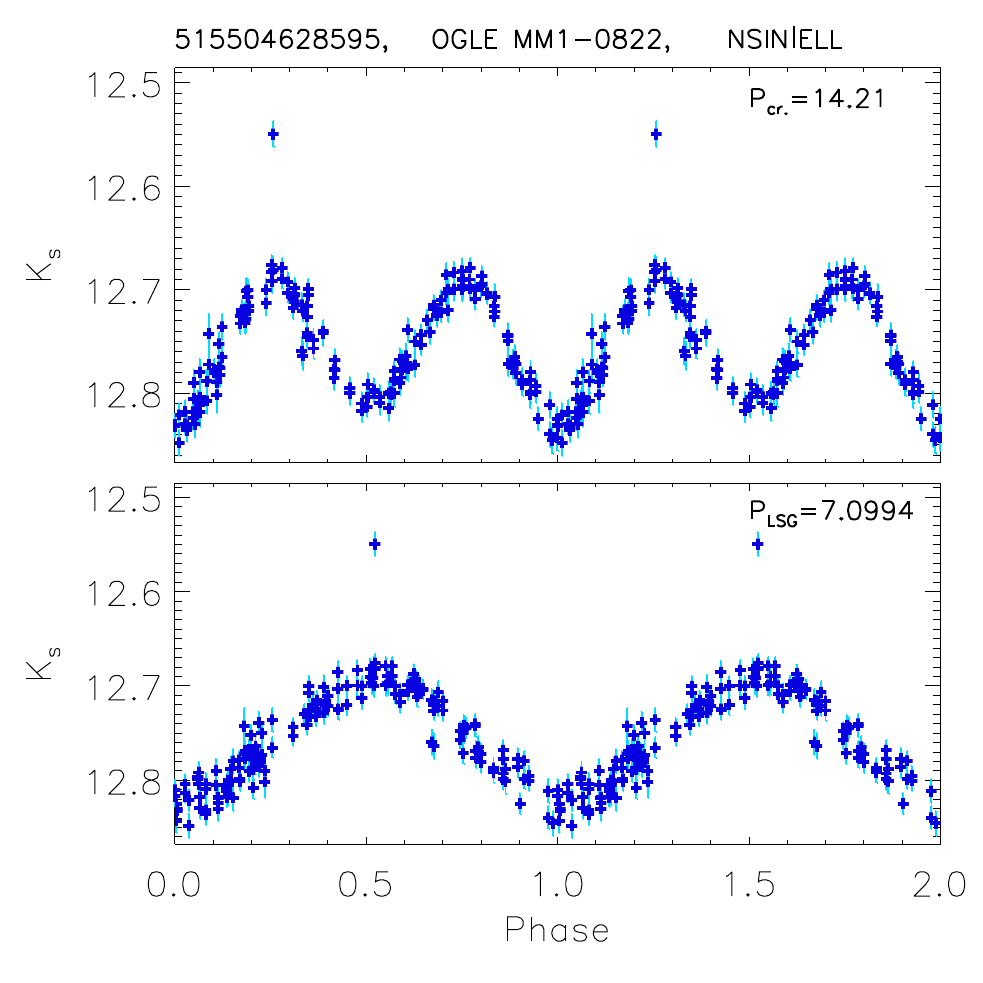} 
  \includegraphics[width=0.33\textwidth,height=0.3\textwidth]{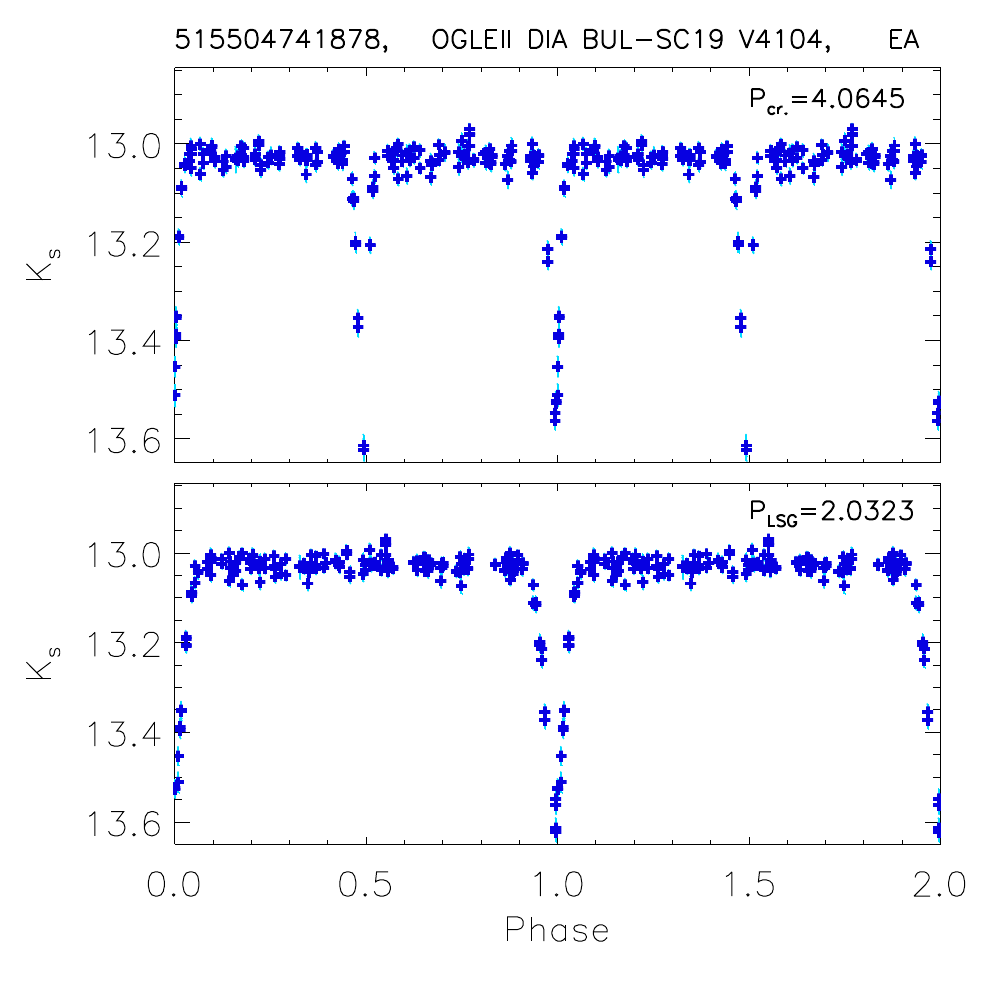} 
  \includegraphics[width=0.33\textwidth,height=0.3\textwidth]{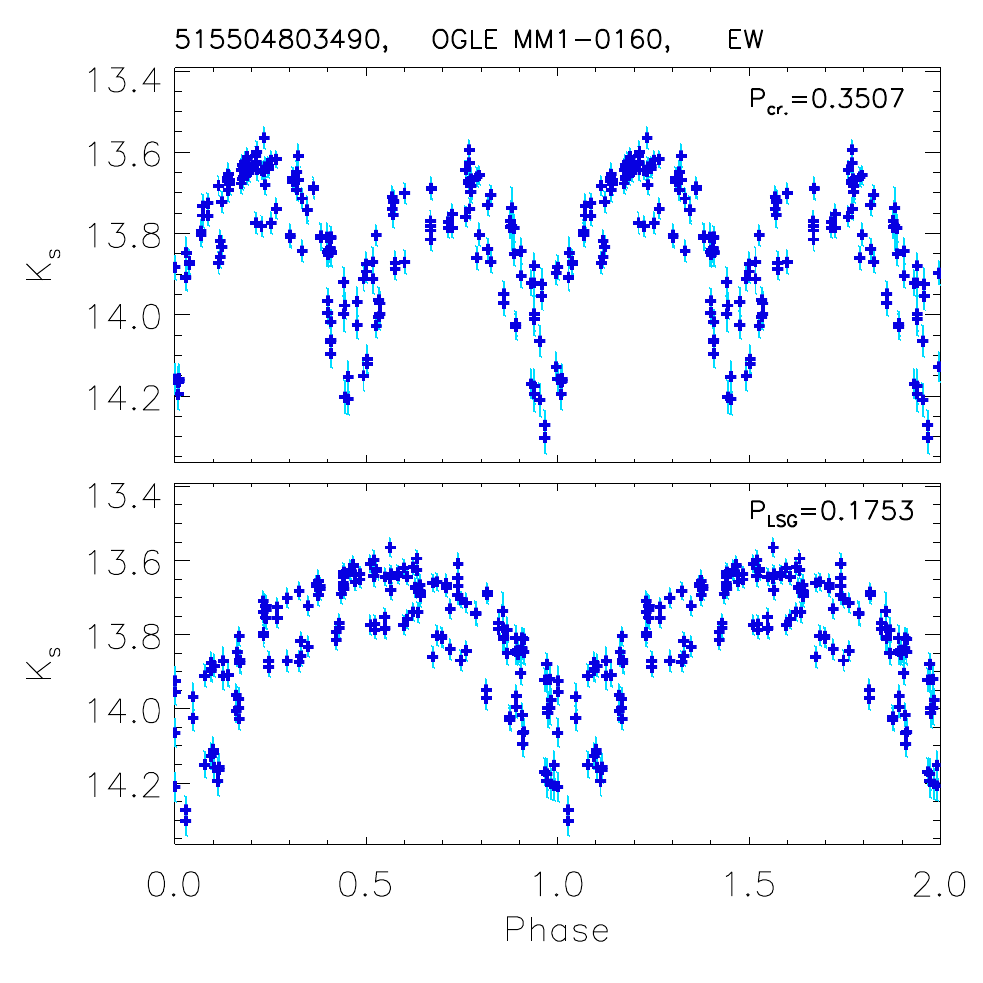} 

  \includegraphics[width=0.33\textwidth,height=0.3\textwidth]{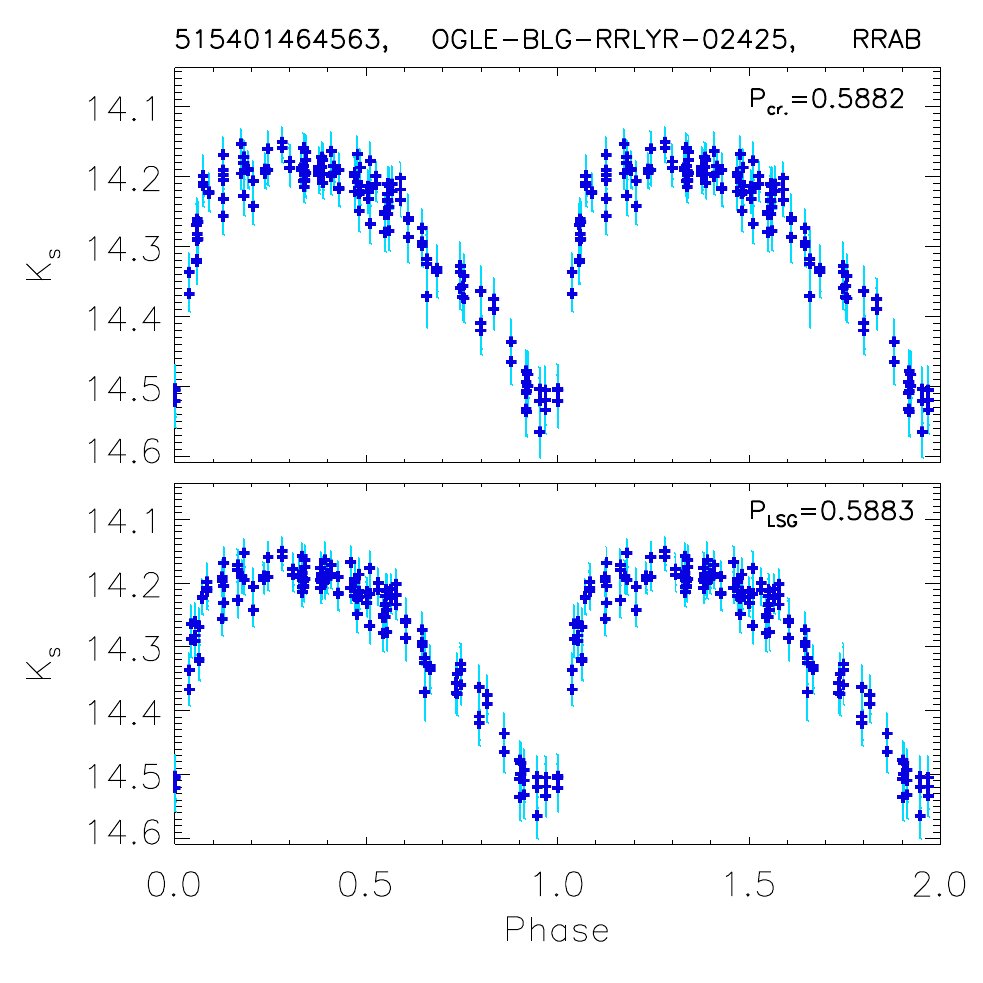} 
  \includegraphics[width=0.33\textwidth,height=0.3\textwidth]{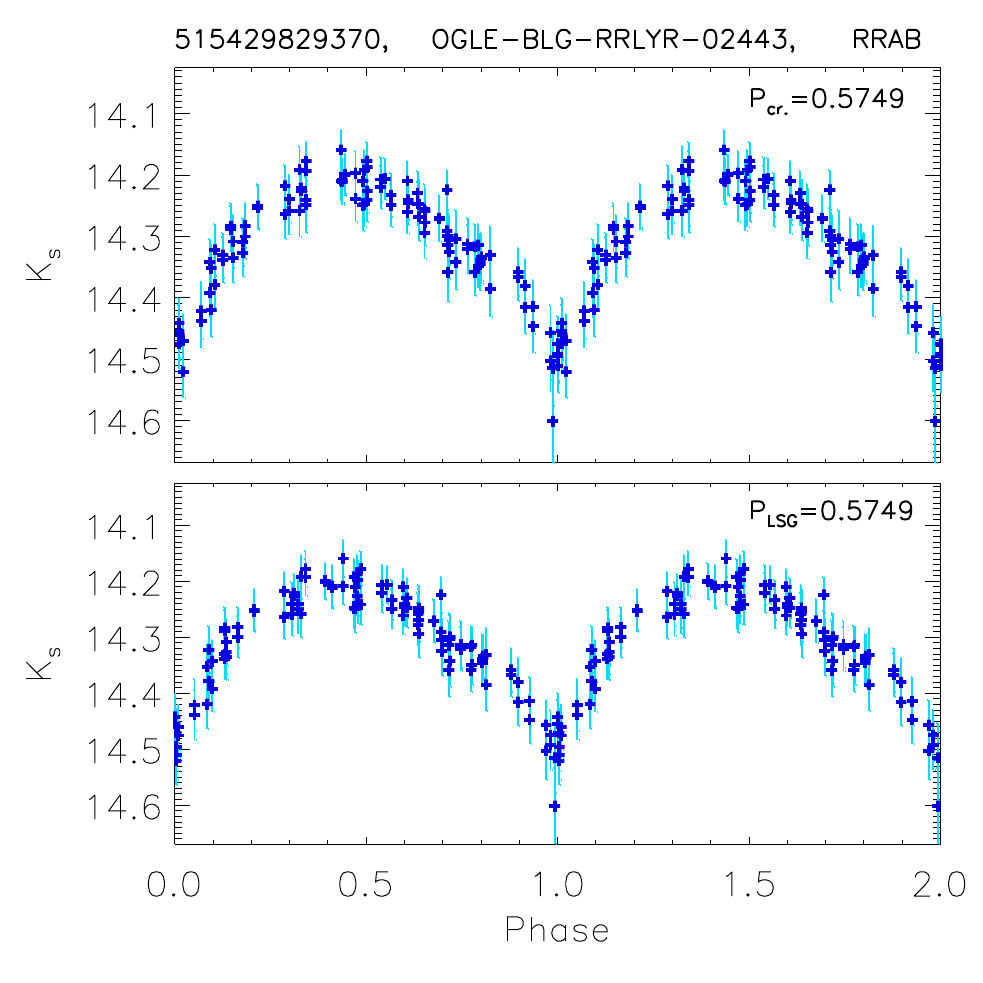} 
  \includegraphics[width=0.33\textwidth,height=0.3\textwidth]{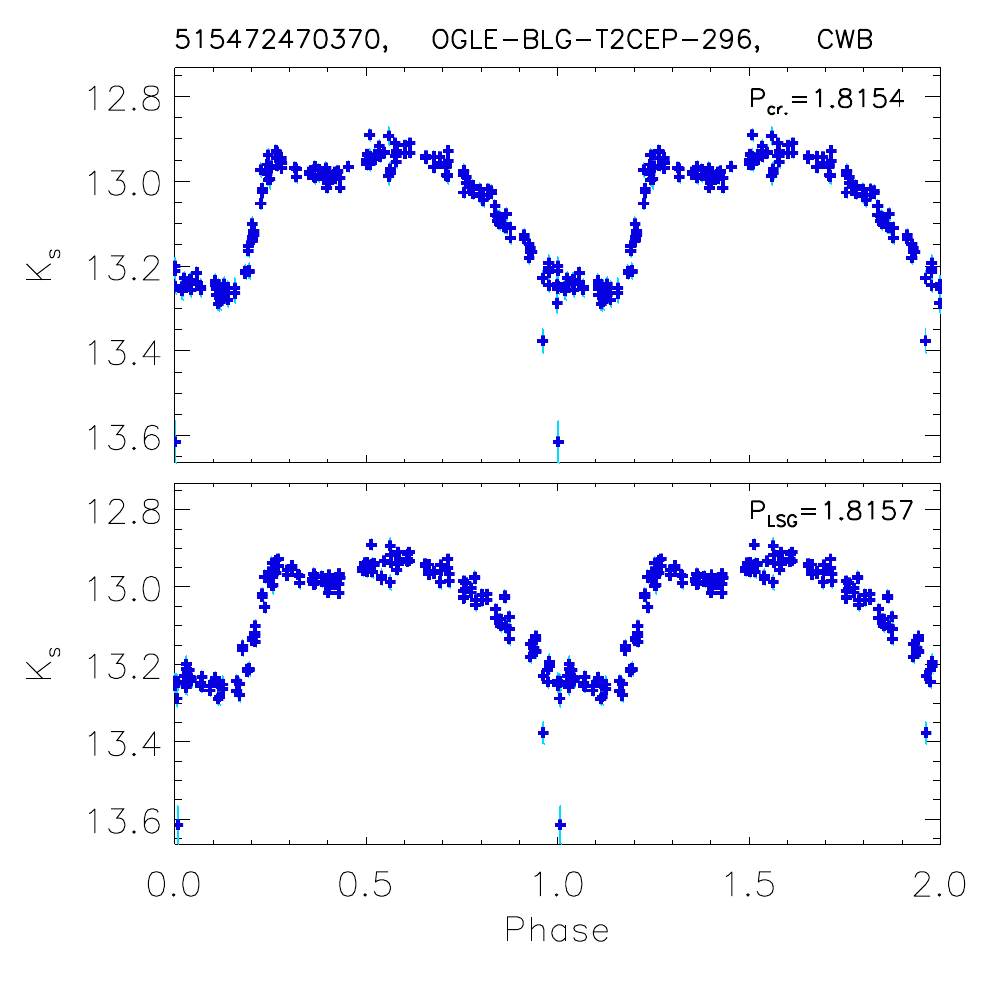} 

  \includegraphics[width=0.33\textwidth,height=0.3\textwidth]{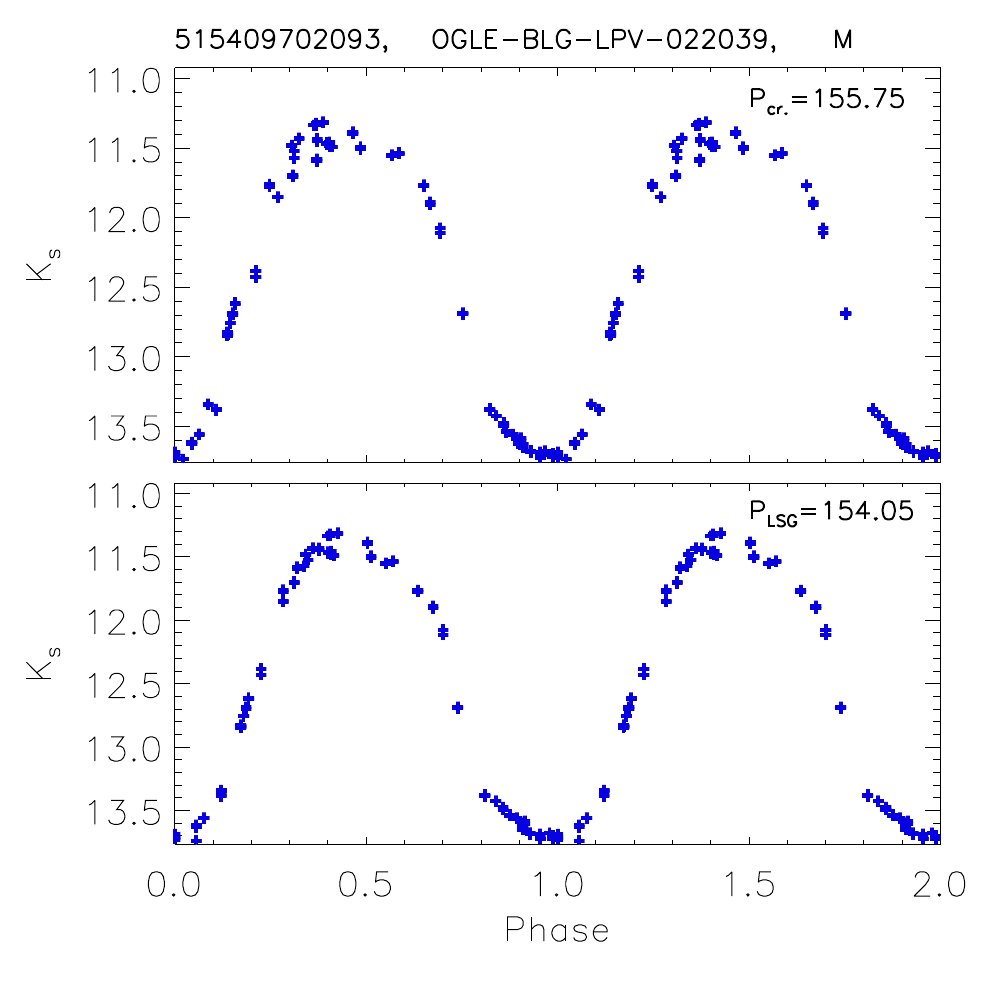} 
  \includegraphics[width=0.33\textwidth,height=0.3\textwidth]{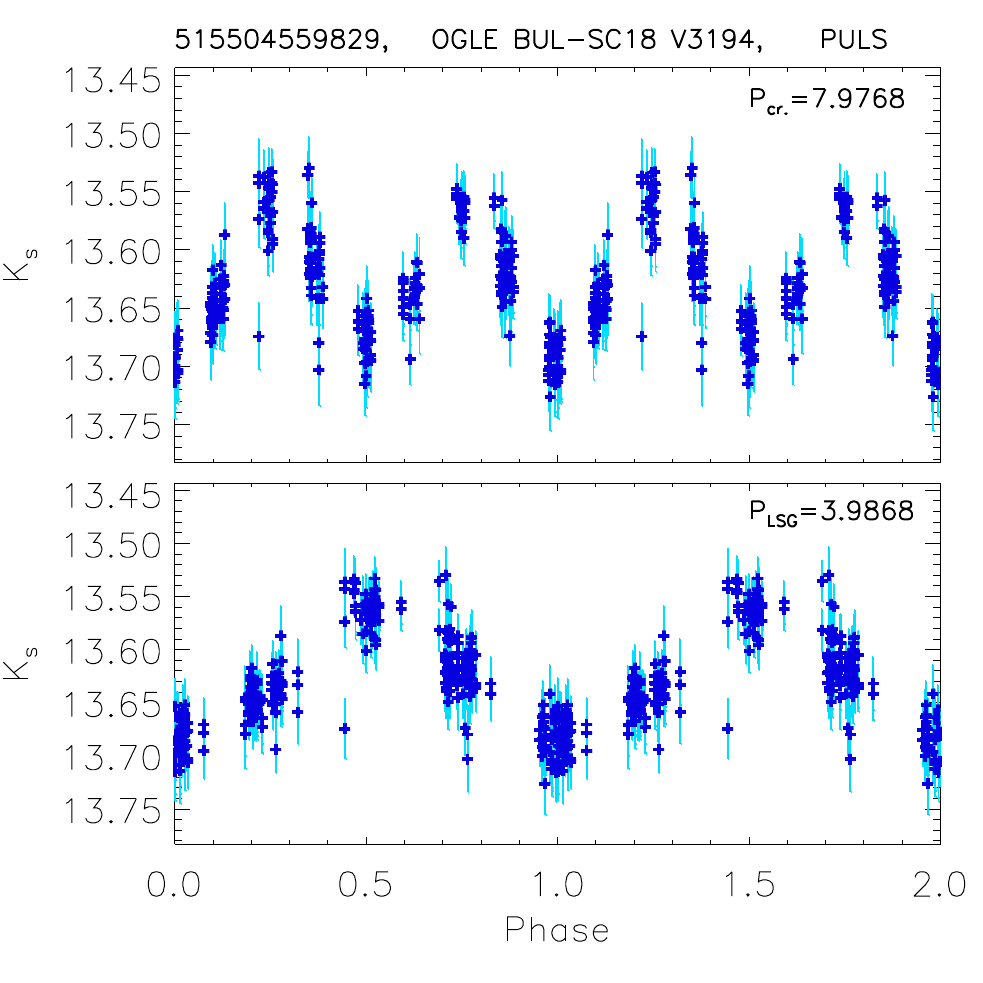} 
  \includegraphics[width=0.33\textwidth,height=0.3\textwidth]{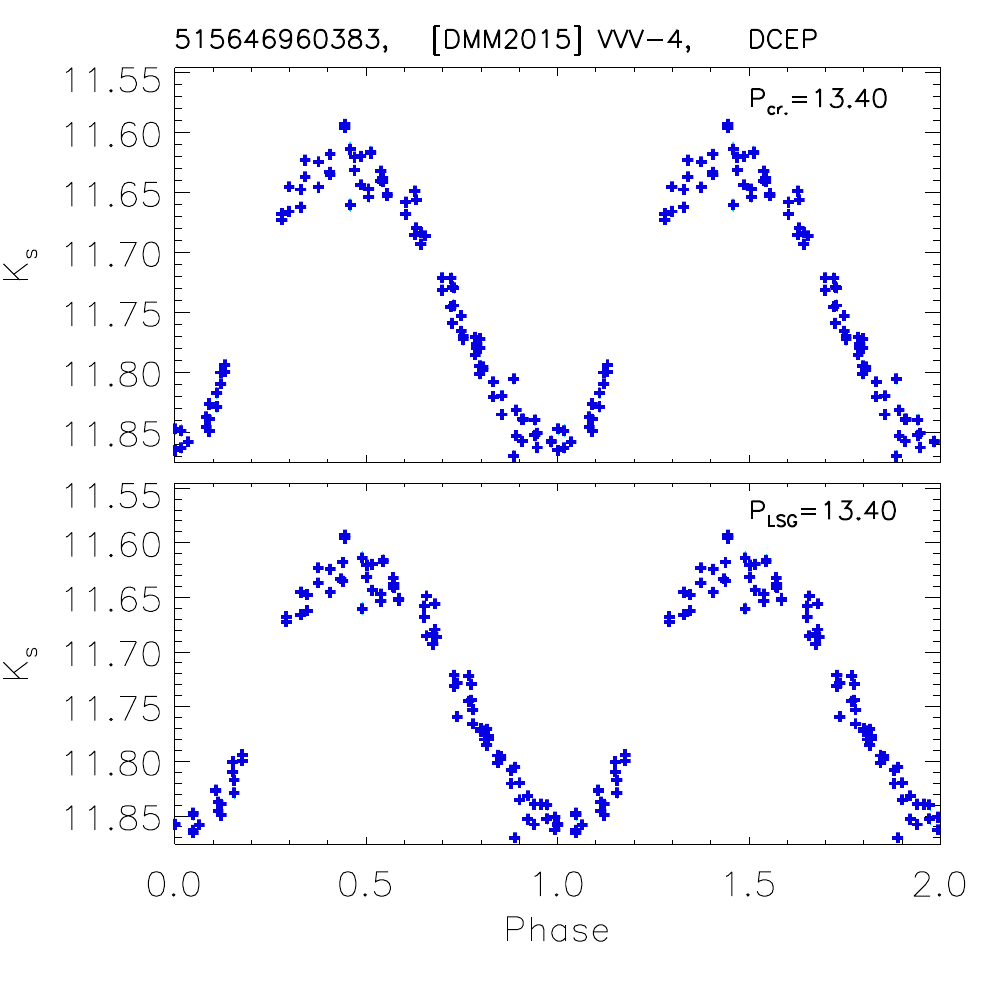} 
    
  \includegraphics[width=0.33\textwidth,height=0.3\textwidth]{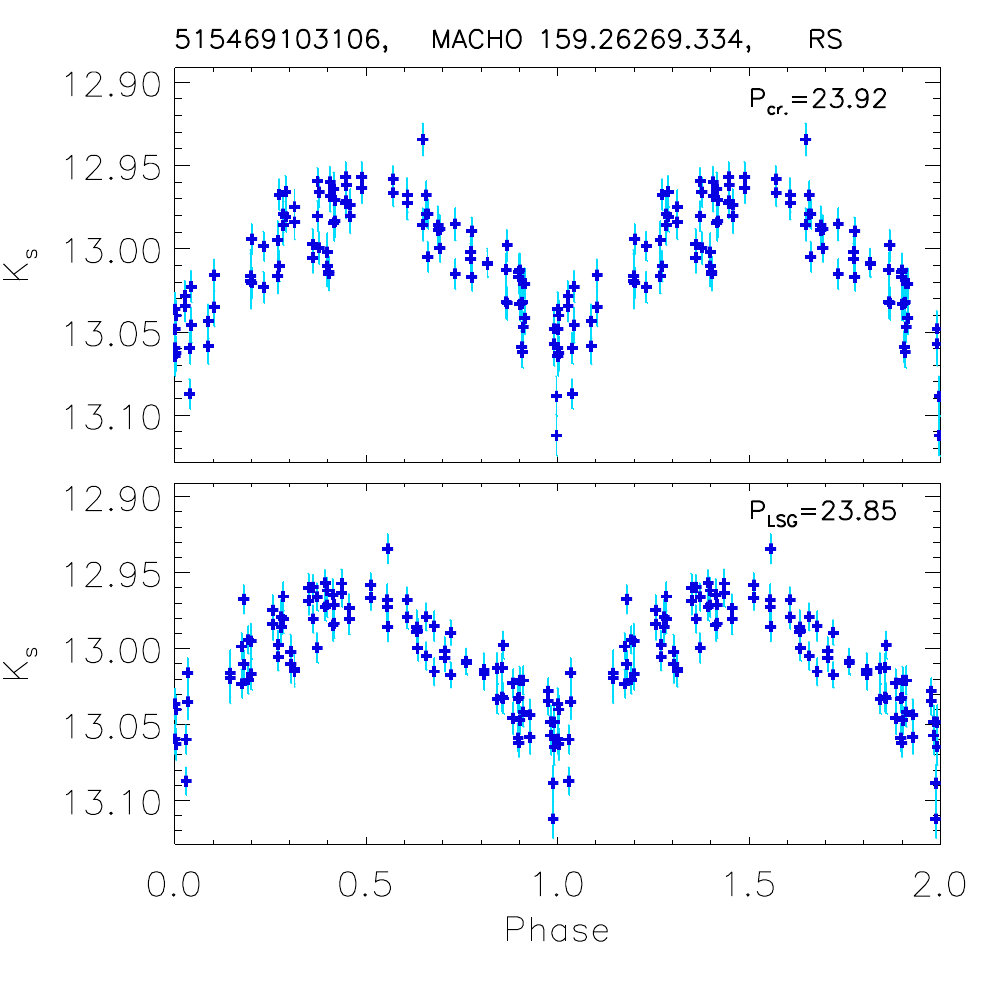} 
  \includegraphics[width=0.33\textwidth,height=0.3\textwidth]{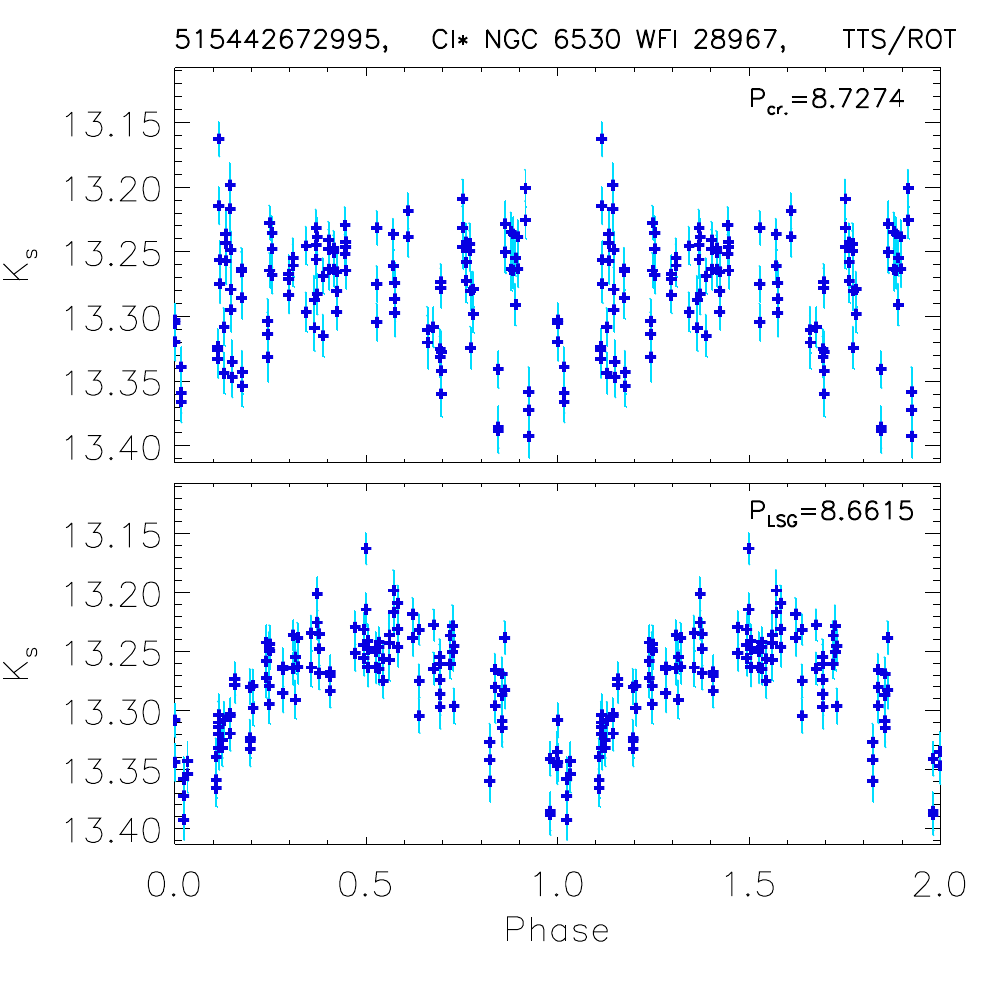} 
  \includegraphics[width=0.33\textwidth,height=0.3\textwidth]{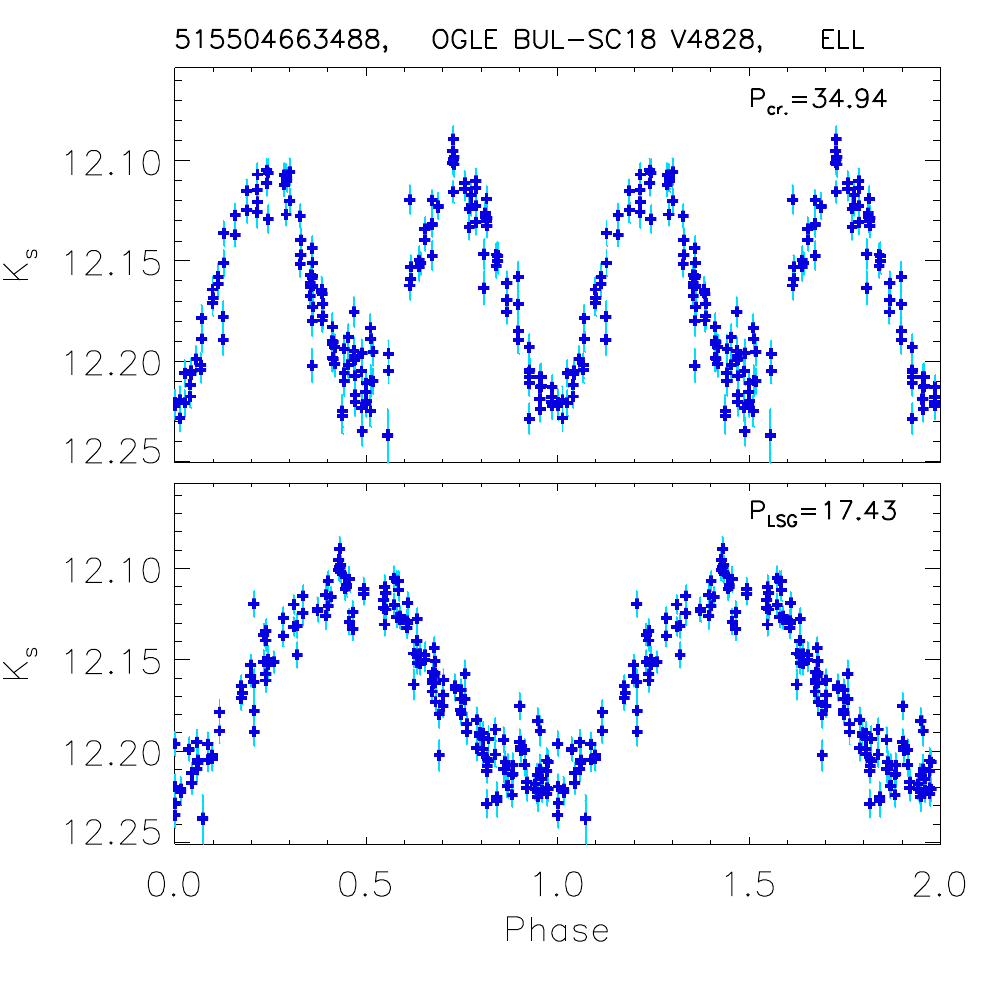} 

  \caption{Typical light curves of \textit{VVV-CVSC} catalog. The phase diagram using the variability periods found in the literature and those estimated in this work are displayed in the upper and lower panels of each plot, respectively.}
  \label{fig_lcscrossbest}
\end{figure*}

The periods estimated by us usually provide equally smooth or even smoother phase diagrams than those found for $P_{cr.}$ when these periods are in agreement (see Fig. \ref{fig_lcscrossbest}). However, our periods can be related with the first harmonic of the true variability period. For instance, the top line of panels shows eclipsing binaries where the OGLE periods are twice those computed by us. The constraints used to determine if the period is double for eclipsing binaries were not considered since a detailed analysis of the symmetry of the eclipses in comparison with pulsating stars is required. Indeed, there are several types of light curves that are very difficult to distinguish: contact binaries with ellipsoidal variations (low inclination eclipsing binaries) and RRc Lyr. Therefore, more information is needed because they shared the same range of periods, amplitudes, shape, and so on. Indeed, sometimes even with visual inspection it is very difficult to determine the variability type if more information is not added. For all these reasons, the harmonics or overtones of the period computed by us were not checked. For instance, the periods of variable stars reported by the Catalina survey where checked and as a result we observe that about $50\%$ of them  are double that found at the highest periodogram peak (Ferreira Lopes et al. submitted). Therefore, a similar or higher rate of matches could be expected in the current catalog since the amplitude and number of observations is smaller that that found in the Catalina data.

\begin{figure*}
  \centering
  \includegraphics[width=0.95\textwidth,height=0.4\textwidth]{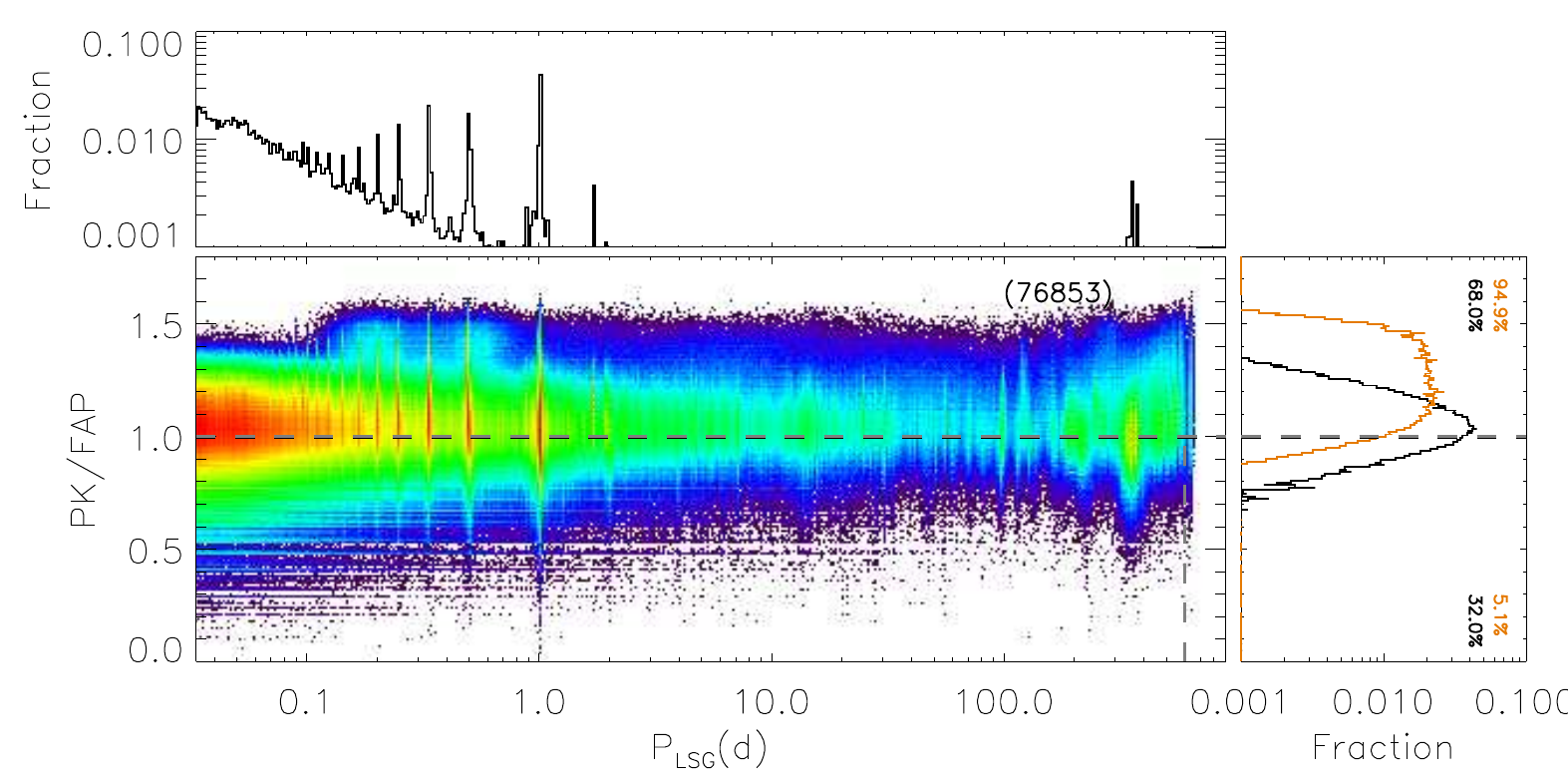} 
  \caption{Relative density plots of the ratio of PK power versus FAP as function of the variability period $P_{LSG}$ in the centre panel. The histograms corresponding to x and y axis are shown at upper and right sides. The histogram corresponding to the crossed sources having matching periods is shown in orange in the right panel. The maximum number of sources per pixel is shown at the top-right of the central panel. The aliasing periods are easily seen at the histogram peaks. On the other hand, the long timescale period cut off is set by the total time span of our observations.}
    \label{fig_periodalias}
\end{figure*}

In summary, the period estimation in this work provides an independent method to check previous estimates, to study the corresponding variations in multi-wavelength data, and to search for new variable stars. The crossmatched sample only corresponds to $\sim 0.5\%$ of the \textit{VVV-CVSC} catalog, i.e. the $\sim99.5\%$ sources of our sample constitutes a number of potentially new objects with variability information for the heavily crowded and reddened regions of the Galactic Plane.

\subsection{Main variability periods}\label{sec_mainperiod}

The main variability period estimated for the five methods are available in the release and hence the user can adopt the one that fits best for his/her purpose. For instance, the STR method is more suitable than other methods for detecting eclipsing binaries since it has the highest yield rate for these kinds of objects. On the other hand, when all variables star types are considered, LSG and PDM method provide better results \citep[e.g.][]{FerreiraLopes-2018papIII}. Indeed, our results also confirm that the highest yield rates are found for LSG and PDM methods (see Sect. \ref{sec_periods}). In order to facilitate the forthcoming discussions, we adopt as the main variability period the one estimated by the LSG method ($P_{LSG}$). In fact, the reliability of the detected signal should be higher when all methods are in agreement.

The period power spectrum heights (PPSH - here labelled just as $H_{method}$, e.g. $H_{LSG}$), found by the five methods can vary with the number of measurements, error bars, and amplitude. In particular, the $PK$ period finding method was designed from the $K_{(fi)}$ index and hence they will have similar properties, i.e weak dependence on the instrumental properties and outliers.  Therefore, $PK$ was chosen to test  the reliability of the signals, which is one of our main concerns.

Figure \ref{fig_periodalias} shows $PK/FAP$ ratio as a function of the main variability period $P_{LSG}$.  The vertical lines found in this diagram are related to seasonal variations, i.e. $1/M$ for all $M \geq 1$ ($1d$, $0.5d$, $0.33d$, $0.25d$, $\cdots$) that are usually known as \textit{"aliasing"}.  Moreover, weak lines are also present that can vary from one tile pointing to another. For instance, the long periods of hundreds of days, i.e. $375.35706$, $333.56015$, $345.76831$, $238.91038$, $193.00734$, $98.301643$ among others are also present in this diagram but they are more evident when the results on each VVV tile are compared individually.
In order to facilitate the identification of spurious periods a flag around these lines was added. We count the number of sources having similar period values with a precision of $10^{-6}$ and $10^{-7}$ in frequency space. As a result, an integer number ranging from $1$ to more than $1000$ giving the number of periods inside in a box with a width of these intervals was set as a flag, i.e. larger numbers indicates spurious periods. These parameters are useful for quality control (for more details see Sect. \ref{sec_reliablesel}). An important note, this flag is calculated in each VVV tile separately and hence the spurious periods can be slightly different from one VVV tile to another.

\subsection{Getting reliable targets}\label{sec_reliablesel}

According to \citet[][]{FerreiraLopes-2016papI} the sample selected using $K_{(fi)}$ returns a contamination ratio, understood as the number of total stars in our sample to the number of true variables, of about $12.6$ to select $\sim90\%$ of the variable stars. The reader should understand contamination rate as a combination of missselection and those ones where  the variability type can not be determined. Therefore, the number of  variable stars where period, amplitude, and light curve shape can be studied will be a fraction of the VIVA catalog. The staset and constraints used by \citet[][]{FerreiraLopes-2016papI} are different to those adopted in this work. Moreover, we are returning a complete sample and hence we must assume there is a contamination rate of at least 10. Therefore, the available parameters should be used to restrict the sample when more reliable samples are required.

Indeed, the fit to the phase diagram can be more easily found using harmonic fits \citep[e.g.][]{Debosscher-2007,DeMedeiros-2013,FerreiraLopes-2015mgiant,FerreiraLopes-2015cycles} and hence many parameters that reduce the misselection rate and are useful for classification can be obtained. Classification will be undertaken in a forthcoming paper of this series. On the other hand, a clue about the reliability of the signal is found straightforwardly from the height or power of the period found by one of the methods.  Indeed, this assumption depends on the signal type for LSG method, for example, i.e. signals mimicking sinusoidal variations have a greater height in the period power-spectrum compared to other signals with the same amplitude. The power or height is greater for light-curves that return a smoother - i.e. less scatter from a simple functional fit - phase diagram when folded on that period. Non-smoothed results such as incorrect periods or aperiodic signals return the expected height for noise. However, peculiarities of each method combined with statistical fluctuations can appear in a non-smoothed phase diagram as a good detection.

The \textit{WFSC1-ZYJHK},  \textit{WFSC1-K},  \textit{CVSC1}, and \textit{GraMi} samples were used as comparison stars (for more details see Sect. \ref{sec_cutoffs}). The period power spectrum heights (PPSH)  were computed for those comparison stars in the same way as for the \textit{VVV-CVSC} data (for more details see Sect. \ref{sec_periods}). However, the $K_{(fi)}$ is computed using multi-wavelength data in order to have correlated measurements, but the $H_{PK}$ is computed for each waveband separately since there is no requirement for correlated measurements. Figure \ref{fig_relibleperiods} shows a comparison of the PPSH for the five period finding methods. For these methods, we found that:

\begin{figure*}
  \centering
  \includegraphics[width=0.98\textwidth,height=0.45\textwidth]{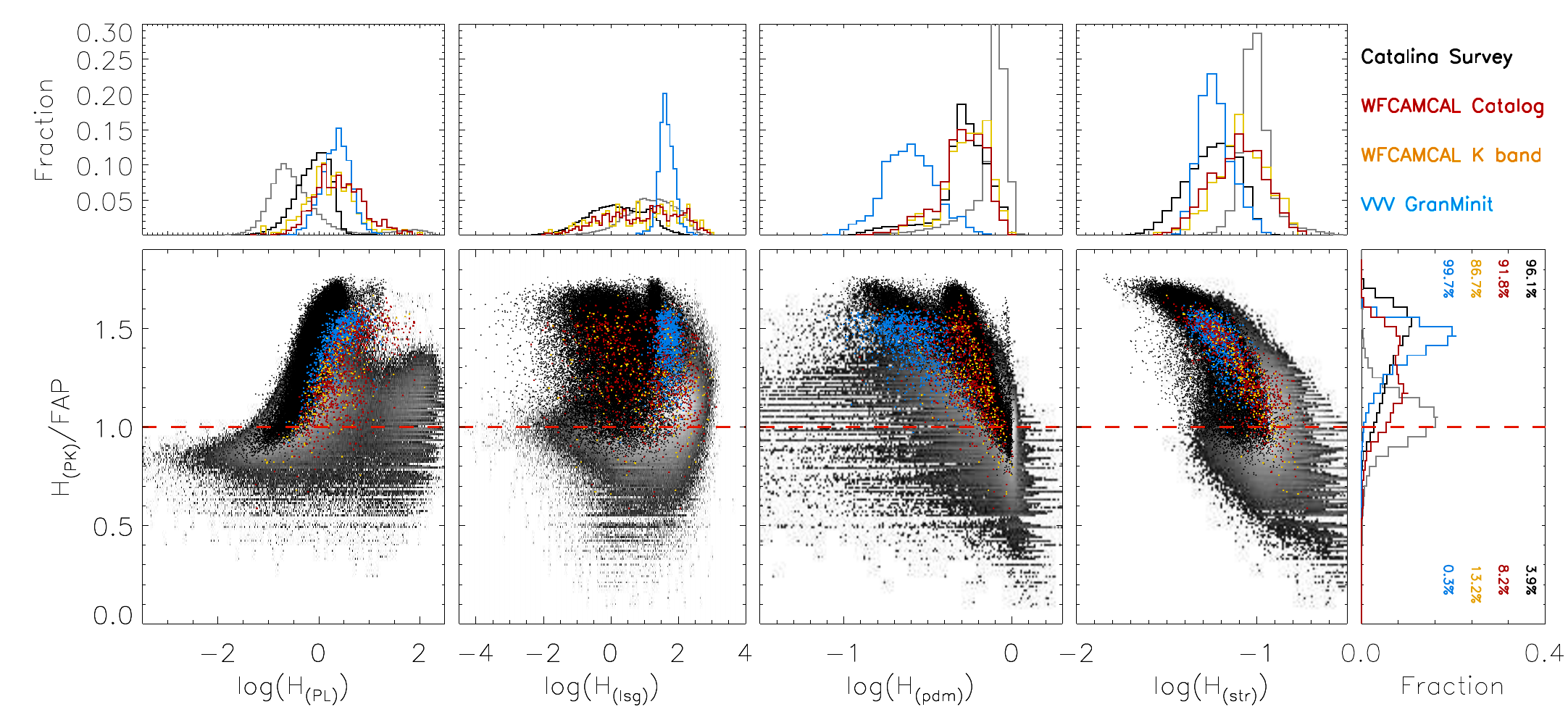} 
  
  \caption{Power spectrum height of PL, LSG, PDM and STR methods as function of ratio of PK power height by its FAP. The grey pixels show the   \textit{CD-CVSC} and \textit{NCD-CVSC} results while the comparison stars are presented in colours. The same colours used in the Fig. \ref{fig_probe} are also used here for the comparison stars.}
  \label{fig_relibleperiods}
\end{figure*}

\begin{itemize}
\item Less than $4\%$ of comparison stars belonging to \textit{CVSC1} have $H_{PK}/FAP$ smaller than 1. However, this is a larger proportion than that found for the $K_{(fi)}/FAP$ statistic. This happens because the folded light curves seem to have a lower signal to noise than those analyzed in time, i.e. cycle by cycle.

\item The same behaviour that is seen for the \textit{CVSC1} and \textit{WFSC1-ZYJHK} and \textit{WFSC1-K} samples, i.e. a higher fraction of sources having $H_{PK}/FAP<1$ than $K_{(fi)}/FAP<1$. The percentage of sources in the \textit{WFSC1-K} group are much higher than the \textit{CVSC1} sample. The reduction in the number of measurements used to compute $H_{PK}$ together with those factors discussed in the previous item are the reasons for the lower yield rate compared with  $K_{(fi)}$ index. 

\item Indeed,  $99.6\%$ of the VVV \textit{GraMi} sample are above this limit. On the other hand, the \textit{WFSC1} and its subsample in the $K_{s}$ waveband have $8.2\%$ and  $13.2\%$  with $H_{PK}/FAP < 1$,respectively. Not all \textit{WFSC1} sources where detected in all wavebands and hence the percentage of sources having $H_{PK}/FAP > 1$ should be bigger. 

\item The \textit{GraMi} and \textit{WFSC1-$K_{s}$} were observed in filters covering a similar wavelength range. Moreover, the yield rate of the \textit{GraMi} is greater than the \textit{CVSC1} sample that is observed in the optical wavelengths. The amplitude, and hence the signal-to-noise ratio, of RR Lyrae stars are usually higher than a heterogeneous sample. Therefore, a higher yield rate found for the \textit{GraMi} sample is expected.

\item The $H_{PL}$ shows a clear separation between \textit{CVSC1} in comparison with \textit{WFSC1} or \textit{GraMi} samples. The $H_{PL}$ depends on the signal amplitude and error bars. Therefore, this difference is related with the combination of higher amplitudes and smaller error bars since, on average, the optical wavelengths have smaller error bars and higher amplitudes than IR wavelengths.  

\item \textit{GraMi} sample has high $H_{LSG}$ values and they are very concentrated at $H_{LSG} \simeq 80$. For example, this happens because the morphology of RR Lyrae stars is closer to a sinusoidal signal \citep[e.g.][]{FerreiraLopes-2015wfcam} than for instance the one from eclipsing binaries. Indeed, a large fraction of the \textit{WFSC1} and  \textit{CVSC1} samples are made up of eclipsing binaries. As expected, the results of \textit{WFSC1} and \textit{CVSC1} data are more spread because they are more heterogeneous samples.

\item The \textit{CVSC1} data seem to form two connected branches in panel 3. The \textit{WFSC1} sources lie along the main branch on the right side ($\log{(H_{pdm})}>-0.5$) of the \textit{CVSC1} data while the \textit{GraMi} sources lie within the left branch of $H_{PDM}$ values. The number of sources in \textit{CVSC1} is $\sim170$ times bigger than \textit{WFSC1}. Therefore, the two branches observed in \textit{CVSC1} are not so evident in \textit{WFSC1} data.  Moreover, the large part of \textit{CVSC1} is composed of eclipsing binaries (usually having high amplitude and signal to noise) and hence the branches can be related to high and low signal to noise data since the first one minimizes the merit figure.

\item The $H_{STR}$ increases seems to have a linear variation with $H_{PK}/FAP$ values. Moreover, the peak of the distribution found for \textit{GraMi} coincides with \textit{CVSC1} despite the last one being less concentrated. The $H_{STR}$ varies with the signal-to-noise ratio and number of measurements where a larger signal-to-noise ratio and a larger number of measurements leads to a smaller $H_{STR}$ value. These aspects explain the differences found among these samples for the same reasons discussed for the other methods.
\end{itemize}

Overall, the height of the power spectrum of methods used in this work can help reduce the number of misselections.  In particular, $H_{PK}/FAP > 1$  includes about $\sim97\%$ of crossmatched sources (see Fig. \ref{fig_relibleperiods}) having crossmatched periods. Moreover, it also results in a yield rate bigger than $\sim90\%$ for  \textit{CVSC1}, \textit{WFSC1}, and \textit{GraMi} samples. These results show that $H_{PK}/FAP$ is a good indicator of the reliable signal with a single cut-off value independent of wavelength observed. Indeed, a small fraction of variable stars will be missed if only one of these methods is used. Hence, the selection criteria can be improved if different methods are combined. Moreover, the results of different methods can be combined to improve the selection criteria. For instance, the furthest left and furthest right panels of Fig. \ref{fig_relibleperiods} have some regions that do not contain reliable signals.  The height for the main period detected by each method is available in the released table where the user can select them as desired.

The flags associated with the variability period and the estimation of the amplitude can help to locate the values above which reliable signals can be found. We use the crossmatched sources having matched periods, named as \textit{VVV-CVSC*}, to analyze these parameters. This consideration ensures that the signal was detected in IR light curves. We discuss how to use these flags to select targets below;

\begin{figure}
  \centering
  \includegraphics[width=0.23\textwidth,height=0.25\textwidth]{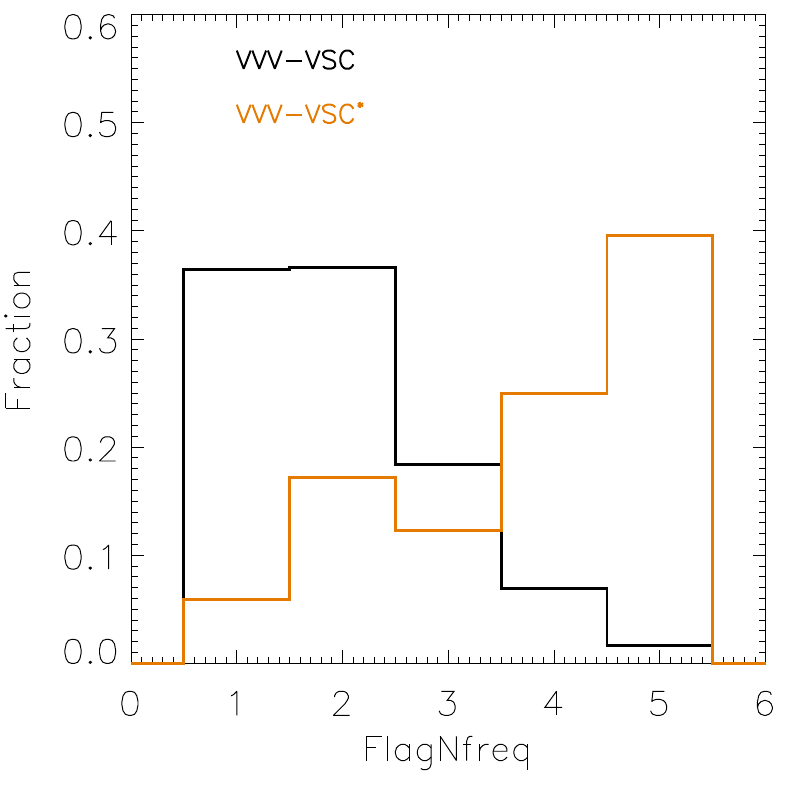} 
  \includegraphics[width=0.23\textwidth,height=0.25\textwidth]{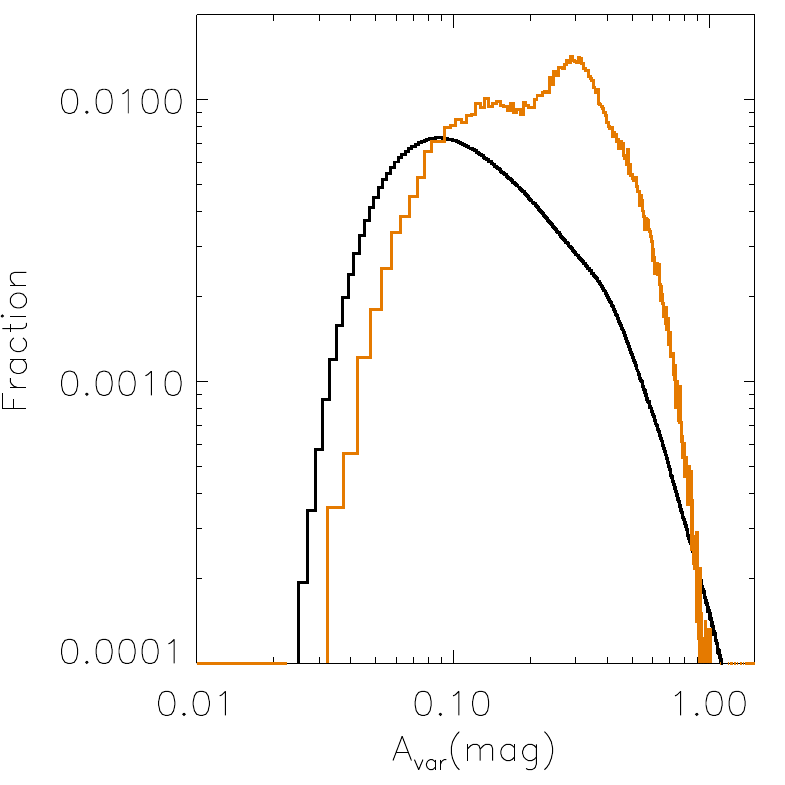}

  \includegraphics[width=0.23\textwidth,height=0.25\textwidth]{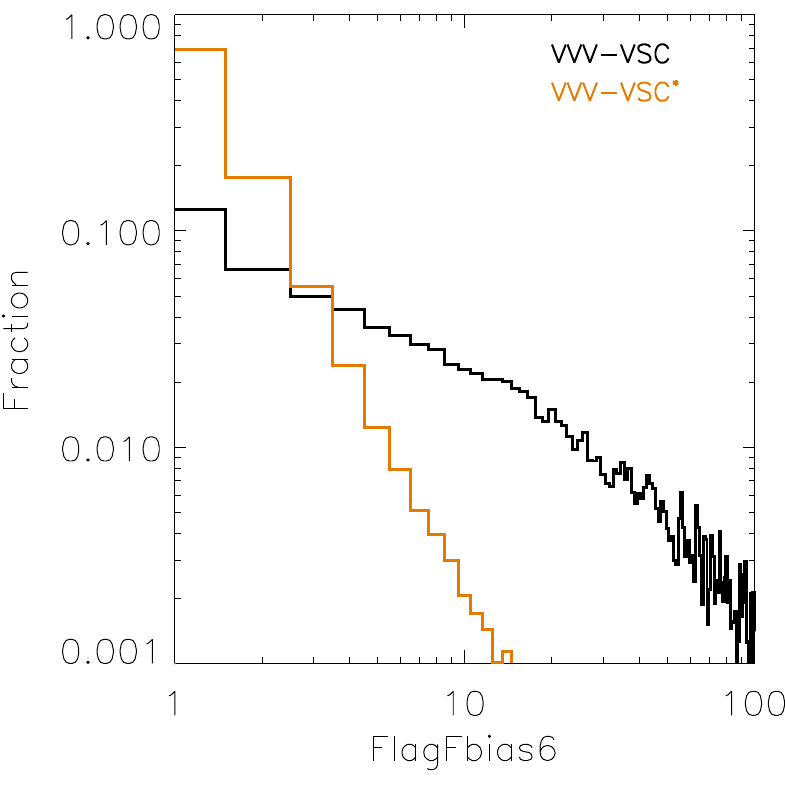} 
  \includegraphics[width=0.23\textwidth,height=0.25\textwidth]{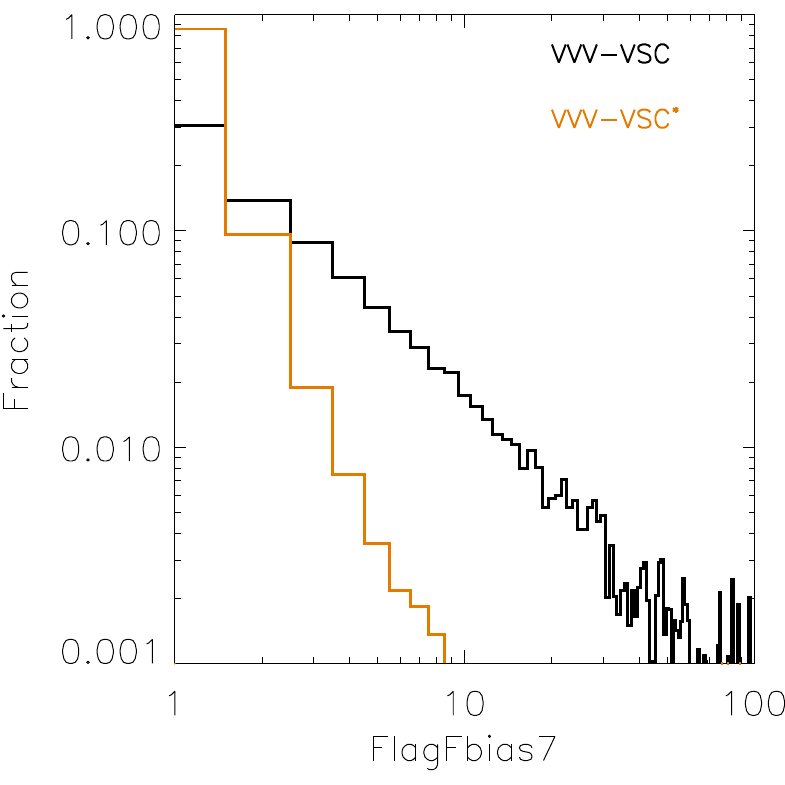} 
  \caption{Histogram of \textbf{\textit{FlagNfreq}} (upper left panel), and \textit{$\bm A_{VAR}$} (upper right panel), \textbf{\textit{FlagFbias6}} (\textbf{lower} left panel), and \textbf{\textit{FlagFbias7}} (lower right panel) for \textit{VVV-CVSC} (black lines) and crossmatched sources having matched periods \textit{VVV-CVSC*} (orange lines).}
  \label{fig_histflags}
\end{figure}

\begin{itemize}
\item \textbf{\textit{FlagNfreq}} gives the number of periods in agreement between the five different methods (see Sect. \ref{sec_columndescription}). We consider that the agreement is found when the period is equal within an accuracy of $10\%$ or when they are matched with the first harmonic or first overtone. The percentage of periods in agreement with the $P_{LSG}$ for \textit{VVV-CVSC} is $\sim36.6\%$, $\sim18.4\%$, $\sim6.9\%$, and $\sim1.7\%$ for two, three, four, and five methods (see up left panel of Fig. \ref{fig_histflags}). This means that there are at least 4 million good detections if four periods in agreement provide trustable parameters. Indeed, $\sim73\%$ of the \textit{VVV-CVSC*} (see orange lines in upper left panel of Fig. \ref{fig_histflags}) meet this criterion. The \textit{FlagNfreq} is the number of different methods that have a large PPSH for the best period (within 10\% or the first harmonic/overtone). Periods that are matched by more methods are more likely to be correct.
However, the efficiency of detection is not the same for all methods and it can vary with the signal type (see Sect. \ref{sec_mainperiod}). For example, about $\sim36.4\%$ of $P_{LSG}$ do not correspond to any other method but that does not necessarily mean that all of these periods are unreliable. Indeed, $P_{LSG}$ and $P_{PDM}$ have similar results as well as efficiency rates (Ferreira Lopes et al. submitted) and therefore the agreement between them can be used to improve the selection criteria.

\item \textbf{\textit{$\bm A_{VAR}$}} denotes the amplitude of the light-curves: calculated by subtracting the $5th$ and $95th$ percentile magnitude measurements (see Sect. \ref{sec_columndescription}). Applying the estimation of amplitude by \textit{$\bm A_{VAR}$} to eclipsing binaries of Algol type and similar morphologies will be biased since these sources usually have few points at the eclipse, and these few will likely be removed in the clipping. These estimations work well for a large majority of variable stars such as those undergoing stellar pulsation or some kind of semi-regular variations. Almost all \textit{VVV-CVSC*} stars have a $K_s$ amplitude greater than $0.1$mag. Indeed, this result is a selection effect.  On the other hand, only about $\sim50\%$ of \textit{VVV-CVSC} stars have amplitudes above this limit (see up right panel of Fig. \ref{fig_histflags}). Indeed, the detection of variability does not necessarily mean a measured variability period, i.e. aperiodic signals or sources having enough variation to be detected by variability indices but not by period finding methods. Therefore, the use of \textit{$\bm A_{VAR}$} will depend of the purpose of users.

\item \textbf{\textit{FlagFbias6}} and \textbf{\textit{FlagFbias7}}: the detection of a signal does not necessarily mean a reliable detection since seasonal variations (or aliases) can also lead to a smooth phase diagram (see Fig. \ref{fig_lcscross} last panels). These variations can be present in a large number of sources. Therefore, we count the number of periods found per VVV tile in bins of $10^{-6}$d$^{-1}$ and $10^{-7}$d$^{-1}$ (see flags \textit{FlagFbias6} and \textit{FlagFbias7}). These parameters indicate the probability of the period be related to instrumental or seasonal variations since on average the number of variable stars with the same period should not be large. For instance, the probability of finding more than $10$ sources in a bin of $10^{-7}$d$^{-1}$  sorted randomly can be easily estimated. The number of sources per VVV tile is typically less than $1.5$ million sources. The probability of it having a frequency in this range will be $10^{-8}$ if we consider that a variable star can assume any value in the interval of periods ranging from zero to $1000$ days We should note, however, that true variable stars also can be flagged if they have the same period as those found to be unreliable signals.

Figure \ref{fig_histflags} shows the histograms of \textit{FlagFbias6} and \textit{FlagFbias7} to \textit{VVV-CVSC}  and \textit{VVV-CVSC*} stars. As expected, the \textit{VVV-CVSC*} stars have flag values smaller than 10. A yield rate bigger than $\sim95\%$ is found if a flag number smaller than 5 is adopted. On the other hand, the \textit{VVV-CVSC} stars have more than $\sim67\%$ of sources with \textit{FlagFbias6}$> 5$. This indicates that a large fraction of these periods can be related with seasonal or instrumental variations since large \textit{FlagFbias6} values are found for these periods. For instance, the \textit{FlagFbias6} for periods of about 1 day (i.e. $1\pm 10^{-6}$) is on average $100$ periods per VVV tile. 
\end{itemize}

In summary, users can select the set of variability indices to reduce the number of stars. Moreover, the probability to detect the correct variability period will increase with the number of measurements and hence a number larger than 10 can be adopted, depending on the user. The PPSH also indicates which sources have reliable signals. Finally, the flags \textit{FlagFbias6-7} indicates the reliability of periods and if they are related with spurious variations.

\begin{figure*}
  \centering
  \includegraphics[width=1.25\textwidth,height=0.85\textwidth,angle =-270]{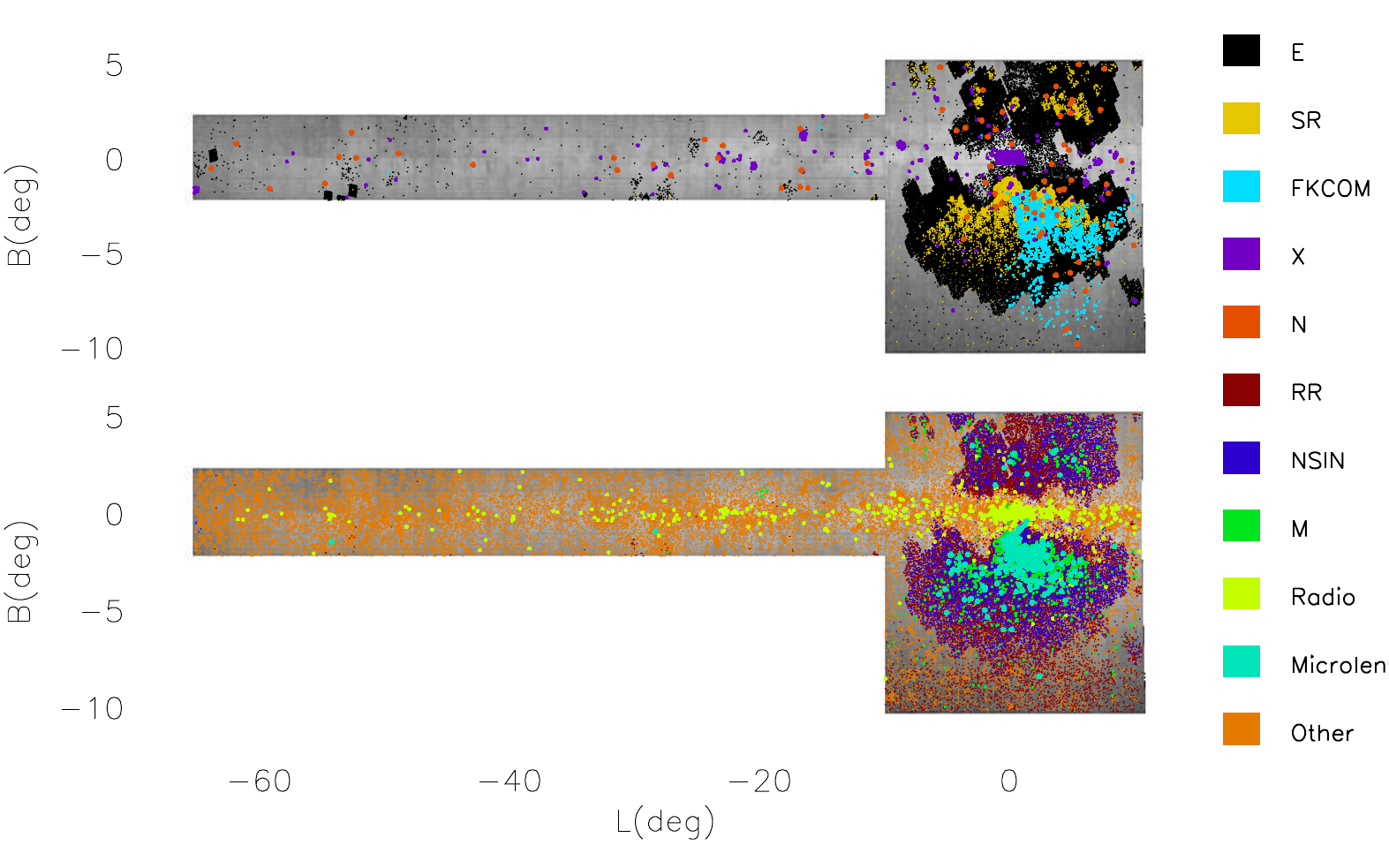} 
  
  \caption{Spatial distribution of \textit{VVV-CVSC} stars (grey colour) in Galactic coordinates for all \textit{VVV-CVSC} (left panel) and for a strict selection considering the flags (right panel - for more details see Sect. \ref{sec_reliablesel}). The crossmatched sources are set by colours (see the labels at the right side).}
  \label{fig_vartipe}
\end{figure*}

\begin{figure*}
  \centering
  \includegraphics[width=0.98\textwidth,height=0.65\textwidth]{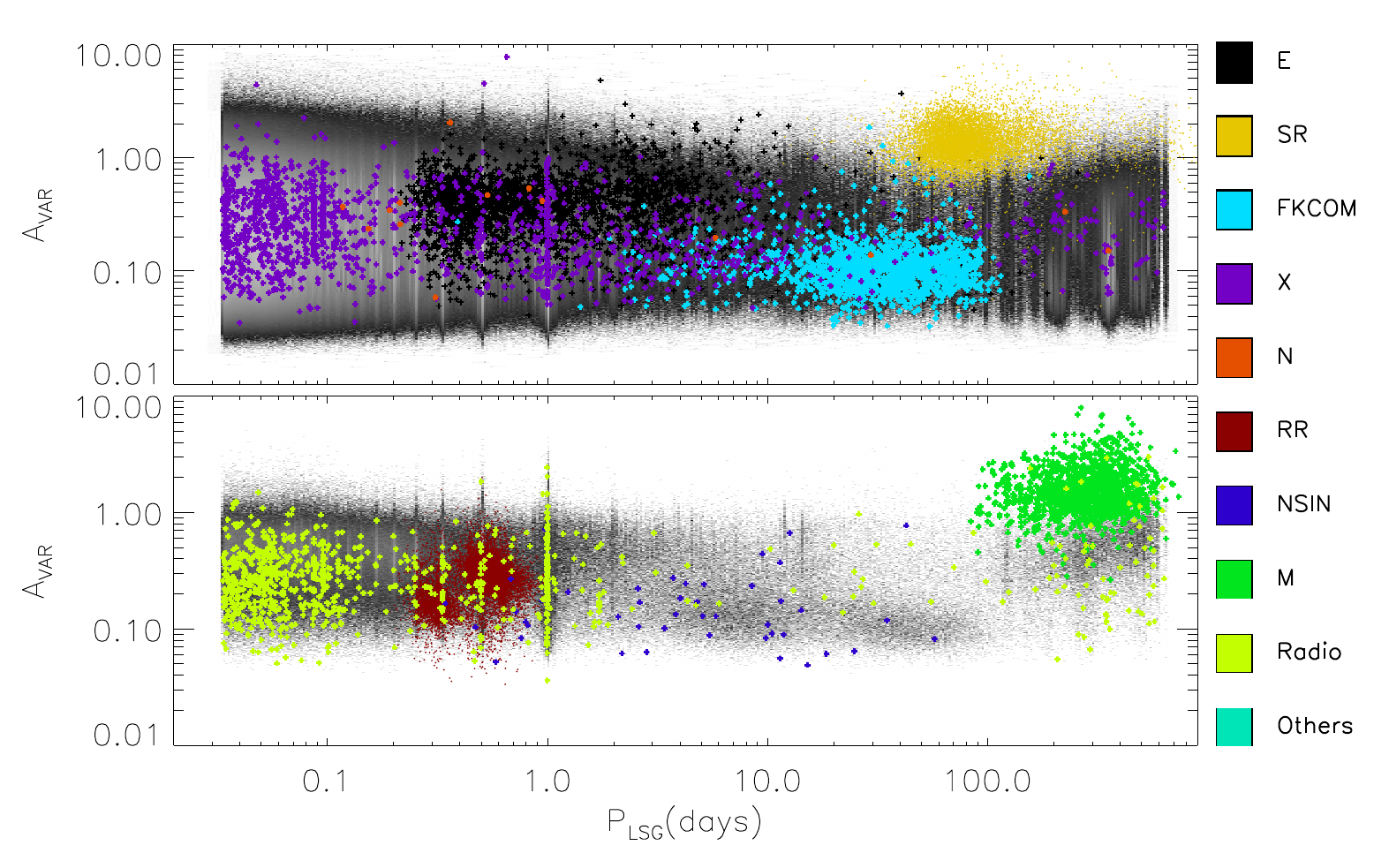} 
  \caption{Amplitude ($\bm A_{var}$) versus variability period ($P_{LSG}$)  for all \textit{VVV-CVSC} (upper panel) and for a strict selection considering the flags (lower panel - for more details see Sect. \ref{sec_reliablesel}). The crossmatched sources are set by colours (see label at right side).} 
  \label{fig_perampdist}
\end{figure*}

\section{Results and discussions}\label{sec_results}

In this work we present an unique near-IR dataset of variable sources based on VVV photometry to investigate different matters of stellar variability. The main goal of this work is to release this variability analysis of the VVV survey. Forthcoming studies will address subjects from classification to peculiar IR variations. In the next sections, we trace an overview of the spatial distribution, colour-colour diagrams, and variability parameters in order to glimpse possibles scientific cases.

\subsection{Spatial distribution}\label{sec_spatialdist}

Figure \ref{fig_vartipe} shows the spatial distribution of \textit{VVV-CVSC} stars. The number of sources is slightly greater for the regions having more measurements. However, the same behaviour is not observed when only the crossmatched sources are considered. These distributions can be understood in terms of Galactic structure and wavelengths observed. Our main remarks are described below;

\begin{itemize}
\item The large majority of orange dots (other - see Sect. \ref{sec_cross}) means detection of unclassified sources having some IR counter-part  (see lower panel). Therefore, these sources cannot be interpreted in terms of the stellar population since no information about stellar evolution is available. However, they are spread along the plane and bulge areas with a concentration about the middle regions observed by VVV. The sources having radio emission (yellow bright dots in lower panel) are concentrated in this mid-plane region. 

\item In terms of variability detection, a smaller number of objects is seen in the innermost bulge area and inner galactic plane. This region is usually avoided by optical surveys and amateur astronomer observations due to the high extinction that hinders the detection of variable stars. This “zone of avoidance” is also present in the distribution of the VVV Novae catalogue \citet[][]{Saito-2013} and is evident in the Gaia-DR2 LPV catalogue release \citep[][]{Mowlavi-2018} where the innermost regions are weakly populated. Indeed, this region is not actively avoided, but Gaia has a limited number of windows that can be assigned at once, so in very crowded regions the incompleteness increases. On the other hand, the highest density of sources are found in the intermediate bulge region ($-3^\circ > b > 3^\circ$)  and caused mostly by eclipsing binaries (E), RR Lyrae (RR), and semi-regular (SR) variable stars detected by variability surveys mainly at optical wavelengths.

\item The largest contribution of crossmatched sources comes from the Optical Gravitation Lensing Experiment (OGLE). OGLE is an optical survey which took many observations for the lower bulge region \citep[see Fig. 1 in][]{Wyrzykowski-2015}. The OGLE observations cover large sky areas where the most overlap with VVV is found in the disk and the outer bulge Milky Way areas. A study using the OGLE and VVV light curves, optical and IR wavelength, will provide clues about interstellar absorption  as well as the stellar physical processes.

\item The density of SR stars found in the southern bulge region ($b < -3$) is much higher than that found in the northern bulge region ($ b > 3$). Similar behaviour is found for Mira type stars (M). SR main sequence stars usually have small amplitude and semi-periodic variations and hence their detection requires more measurements in comparison with RR stars, for example. On the other hand, M stars need a large coverage time to be detected. The numbers of detected SR stars is growing quickly with dedicated surveys like the CoRoT and Kepler surveys \citep{DeMedeiros-2013,McQuillan-2013,FerreiraLopes-2015mgiant}. These results indicate that the population of SR stars is much larger than that found in Fig. \ref{fig_vartipe} and the spatial difference is not real, i.e. the population studies are limited in terms of total time span and the cadence of observations.

\item We expect that metal-rich RR Lyrae should be located in the Galactic disk while metal-poor RR Lyrae should be located in the bulge region \citep[e.g.][]{Binney-1998}.  A large number of VSC stars in the Galactic disk give a unique opportunity to significantly increase the numbers of RR type I stars at this region since we have a limited presence of crossmatched sources in this region.

\item  The eclipsing binaries are mainly found in larger numbers in the Galactic bulge. The \textit{VVV-CVSC} provides an opportunity to fill the empty areas of the disk since a large number of these objects are expected along all Galactic regions. 

\item  A large number of X-ray sources were found at the Galactic centre. The variability behaviour of many of these stars has not been addressed so far. Indeed, the X-ray and XMM observations are mainly taken towards the Galactic centre and hence the large numbers of sources found in this region. The precision of X-ray coordinates are much worse than Optical or IR observations. Hence the X-ray crossmatched sources must be verified carefully. The stellar physical process related with these stars can be explored using spectroscopic follow-up together with IR light curves. 
\end{itemize}

To summarize, from the spatial distribution viewpoint, the VSC catalogue offers a unique opportunity to cover regions under-explored by previous missions as well as to give new insights into those stars where the variability nature is unknown.

\subsection{General variability properties}\label{sec_varproperties}	

Figure \ref{fig_perampdist} shows the variability periods as a function of Ks-band amplitudes found in the IR light curves. The crossmatched data having previous variability periods are labelled by colour. The upper panel shows results for the entire \textit{VVV-CVSC} (grey colour) while the lower panel only shows that for those sources having (A) $X > 2$, (B) $N > 30$, (C) \textit{FlagNfreq}$\geq 2$, (D) \textit{FlagFbias6} $\leq 2$, and (E) $H_{PK}/FAP > 1.0$. Criteria (A) removes low signal-to-noise ratio data and misselected sources, (B) removes the sources where there is a low probability to estimate good periods, (C) and (D) remove the sources where the periods are not in agreement or they are probably related with a dubious period from aliasing or seasonal effects, while (E) keeps only those sources where the strength of variability period is greater than the white noise value considering a sinusoidal variation. Different astronomers can use these parameters or  other combinations of criteria to select samples that suit their science. Publishing a more complete catalogue with parameters to select reliable samples save time of all users. These constraints reduce the sample to about one million sources. A large fraction of sources outside of these limits are not reliable signals (for more details see Sect. \ref{sec_reliablesel}). The periods plotted for the crossmatched sources are those found in the literature when available otherwise those ones computed by us are used. Indeed, detection of variability does not mean that periodic features will be present or measurable. The main concerns about the period versus amplitude distribution can be summarized as follows:

\begin{itemize}
    \item  The \textit{VVV-CVSC} sources show a lower limit of $\bm A_{VAR} \simeq 0.01$ magnitudes in $K_s$ considering the entire sample.  On the other hand, the strict selection performs a lower limit of $\bm A_{VAR} \simeq 0.05$ magnitudes in $K_s$. It seems that this is the lower detection limit of the VVV survey. Indeed, we are looking at the Rayleigh Jeans tail of the stellar fluxes and hence the amplitudes are smaller than in the optically selected variable stars. Therefore, FKCOM, NSIM, and other sources having $\bm A_{VAR}$ smaller than this limit will be missed, for instance. Indeed, the crossmatched sources (lower panel) have $\bm A_{VAR}$ value distributed along the whole range of amplitudes detected by VVV observations. Moreover, for this sub-sample, the number of sources with periods equal to seasonal periods are reduced.
    \item The peaks in the distribution due to seasonal variations also appear in the strict selection. This happens because seasonal variations and true signals can have periods around 1 day and aliased phase diagrams (see OGLE II Dia BUL-SC12 V0700 in Fig.  \ref{fig_lcscross}). Signals about these peaks must be considered carefully. On the other hand, "data mining" ~of signals having amplitudes smaller than $\bm A_{VAR} \simeq 0.05$ is hindered since sources with these amplitudes are dominated by a large number of noisy or unreliable signals.
    \item The limits on the range of periods used to discriminate different variable stars types are not well defined, as expected. On the other hand, the mean amplitude for M ($\bm A_{VAR} \simeq 0.96$mags) and SR  ($\bm A_{VAR} \simeq 0.77$mags) type stars are much larger than other ones since they have long variability timescales. The FKCOM variable stars have mean variability periods of $\sim37$ days and an amplitude of about $ \sim 0.06$ mags.
    \item Aperiodic variable stars, long period variables (LPVs), low amplitude variables, and all other variable stars where the complete variability phases was not covered by VVV observation can have $H_{PK}/FAP < 1$ and this will reduce their completeness in strictly selected samples. 
    \item Radio and X-ray sources have no variability periods previously estimated. Many of them are related with the aliases of one day. The other ones must be checked in order to determine the IR variability counterparts to these detections.
    \item The variability indices indicate an intrinsic variation while the amplitude shows the signal strength at 2 microns. Indeed, the amplitude is helpful to discriminate those sources having characteristic amplitudes like Miras (M) stars. 
    \item For periodic variable stars, the light curve shape can be easily accessed  from the phase diagram folded by its variability period in order to facilitate its classification.
\end{itemize}

  This catalog is a unique tool to identify variable types in terms of amplitude and variability periods from already available data. Indeed, the limits that are required to create a reliable or complete selection depend on the purpose of each user. Once this has been decided, the users can download the light curves and tables in order to combine colour information and shape parameters that can be easily  computed from the light curves. Figure \ref{fig_lcscrossbest} shows some examples of data quality and a wide number of variability types that can be accessed from the \textit{VVV-CVSC}. Users should realize that the variability periods found by us correspond to the first harmonic of a large minority of sources (see Sect. \ref{sec_periods}). Therefore, the analysis of the harmonics must be addressed before fully analysing the data.

\begin{figure}
  \centering
  \includegraphics[width=0.48\textwidth,height=0.55\textwidth]{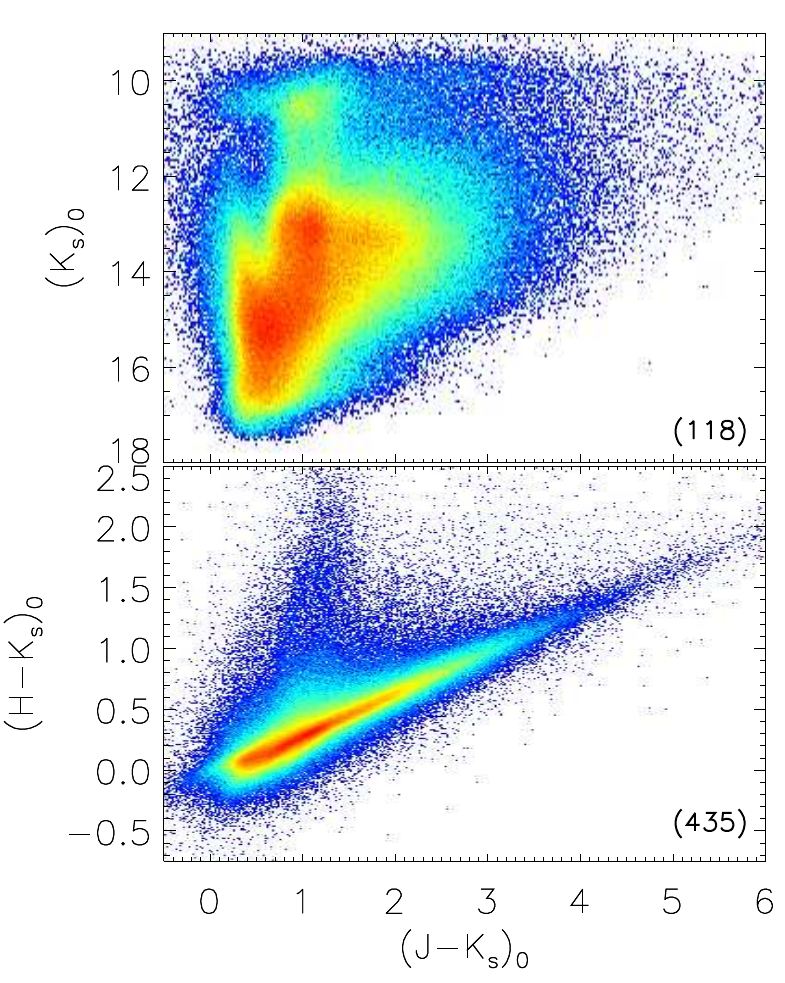} 
  
  \caption{$(J - K)_{0}$ versus $(K)_{0}$ colour-magnitude diagram (upper panel) and $(H - K)_{0}$ vs $(J - K)_{0}$ colour-colour diagram (lower panel). The higher number of sources per pixel is shown in the right corner in each diagram. We should probably note that the brightest sources with $Ks<11$ mag are saturated.}
  \label{fig_colorcolor}
\end{figure}

\subsection{Colour-colour and colour-magnitude diagrams}\label{sec_colorcolor}

Sections \ref{sec_spatialdist} and \ref{sec_varproperties} discuss the \textit{VVV-CVSC} catalog from a framework of spatial distribution and variability parameters (amplitudes and periods). The spatial distribution of \textit{VVV-CVSC} is important because it is not possible to obtain the variability parameters of the entire \textit{VVV-CVSC}. Aperiodic variable stars, low signal-to-noise ratio data, saturated stars, reduced number of measurements among other things hinder this achievement, i.e. variability periods, amplitude, and morphology of variation of a large fraction of the variable stars included in the  \textit{VVV-CVSC}  are not measurable despite the detection of reliable variability for many of these sources. On the other hand, colour-colour and colour-magnitude diagrams provide additional clues about the stellar evolution stages and hence allow us to speculate about the reasons why the variability periods are not accessible. The VVV area overlaps with many other surveys at optical and mid-IR wavelengths, see Sec. ~\ref{sec_introduction} which will also provide additional constraints on each star.

\begin{figure*}
  \centering
    \includegraphics[width=0.23\textwidth,height=0.25\textwidth]{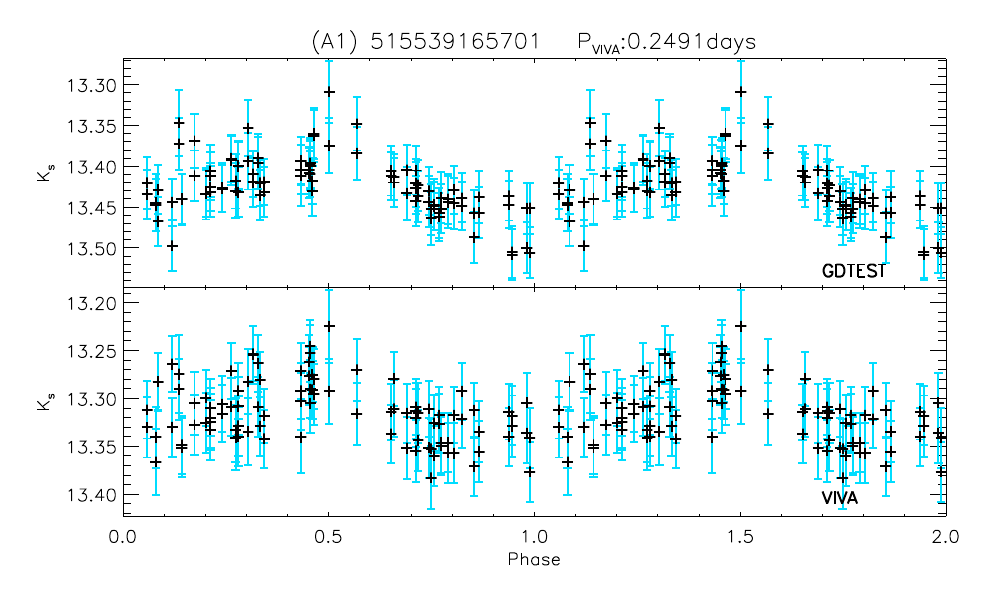}
  \includegraphics[width=0.23\textwidth,height=0.25\textwidth]{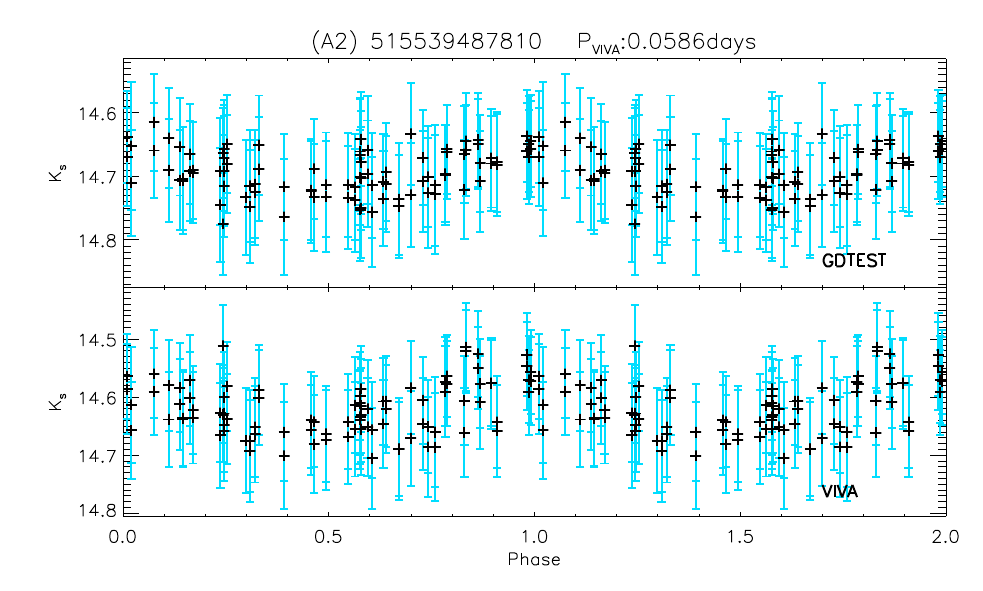}
  \includegraphics[width=0.23\textwidth,height=0.25\textwidth]{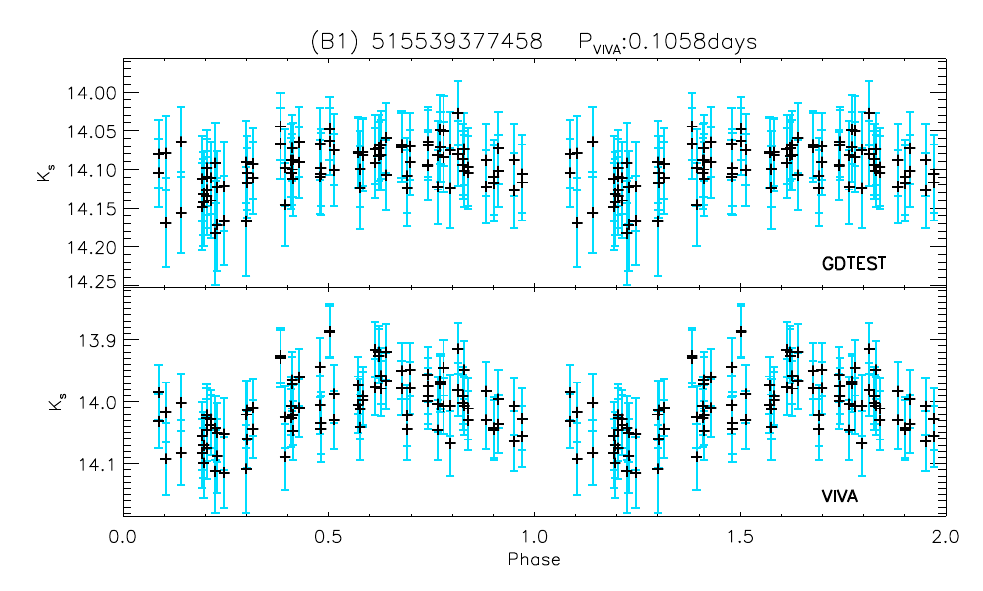}
  \includegraphics[width=0.23\textwidth,height=0.25\textwidth]{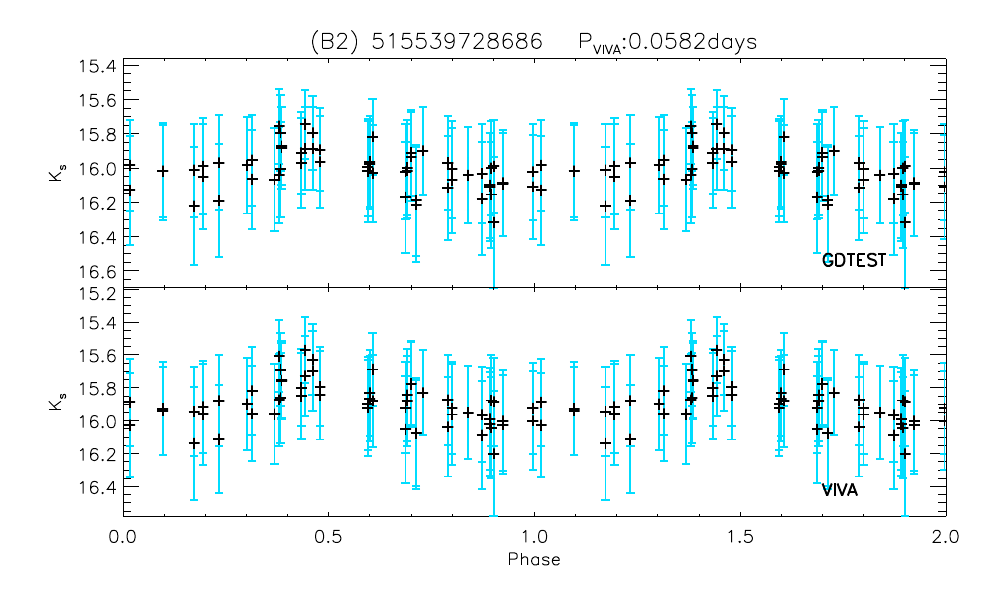}
  
  \includegraphics[width=0.23\textwidth,height=0.25\textwidth]{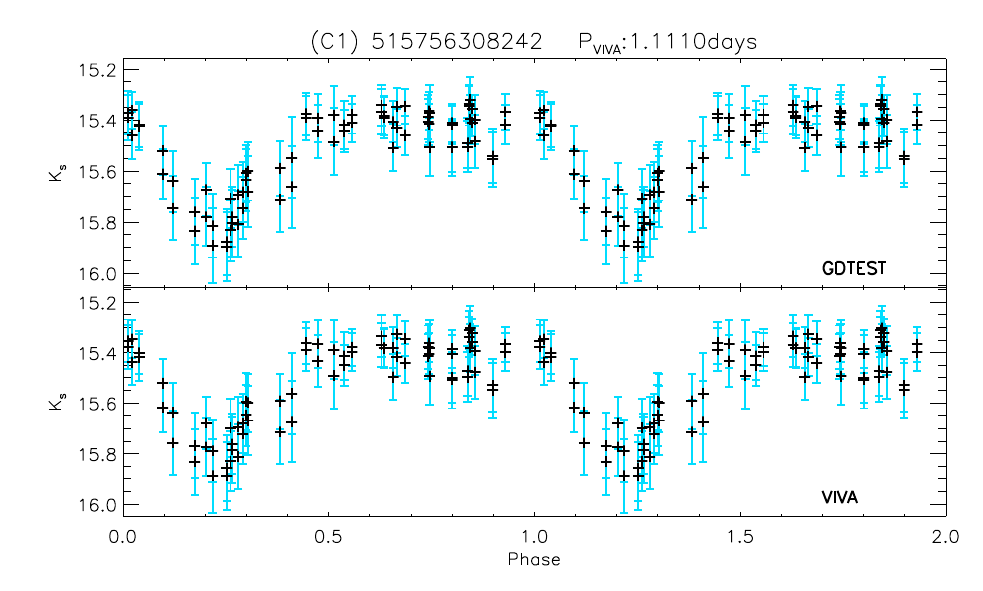}
  \includegraphics[width=0.23\textwidth,height=0.25\textwidth]{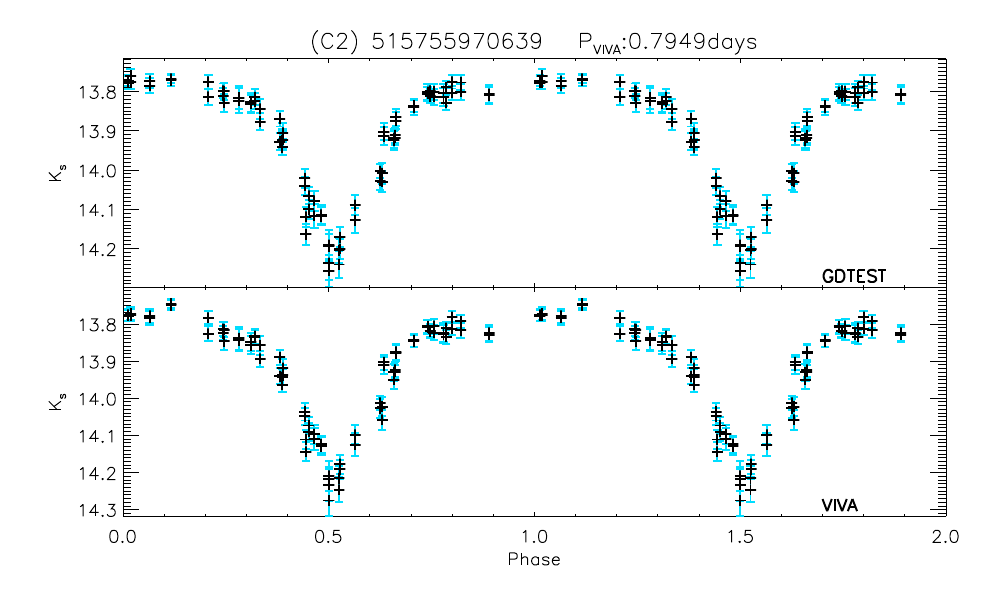}
  \includegraphics[width=0.23\textwidth,height=0.25\textwidth]{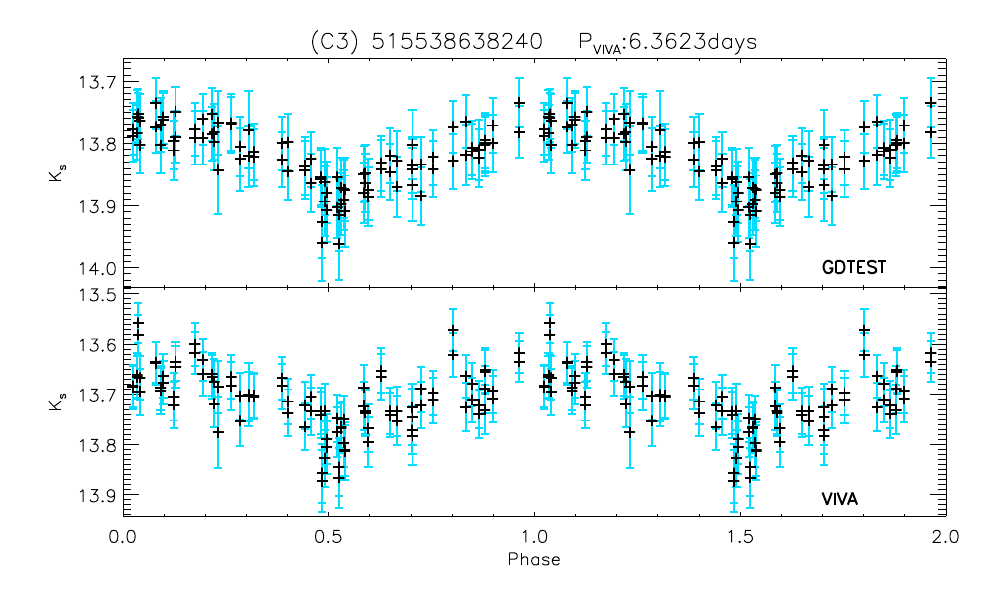}
  \includegraphics[width=0.23\textwidth,height=0.25\textwidth]{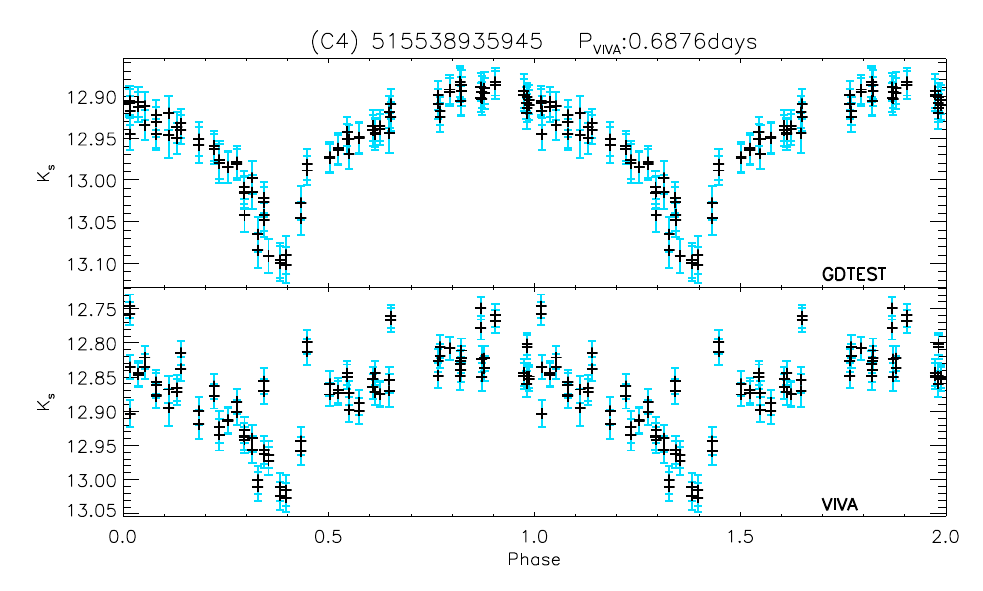}
  
  \includegraphics[width=0.32\textwidth,height=0.25\textwidth]{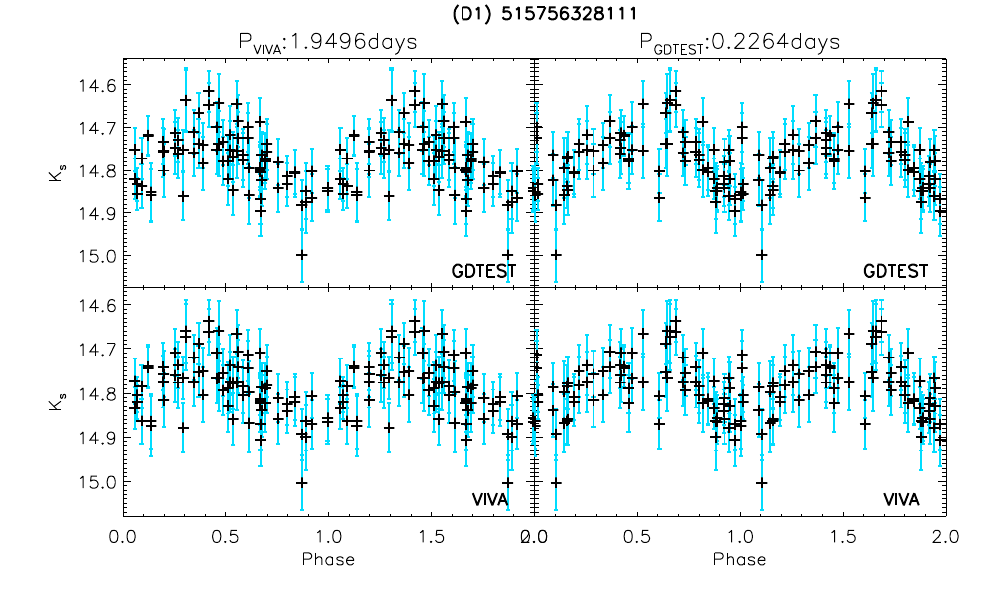}
  \includegraphics[width=0.32\textwidth,height=0.25\textwidth]{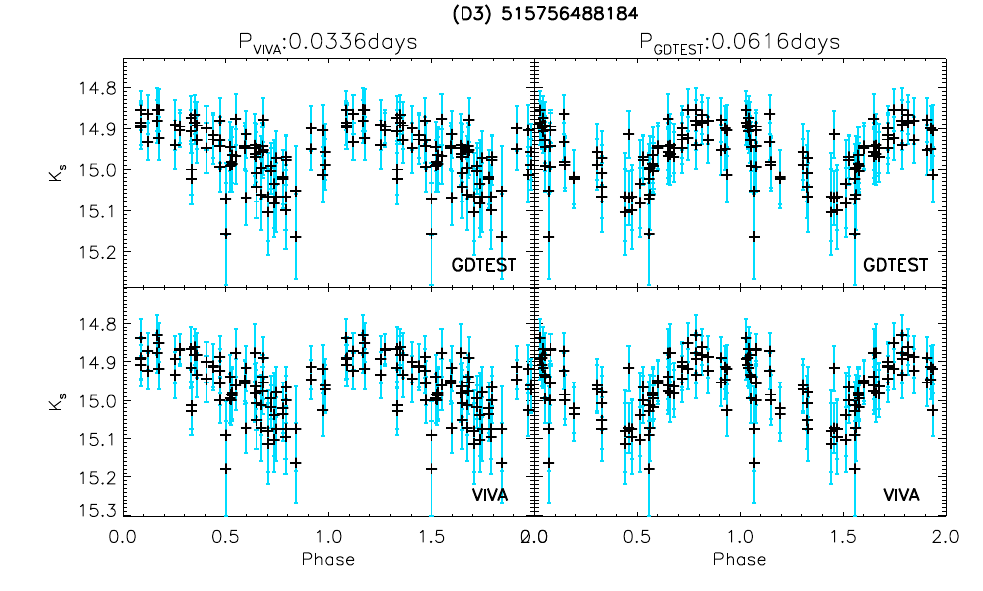}
  \includegraphics[width=0.32\textwidth,height=0.25\textwidth]{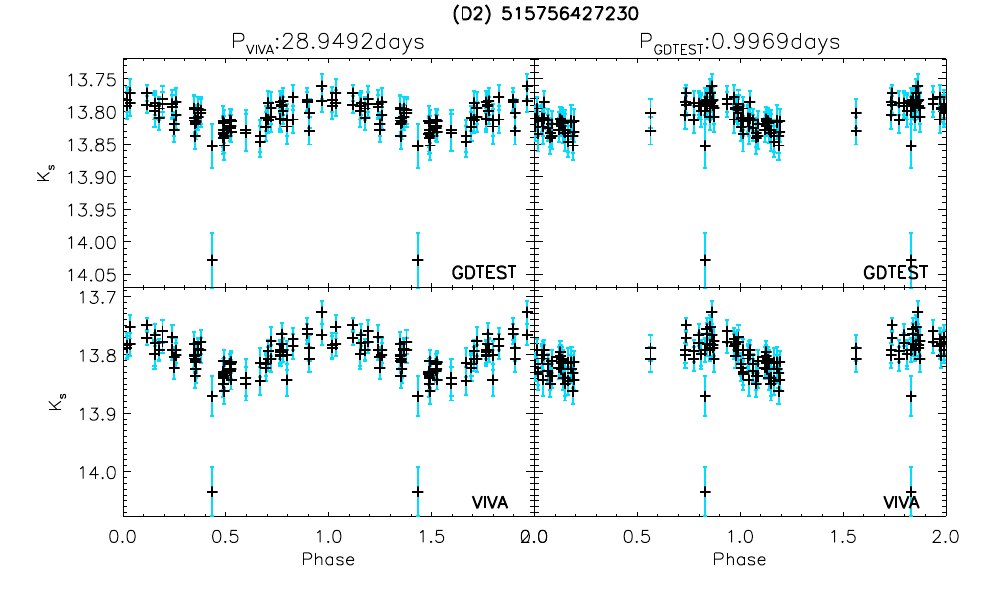}
  
  \caption{Phase light curves of GTEST (upper panel) and VIVA (lower panel) data. (A) and (B) panels include the sources missed in the GTEST or VIVA datasets, respectively. (C) panels shows sources having periods in agreement in both datasets while (D) panels shows variables selected in both with inconsistent periods. The phased light curves in the last line of panels are split in two columns where the first column shows the light-curve folded using the period found in the VIVA catalog (first column) and using the period in the GDTEST data (second column).}
  \label{fig_lcagree}
\end{figure*}

Figure \ref{fig_colorcolor} shows the colour-colour diagram for the \textit{VVV-CVSC} dataset. The colour-colour diagram (lower panel) covers all stellar stages, i.e. the whole HR diagram. Therefore, a study of variability related to IR variations can be made using the present catalogue. On the other hand, the magnitude versus colour diagram also is quite similar to  the colour-colour diagram in terms of stellar evolution. Indeed, we notice a strong reduction in the number of sources at $K_s \simeq 12$.  This effect was also observed when the initial sample was analyzed (see Fig. \ref{fig_initialdata}). The cut-off value chosen (i.e. $X > 1.5$) for $K_s < 11.5$ for NC data is twice that used for fainter $K_s$ values. Indeed, the NCD data only corresponds to $18\%$ of the initial data. On the other hand, the CD data does not use any consideration about the magnitude and it corresponds to $82\%$ of initial data. Therefore, the gap that we are observing is related to the initial data. Users should note that, saturated objects, which may include nearby stars are probably not included in the \textit{VVV-CVSC} catalogue.  However, this does not explain this gap. On the other hand, the increased number of objects at $K_s\sim11.0$ occurs because of an increase in the false positive rate as non-linearities and saturation effects the magnitudes despite an increased cutoff in X index (see lower left panel of Fig.  \ref{fig_initialdata}).

Herpich et al. (submitted) have also presented a catalogue of VVV cross-match sources with the VSX-AAVSO catalogue. The authors analyse near-IR CMDs and spatial distributions for the different types of variables that enable them to discuss our current knowledge about variability in the Galaxy. The current knowledge about variability in the Galaxy is biased to the nearby and low extincted stars according the authors. The results of our cross-match is about four times larger than that found by the authors since we used other databases as well as VSX-AAVSO (see Sect. \ref{sec_cross}). A deep analysis on the near-IR CMDs and spatial distributions from a larger cross-match sample is beyond the current project. Indeed, the study of open questions about the empirical relationship of the stellar and variability parameters of RR Lyrae stars can be assessed already with the available data. All of these aspects can be better explored when the classification of  \textit{VVV-CVSC} takes place.

\section{New Calibration of VVV Photometry}\label{sec_newcalibration}

Recently \citet[][]{Hajdu2019} identified two independent kinds of bias in the photometric zero-points on the VVV data: intra-array variations in the detector’s response, and the blending of local secondary standard stars. According to the authors the combination of these effects provide a space-varying bias in the absolute photometric calibration, and a time-varying error in the photometric zero-points on various time-scales. The authors also show that the first effect affects the absolute magnitude, while the latter can also affect the shape and amount of scatter in the light curve. These problems mainly affect crowded VVV regions.

We perform our own tests in three VVV tiles B306, B201, and D068 having $971093$, $294696$, and $934953$ sources, respectively. This subset of data was labelled as GDTEST. The two first VVV tiles are in the Galactic Bulge while the third one in the Galactic Disk. The comparison between B306 and B201 allows us to measure the bias related with the density of detected sources per field while the comparison between B306 and D068 investigates reddening effects. Indeed, the same algorithm and constraints were applied to the GDTEST data as the VIVA data, so that a straightforward comparison can be made.

 \begin{table}
 \centering     
 \caption[]{Total number of sources (N), along with the selected targets found in  VIVA (N$_{VIVA}$) and GDTEST (N$_{GDTEST}$) datasets, as well as the number of matched sources between them (N$_{BOTH}$).} 
 \begin{tabular}{c c c c c} 
   \hline\hline           
VVV Tile & N & N$_{GDTEST}$ & N$_{VIVA}$ & N$_{BOTH}$  \\   
   \hline 
b306 & 971093 & 200177 & 501472 &  153869 \\ 
b201 & 294696 &   7029 &   8326 &    5663 \\ 
d068 & 934953 & 116200 & 141885 &   91830 \\ 
\hline                                   
\end{tabular}
\label{tab_gbtest}
\end{table}

From the viewpoint of selection criteria, the number of sources selected in the B306, B201, and D068 fields are 2.5, 1.2, and 1.2, larger than $N_{GDTEST}$ respectively. Table \ref{tab_gbtest} shows the number of selected targets in the VIVA and GDTEST datasets of the anlayzed VVV fields. As expected the largest difference in the selected samples is found in B306. On the other hand, the number of sources found in B306 is almost the same as that found in D068 however the number of selected sources is $2.5$ times larger. Moreover, the number of selected sources found in D068 using VIVA and GBTEST differs by a factor of $1.2$. This indicates the problems related with the VVV photometric reduction are more strong related with extinction than density of stars. This indicates that B306 includes a large number of misselected sources if we consider that the number of true variable stars included in these fields is likely to be similar. However, the stellar populations are a bit different and hence a direct comparison of fraction of variables is too simplistic. Statistical fluctuations provided by the \citet[][]{Hajdu2019} approach can either include (see (A) panel Fig. \ref{fig_lcagree}) or exclude (see (B) panel Fig. \ref{fig_lcagree}) sources with small amplitudes, those smaller than $\sim0.03$ mag. Indeed, the large majority of sources not included in both datasets do not present a clear signal in the folded phase diagram. 
 
The mean magnitudes found in  B306, B201, and D068 GDTEST corrected data are about $0.12\%$, $0.03\%$, and $0.003\%$ brighter than the current VVV data, respectively. On the other hand, we also test the common selected sources in VIVA and GDTEST in order to check the period detection. We considered as matched periods those having a relative error smaller than 10\% of the main period or its first harmonic for the LSG method. An agreement on period estimation of $50\%$,  $87\%$, and  $70\%$ was found for each field respectively. Indeed, more than $90\%$ of periods match directly and do not match via a harmonic or overtone. (C) panels of Fig. \ref{fig_lcagree} show some examples where the period estimations are in agreement. The C4 panel shows a particularly striking example with large corrections. Very few stars have such strong modifications as those provided by \citet[][]{Hajdu2019}. 

On the other hand, we also found sources where the period estimations are different or have a relative error bigger than $10\%$ (see (D) panel Fig. \ref{fig_lcagree}). For these sources, we can find period estimations where the periods estimated in GDTEST datasets seems better that the VIVA catalog (D1 panel), the opposite (D2 panel), and those ones where both estimations must be more carefully analysed (D3 panel). This indicates that the phase diagram by itself is not always enough to settle the best period, particularly for those sources having small amplitude. The results found for these sources in terms of variability indices and period estimation must be used carefully.

The comments above were created from a visual inspection on some thousand sources in order to provide a check of the period detection and data quality in three VVV tiles. The sources where the variability indices or period estimations are different are mainly related with sources having small dispersion values ($ED < 0.03$) and a small number of observations (typically fewer than 40) where statistical fluctuations will be more important. In summary, the analysis performed in this work can be strongly affected, mainly for sources having sigma value smaller than $0.03$ mag or for those sources where the \citet[][]{Hajdu2019} corrections are larger, e.g. where there is a higher source density and more blending, and where the extinction is higher. 

\section{Conclusions and Discussions}\label{sec_conclusion}

Data-mining of near-IR surveys is a good opportunity to test our capability to efficiently explore future variability datasets as well as investigating Galaxy regions that cannot be observed in the optical and have been explored less by previous surveys and other open scientific matters. This paper addresses the variability analysis of all VVV point sources having more than $10$ measurements using a novel approach proposed in the NITSA project. That project provided new variability indices to detect reliable signals, constraints to detect periodic signals as well as new period finding methods. These works give reliable constraints to select and detect signals in big-data sets. 

In total, $288,378,769$ near-IR light curves were analyzed and as a result, we have produced a catalog $44,998,752$ of variable stars candidates (\textit{VVV-CVSC}). The contamination ratio of VIVA catalog could be higher than 10 (for more details see \ref{sec_reliablesel}). Five period finding methods were used to estimate the main variability periods. Moreover, our final catalog includes accurate individual coordinates, near-IR magnitudes ($ZYJHK_s$), extinctions $A(K_s)$, variability indices, periods, near-IR amplitudes, among other parameters to access the science in \textit{VVV-CVSC}, and is linked into the VSA where it can be used with the other VVV data and cross-matched catalogues, see \S~\ref{app:sql}. Users can discriminate among these parameters to select their targets of interest. Indeed, the variability detection does not necessarily mean period detection since sometimes there is not enough available data to do that or the source may not be periodically varying. Therefore, the current catalogue also can be used to select sources to be followed-up for current or ongoing surveys.

\citet[][]{Hajdu2019} reported some problems related to the photometric calibration found in VVV dataset. We perform our own analysis in three VVV tiles in order to measure the weight of these corrections in our analysis. As expected the greatesT bias were found in the most crowded and highly-extincted VVV regions. In the future, PSF photometry of each pawprint epoch will be more suitable than the aperture photometry in the most crowded regions. Therefore, the VIVA catalog will be updated using PSF photometry in these regions.

\textit{VVV-CVSC} was crossmatched with the SIMBAD and VSX-AAVSO catalogs, and a total of $339, 601$ sources were in common. This subsample is a unique dataset to study the corresponding near-IR variability of known sources as well as to verify which sources did not have detected periods. Moreover, the near-IR amplitude used to select a certain classes of variable stars can also be determined from this sub-sample. On the other hand, the non-crossmatched sources is a matchless data that can be used to explore the heavily crowded and reddened regions of the Galactic plane, including stellar populations on the far side of the Galaxy. The present result also provides an important query source to perform variability analysis and characterize ongoing and future surveys like TESS and LSST.
    
\section*{Acknowledgements}
C.E.F.L. acknowledges a PCI/CNPQ/MCTIC post-doctoral support. N.J.G.C. acknowledges support from the UK Science and Technology Facilities Council.D.M. and C.M.P.R. are supported by the BASAL Center for Astrophysics and Associated Technologies (CATA) through grant A.F.B. 170002.  D.M. is supported by the Programa Iniciativa Científica Milenio grant IC120009, awarded to the Millennium Institute of Astrophysics (MAS), and by Proyecto FONDECYT No. 1170121. J.A.-G. acknowledges support by Proyecto Fondecyt Regular 1201490 and by the Chilean Ministry for the Economy, Development, and Tourism's Programa Iniciativa Cient\'ifica Milenio through grant IC120009, awarded to the Millennium Institute of Astrophysics (MAS) J.C.B. acknowledge support from FONDECYT (grant 3180716). The authors thank MCTIC/FINEP (CT-INFRA grant 0112052700) and the Embrace Space Weather Program for the computing facilities at INPE. We gratefully acknowledge data from the ESO Public Survey program ID 179.B-2002 taken with the VISTA telescope, and products from the Cambridge Astronomical Survey Unit (CASU). K.P acknowledge the support from CONICYT-Chile, through the FONDECYT Regular project number 1180054. C.M.P.R. acknowledges support from FONDECYT grant 3170870. The authors give thanks to G. Hajdu and I. Dekany by the support with the new calibration of VVV photometry. JRM and ICL acknowledge continuous support from CNPq and FAPERN brazilian agencies. C.M.P.R. acknowledges support from FONDECYT grant 3170870 and from the Max Planck Society through an MPE "Partner Group" grant.

\bibliographystyle{mnras}
\bibliography{mylib_1611.bib}

\begin{thebibliography}{}
\makeatletter
\relax
\def\mn@urlcharsother{\let\do\@makeother \do\$\do\&\do\#\do\^\do\_\do\%\do\~}
\def\mn@doi{\begingroup\mn@urlcharsother \@ifnextchar [ {\mn@doi@}
  {\mn@doi@[]}}
\def\mn@doi@[#1]#2{\def\@tempa{#1}\ifx\@tempa\@empty \href
  {http://dx.doi.org/#2} {doi:#2}\else \href {http://dx.doi.org/#2} {#1}\fi
  \endgroup}
\def\mn@eprint#1#2{\mn@eprint@#1:#2::\@nil}
\def\mn@eprint@arXiv#1{\href {http://arxiv.org/abs/#1} {{\tt arXiv:#1}}}
\def\mn@eprint@dblp#1{\href {http://dblp.uni-trier.de/rec/bibtex/#1.xml}
  {dblp:#1}}
\def\mn@eprint@#1:#2:#3:#4\@nil{\def\@tempa {#1}\def\@tempb {#2}\def\@tempc
  {#3}\ifx \@tempc \@empty \let \@tempc \@tempb \let \@tempb \@tempa \fi \ifx
  \@tempb \@empty \def\@tempb {arXiv}\fi \@ifundefined
  {mn@eprint@\@tempb}{\@tempb:\@tempc}{\expandafter \expandafter \csname
  mn@eprint@\@tempb\endcsname \expandafter{\@tempc}}}

\bibitem[\protect\citeauthoryear{{Akerlof} et~al.,}{{Akerlof}
  et~al.}{2000}]{Akerlof-2000}
{Akerlof} C.,  et~al., 2000, \mn@doi [\aj] {10.1086/301321}, \href
  {http://adsabs.harvard.edu/abs/2000AJ....119.1901A} {119, 1901}

\bibitem[\protect\citeauthoryear{{Almeida} et~al.,}{{Almeida}
  et~al.}{2019}]{Almeida-2019}
{Almeida} L.~A.,  et~al., 2019, \mn@doi [\aj] {10.3847/1538-3881/ab0963}, \href
  {https://ui.adsabs.harvard.edu/abs/2019AJ....157..150A} {157, 150}

\bibitem[\protect\citeauthoryear{{Alonso-Garcia}}{{Alonso-Garcia}}{2018}]{Alonso-Garcia-2018B}
{Alonso-Garcia} J.,  2018, in The Galactic Bulge at the Crossroads (GBX2018.
  p.~1, \mn@doi{10.5281/zenodo.2595299}

\bibitem[\protect\citeauthoryear{{Alonso-Garc{\'\i}a}
  et~al.,}{{Alonso-Garc{\'\i}a} et~al.}{2018}]{Alonso-Garcia-2018}
{Alonso-Garc{\'\i}a} J.,  et~al., 2018, \mn@doi [\aap]
  {10.1051/0004-6361/201833432}, \href
  {https://ui.adsabs.harvard.edu/abs/2018A&A...619A...4A} {619, A4}

\bibitem[\protect\citeauthoryear{{Andersson} \& {Kokkotas}}{{Andersson} \&
  {Kokkotas}}{1996}]{Andersson-1996}
{Andersson} N.,  {Kokkotas} K.~D.,  1996, \mn@doi [\prl]
  {10.1103/PhysRevLett.77.4134}, \href
  {https://ui.adsabs.harvard.edu/abs/1996PhRvL..77.4134A} {77, 4134}

\bibitem[\protect\citeauthoryear{{Angeloni}, {Di Mille}, {Ferreira Lopes}  \&
  {Masetti}}{{Angeloni} et~al.}{2012}]{Angeloni-2012}
{Angeloni} R.,  {Di Mille} F.,  {Ferreira Lopes} C.~E.,   {Masetti} N.,  2012,
  \mn@doi [\apjl] {10.1088/2041-8205/756/1/L21}, \href
  {http://adsabs.harvard.edu/abs/2012ApJ...756L..21A} {756, L21}

\bibitem[\protect\citeauthoryear{{Angeloni} et~al.,}{{Angeloni}
  et~al.}{2014a}]{Angeloni-2014ogle}
{Angeloni} R.,  et~al., 2014a, \mn@doi [\mnras] {10.1093/mnras/stt1823}, \href
  {http://adsabs.harvard.edu/abs/2014MNRAS.438...35A} {438, 35}

\bibitem[\protect\citeauthoryear{{Angeloni} et~al.,}{{Angeloni}
  et~al.}{2014b}]{Angeloni-2014vvv}
{Angeloni} R.,  et~al., 2014b, \mn@doi [\aap] {10.1051/0004-6361/201423904},
  \href {http://adsabs.harvard.edu/abs/2014A%26A...567A.100A} {567, A100}

\bibitem[\protect\citeauthoryear{{Athanassoula}}{{Athanassoula}}{2005}]{Athanassoula-2005}
{Athanassoula} E.,  2005, \mn@doi [\mnras] {10.1111/j.1365-2966.2005.08872.x},
  \href {http://adsabs.harvard.edu/abs/2005MNRAS.358.1477A} {358, 1477}

\bibitem[\protect\citeauthoryear{{Balona} et~al.,}{{Balona}
  et~al.}{2019}]{Balona-2019}
{Balona} L.~A.,  et~al., 2019, \mn@doi [\mnras] {10.1093/mnras/stz586}, \href
  {https://ui.adsabs.harvard.edu/abs/2019MNRAS.485.3457B} {485, 3457}

\bibitem[\protect\citeauthoryear{{Banerjee} et~al.,}{{Banerjee}
  et~al.}{2018}]{Banerjee-2018}
{Banerjee} D.~P.~K.,  et~al., 2018, \mn@doi [\apj] {10.3847/1538-4357/aae5d3},
  \href {https://ui.adsabs.harvard.edu/abs/2018ApJ...867...99B} {867, 99}

\bibitem[\protect\citeauthoryear{{Bellm} et~al.,}{{Bellm}
  et~al.}{2019}]{Bellm-2019}
{Bellm} E.~C.,  et~al., 2019, \mn@doi [\pasp] {10.1088/1538-3873/aaecbe}, \href
  {http://adsabs.harvard.edu/abs/2019PASP..131a8002B} {131, 018002}

\bibitem[\protect\citeauthoryear{{Benavente}, {Protopapas}  \&
  {Pichara}}{{Benavente} et~al.}{2017}]{Benavente-2017}
{Benavente} P.,  {Protopapas} P.,   {Pichara} K.,  2017, \mn@doi [\apj]
  {10.3847/1538-4357/aa7f2d}, \href
  {http://adsabs.harvard.edu/abs/2017ApJ...845..147B} {845, 147}

\bibitem[\protect\citeauthoryear{{Bhatti}, {Richmond}, {Ford}  \&
  {Petro}}{{Bhatti} et~al.}{2010}]{Bhatti-2010}
{Bhatti} W.~A.,  {Richmond} M.~W.,  {Ford} H.~C.,   {Petro} L.~D.,  2010,
  \mn@doi [\apjs] {10.1088/0067-0049/186/2/233}, \href
  {http://adsabs.harvard.edu/abs/2010ApJS..186..233B} {186, 233}

\bibitem[\protect\citeauthoryear{{Binney} \& {Merrifield}}{{Binney} \&
  {Merrifield}}{1998}]{Binney-1998}
{Binney} J.,  {Merrifield} M.,  1998, {Galactic Astronomy}.
Princeton University Press

\bibitem[\protect\citeauthoryear{{Bloom} et~al.,}{{Bloom}
  et~al.}{2012}]{Bloom-2012}
{Bloom} J.~S.,  et~al., 2012, \mn@doi [\pasp] {10.1086/668468}, \href
  {http://adsabs.harvard.edu/abs/2012PASP..124.1175B} {124, 1175}

\bibitem[\protect\citeauthoryear{{Braga}, {Contreras Ramos}, {Minniti},
  {Ferreira Lopes}, {Catelan}, {Minniti}, {Nikzat}  \& {Zoccali}}{{Braga}
  et~al.}{2019}]{Braga-2019}
{Braga} V.~F.,  {Contreras Ramos} R.,  {Minniti} D.,  {Ferreira Lopes} C.~E.,
  {Catelan} M.,  {Minniti} J.~H.,  {Nikzat} F.,   {Zoccali} M.,  2019, \mn@doi
  [\aap] {10.1051/0004-6361/201935103}, \href
  {https://ui.adsabs.harvard.edu/abs/2019A&A...625A.151B} {625, A151}

\bibitem[\protect\citeauthoryear{{Cabrera-Vives}, {Reyes}, {F{\"o}rster},
  {Est{\'e}vez}  \& {Maureira}}{{Cabrera-Vives}
  et~al.}{2017}]{Cabrera-Vives-2017}
{Cabrera-Vives} G.,  {Reyes} I.,  {F{\"o}rster} F.,  {Est{\'e}vez} P.~A.,
  {Maureira} J.-C.,  2017, \mn@doi [\apj] {10.3847/1538-4357/836/1/97}, \href
  {http://adsabs.harvard.edu/abs/2017ApJ...836...97C} {836, 97}

\bibitem[\protect\citeauthoryear{{Cardelli}, {Clayton}  \& {Mathis}}{{Cardelli}
  et~al.}{1989}]{Cardelli-1989}
{Cardelli} J.~A.,  {Clayton} G.~C.,   {Mathis} J.~S.,  1989, \mn@doi [\apj]
  {10.1086/167900}, \href {http://adsabs.harvard.edu/abs/1989ApJ...345..245C}
  {345, 245}

\bibitem[\protect\citeauthoryear{{Catelan} \& {Smith}}{{Catelan} \&
  {Smith}}{2015}]{Catelan-2015book}
{Catelan} M.,  {Smith} H.~A.,  2015, {Pulsating Stars}.
Wiley-VCH

\bibitem[\protect\citeauthoryear{{Contreras Pe{\~n}a} et~al.,}{{Contreras
  Pe{\~n}a} et~al.}{2017}]{ContrerasPenaKU-2017}
{Contreras Pe{\~n}a} C.,  et~al., 2017, \mn@doi [\mnras]
  {10.1093/mnras/stw2802}, \href
  {https://ui.adsabs.harvard.edu/abs/2017MNRAS.465.3039C} {465, 3039}

\bibitem[\protect\citeauthoryear{{Contreras Pena} et~al.,}{{Contreras Pena}
  et~al.}{2017}]{ContrerasPenaDM-2017}
{Contreras Pena} C.,  et~al., 2017, VizieR Online Data Catalog, \href
  {https://ui.adsabs.harvard.edu/abs/2017yCat..74653011C} {p. J/MNRAS/465/3011}

\bibitem[\protect\citeauthoryear{{Contreras Ramos} et~al.,}{{Contreras Ramos}
  et~al.}{2017}]{ContrerasRamos-2017}
{Contreras Ramos} R.,  et~al., 2017, \mn@doi [\aap]
  {10.1051/0004-6361/201731462}, \href
  {https://ui.adsabs.harvard.edu/abs/2017A&A...608A.140C} {608, A140}

\bibitem[\protect\citeauthoryear{{Cort{\'e}s} et~al.,}{{Cort{\'e}s}
  et~al.}{2015}]{Cortes-2015}
{Cort{\'e}s} C.,  et~al., 2015, \mn@doi [\aap] {10.1051/0004-6361/201424155},
  \href {http://adsabs.harvard.edu/abs/2015A%26A...581A..68C} {581, A68}

\bibitem[\protect\citeauthoryear{{Cort{\'e}s}, {Minniti}  \&
  {Villanova}}{{Cort{\'e}s} et~al.}{2019}]{CortesDM-2019}
{Cort{\'e}s} C.~C.,  {Minniti} D.,   {Villanova} S.,  2019, \mn@doi [\mnras]
  {10.1093/mnras/sty3224}, \href
  {https://ui.adsabs.harvard.edu/abs/2019MNRAS.485.4502C} {485, 4502}

\bibitem[\protect\citeauthoryear{{Cross}, {Collins}, {Hambly}, {Blake}, {Read},
  {Sutorius}, {Mann}  \& {Williams}}{{Cross} et~al.}{2009}]{Cross-2009}
{Cross} N.~J.~G.,  {Collins} R.~S.,  {Hambly} N.~C.,  {Blake} R.~P.,  {Read}
  M.~A.,  {Sutorius} E.~T.~W.,  {Mann} R.~G.,   {Williams} P.~M.,  2009,
  \mn@doi [\mnras] {10.1111/j.1365-2966.2009.15396.x}, \href
  {http://adsabs.harvard.edu/abs/2009MNRAS.399.1730C} {399, 1730}

\bibitem[\protect\citeauthoryear{{Cross} et~al.,}{{Cross}
  et~al.}{2012}]{Cross-2012}
{Cross} N.~J.~G.,  et~al., 2012, \mn@doi [\aap] {10.1051/0004-6361/201219505},
  \href {http://adsabs.harvard.edu/abs/2012A%26A...548A.119C} {548, A119}

\bibitem[\protect\citeauthoryear{{Damerdji}, {Klotz}  \& {Bo{\"e}r}}{{Damerdji}
  et~al.}{2007}]{Damerdji-2007}
{Damerdji} Y.,  {Klotz} A.,   {Bo{\"e}r} M.,  2007, \mn@doi [\aj]
  {10.1086/511747}, \href {http://adsabs.harvard.edu/abs/2007AJ....133.1470D}
  {133, 1470}

\bibitem[\protect\citeauthoryear{{De Medeiros} et~al.,}{{De Medeiros}
  et~al.}{2013}]{DeMedeiros-2013}
{De Medeiros} J.~R.,  et~al., 2013, \mn@doi [\aap]
  {10.1051/0004-6361/201219415}, \href
  {http://adsabs.harvard.edu/abs/2013A\%26A...555A..63D} {555, A63}

\bibitem[\protect\citeauthoryear{{Debosscher}, {Sarro}, {Aerts}, {Cuypers},
  {Vandenbussche}, {Garrido}  \& {Solano}}{{Debosscher}
  et~al.}{2007}]{Debosscher-2007}
{Debosscher} J.,  {Sarro} L.~M.,  {Aerts} C.,  {Cuypers} J.,  {Vandenbussche}
  B.,  {Garrido} R.,   {Solano} E.,  2007, \mn@doi [\aap]
  {10.1051/0004-6361:20077638}, \href
  {http://adsabs.harvard.edu/abs/2007A\%26A...475.1159D} {475, 1159}

\bibitem[\protect\citeauthoryear{{Deleuil} et~al.,}{{Deleuil}
  et~al.}{2018}]{Deleuil-2018}
{Deleuil} M.,  et~al., 2018, \mn@doi [\aap] {10.1051/0004-6361/201731068},
  \href {https://ui.adsabs.harvard.edu/abs/2018A&A...619A..97D} {619, A97}

\bibitem[\protect\citeauthoryear{{Drake} et~al.,}{{Drake}
  et~al.}{2014}]{Drake-2014}
{Drake} A.~J.,  et~al., 2014, \mn@doi [\apjs] {10.1088/0067-0049/213/1/9},
  \href {http://adsabs.harvard.edu/abs/2014ApJS..213....9D} {213, 9}

\bibitem[\protect\citeauthoryear{{Dubath} et~al.,}{{Dubath}
  et~al.}{2011}]{Dubath-2011}
{Dubath} P.,  et~al., 2011, \mn@doi [\mnras]
  {10.1111/j.1365-2966.2011.18575.x}, \href
  {http://adsabs.harvard.edu/abs/2011MNRAS.414.2602D} {414, 2602}

\bibitem[\protect\citeauthoryear{{Dubath} et~al.,}{{Dubath}
  et~al.}{2012}]{Dubath-2012}
{Dubath} P.,  et~al., 2012, VizieR Online Data Catalog, \href
  {http://adsabs.harvard.edu/abs/2012yCat..74142602D} {741, 42602}

\bibitem[\protect\citeauthoryear{{Dupuy} \& {Hoffman}}{{Dupuy} \&
  {Hoffman}}{1985}]{Dupuy-1985}
{Dupuy} D.~L.,  {Hoffman} G.~A.,  1985, International Amateur-Professional
  Photoelectric Photometry Communications, \href
  {http://adsabs.harvard.edu/abs/1985IAPPP..20....1D} {20, 1}

\bibitem[\protect\citeauthoryear{{Dworetsky}}{{Dworetsky}}{1983}]{Dworetsky-1983}
{Dworetsky} M.~M.,  1983, \mnras, \href
  {http://adsabs.harvard.edu/abs/1983MNRAS.203..917D} {203, 917}

\bibitem[\protect\citeauthoryear{{Fern{\'a}ndez}, {Minniti}, {Pietrzynski},
  {Gieren}, {Ru{\'\i}z}, {Zoccali}, {Udalski}  \& {Szeifert}}{{Fern{\'a}ndez}
  et~al.}{2006}]{Fernandez-2006}
{Fern{\'a}ndez} J.~M.,  {Minniti} D.,  {Pietrzynski} G.,  {Gieren} W.,
  {Ru{\'\i}z} M.~T.,  {Zoccali} M.,  {Udalski} A.,   {Szeifert} T.,  2006,
  \mn@doi [\apj] {10.1086/500921}, \href
  {https://ui.adsabs.harvard.edu/abs/2006ApJ...647..587F} {647, 587}

\bibitem[\protect\citeauthoryear{{Ferreira Lopes} \& {Cross}}{{Ferreira Lopes}
  \& {Cross}}{2016}]{FerreiraLopes-2016papI}
{Ferreira Lopes} C.~E.,  {Cross} N.~J.~G.,  2016, \mn@doi [\aap]
  {10.1051/0004-6361/201526733}, \href
  {http://adsabs.harvard.edu/abs/2016A%26A...586A..36F} {586, A36}

\bibitem[\protect\citeauthoryear{{Ferreira Lopes} \& {Cross}}{{Ferreira Lopes}
  \& {Cross}}{2017}]{FerreiraLopes-2017papII}
{Ferreira Lopes} C.~E.,  {Cross} N.~J.~G.,  2017, \mn@doi [\aap]
  {10.1051/0004-6361/201630109}, \href
  {http://adsabs.harvard.edu/abs/2017A%26A...604A.121F} {604, A121}

\bibitem[\protect\citeauthoryear{{Ferreira Lopes}, {Dekany}, {Catelan},
  {Cross}, {Angeloni}, {Leao}  \& {De Medeiros}}{{Ferreira Lopes}
  et~al.}{2015a}]{FerreiraLopes-2015wfcam}
{Ferreira Lopes} C.~E.,  {Dekany} I.,  {Catelan} C.,  {Cross} N. J. G.~C.,
  {Angeloni} R.,  {Leao} I.~C.,   {De Medeiros} J.~R.,  2015a, \mn@doi [\aap]
  {10.1051/0004-6361/201423793}, \href
  {http://www.aanda.org/articles/aa/abs/2015/01/aa23793-14/aa23793-14.html}
  {573, A100}

\bibitem[\protect\citeauthoryear{{Ferreira Lopes} et~al.,}{{Ferreira Lopes}
  et~al.}{2015b}]{FerreiraLopes-2015mgiant}
{Ferreira Lopes} C.~E.,  et~al., 2015b, \mn@doi [\aap]
  {10.1051/0004-6361/201425503}, \href
  {http://adsabs.harvard.edu/abs/2015A%26A...583A.122F} {583, A122}

\bibitem[\protect\citeauthoryear{{Ferreira Lopes}, {Le{\~a}o}, {de Freitas},
  {Canto Martins}, {Catelan}  \& {De Medeiros}}{{Ferreira Lopes}
  et~al.}{2015c}]{FerreiraLopes-2015cycles}
{Ferreira Lopes} C.~E.,  {Le{\~a}o} I.~C.,  {de Freitas} D.~B.,  {Canto
  Martins} B.~L.,  {Catelan} M.,   {De Medeiros} J.~R.,  2015c, \mn@doi [\aap]
  {10.1051/0004-6361/201424900}, \href
  {http://adsabs.harvard.edu/abs/2015A%26A...583A.134F} {583, A134}

\bibitem[\protect\citeauthoryear{Ferreira Lopes, Cross  \&
  Jablonski}{Ferreira Lopes et~al.}{2018a}]{FerreiraLopes-2018papV}
Ferreira Lopes C.~E.,  Cross N. J.~G.,   Jablonski F.,  2018a, \mn@doi
  [Monthly Notices of the Royal Astronomical Society] {10.1093/mnras/sty2469},
  0, 0

\bibitem[\protect\citeauthoryear{Ferreira Lopes, Cross  \&
  Jablonski}{Ferreira Lopes et~al.}{2018b}]{FerreiraLopes-2018papIII}
Ferreira Lopes C.~E.,  Cross N. J.~G.,   Jablonski F.,  2018b, \mn@doi
  [Monthly Notices of the Royal Astronomical Society] {10.1093/mnras/sty2469},
  481, 3083

\bibitem[\protect\citeauthoryear{{Garc{\'\i}a} et~al.,}{{Garc{\'\i}a}
  et~al.}{2014}]{Garcia-2014}
{Garc{\'\i}a} R.~A.,  et~al., 2014, \mn@doi [\aap]
  {10.1051/0004-6361/201423888}, \href
  {https://ui.adsabs.harvard.edu/abs/2014A&A...572A..34G} {572, A34}

\bibitem[\protect\citeauthoryear{{Gillon} et~al.,}{{Gillon}
  et~al.}{2017}]{Gillon-2017}
{Gillon} M.,  et~al., 2017, \mn@doi [\nat] {10.1038/nature21360}, \href
  {https://ui.adsabs.harvard.edu/abs/2017Natur.542..456G} {542, 456}

\bibitem[\protect\citeauthoryear{{Gonzalez} \& {Gadotti}}{{Gonzalez} \&
  {Gadotti}}{2016}]{Gonzalez-2016}
{Gonzalez} O.~A.,  {Gadotti} D.,  2016, in {Laurikainen} E.,  {Peletier} R.,
  {Gadotti} D.,  eds,  Astrophysics and Space Science Library Vol. 418,
  Galactic Bulges. p.~199 (\mn@eprint {arXiv} {1503.07252}),
  \mn@doi{10.1007/978-3-319-19378-6_9}

\bibitem[\protect\citeauthoryear{{Gonzalez}, {Rejkuba}, {Zoccali}, {Valenti},
  {Minniti}, {Schultheis}, {Tobar}  \& {Chen}}{{Gonzalez}
  et~al.}{2012}]{Gonzalez-2012}
{Gonzalez} O.~A.,  {Rejkuba} M.,  {Zoccali} M.,  {Valenti} E.,  {Minniti} D.,
  {Schultheis} M.,  {Tobar} R.,   {Chen} B.,  2012, \mn@doi [\aap]
  {10.1051/0004-6361/201219222}, \href
  {http://adsabs.harvard.edu/abs/2012A%26A...543A..13G} {543, A13}

\bibitem[\protect\citeauthoryear{{Gonzalez} et~al.,}{{Gonzalez}
  et~al.}{2018}]{Gonzalez-2018}
{Gonzalez} O.~A.,  et~al., 2018, \mn@doi [\mnras] {10.1093/mnrasl/sly171},
  \href {https://ui.adsabs.harvard.edu/abs/2018MNRAS.481L.130G} {481, L130}

\bibitem[\protect\citeauthoryear{{Graham}, {Drake}, {Djorgovski}, {Mahabal}  \&
  {Donalek}}{{Graham} et~al.}{2017}]{Graham-2017}
{Graham} M.,  {Drake} A.,  {Djorgovski} S.~G.,  {Mahabal} A.,   {Donalek} C.,
  2017, in European Physical Journal Web of Conferences. p. 03001,
  \mn@doi{10.1051/epjconf/201715203001}

\bibitem[\protect\citeauthoryear{{Gran}, {Minniti}, {Saito}, {Navarrete},
  {D{\'e}k{\'a}ny}, {McDonald}, {Contreras Ramos}  \& {Catelan}}{{Gran}
  et~al.}{2015}]{Gran-2015}
{Gran} F.,  {Minniti} D.,  {Saito} R.~K.,  {Navarrete} C.,  {D{\'e}k{\'a}ny}
  I.,  {McDonald} I.,  {Contreras Ramos} R.,   {Catelan} M.,  2015, \mn@doi
  [\aap] {10.1051/0004-6361/201424333}, \href
  {https://ui.adsabs.harvard.edu/abs/2015A&A...575A.114G} {575, A114}

\bibitem[\protect\citeauthoryear{{Gran} et~al.,}{{Gran}
  et~al.}{2016}]{Gran-2016}
{Gran} F.,  et~al., 2016, \mn@doi [\aap] {10.1051/0004-6361/201527511}, \href
  {http://adsabs.harvard.edu/abs/2016A%26A...591A.145G} {591, A145}

\bibitem[\protect\citeauthoryear{{Guo} et~al.,}{{Guo} et~al.}{2019}]{Guo-2019}
{Guo} Z.,  et~al., 2019, arXiv e-prints, \href
  {https://ui.adsabs.harvard.edu/abs/2019arXiv191200729G} {p. arXiv:1912.00729}

\bibitem[\protect\citeauthoryear{{Hajdu}, {D{\'e}k{\'a}ny}, {Catelan}  \&
  {Grebel}}{{Hajdu} et~al.}{2019}]{Hajdu2019}
{Hajdu} G.,  {D{\'e}k{\'a}ny} I.,  {Catelan} M.,   {Grebel} E.~K.,  2019, arXiv
  e-prints, \href {https://ui.adsabs.harvard.edu/abs/2019arXiv190806160H} {p.
  arXiv:1908.06160}

\bibitem[\protect\citeauthoryear{{Hall}}{{Hall}}{1932}]{Hall-1932}
{Hall} J.~S.,  1932, \mn@doi [Proceedings of the National Academy of Science]
  {10.1073/pnas.18.5.365}, \href
  {http://adsabs.harvard.edu/abs/1932PNAS...18..365H} {18, 365}

\bibitem[\protect\citeauthoryear{{Hall}}{{Hall}}{1934}]{Hall-1934}
{Hall} J.~S.,  1934, \mn@doi [\apj] {10.1086/143527}, \href
  {http://adsabs.harvard.edu/abs/1934ApJ....79..145H} {79, 145}

\bibitem[\protect\citeauthoryear{{He{\l}miniak}, {Devor}, {Minniti}  \&
  {Sybilski}}{{He{\l}miniak} et~al.}{2013}]{Helminiak-2013}
{He{\l}miniak} K.~G.,  {Devor} J.,  {Minniti} D.,   {Sybilski} P.,  2013,
  \mn@doi [\mnras] {10.1093/mnras/stt675}, \href
  {https://ui.adsabs.harvard.edu/abs/2013MNRAS.432.2895H} {432, 2895}

\bibitem[\protect\citeauthoryear{{Hoffleit}}{{Hoffleit}}{1987}]{Hoffleit-1987}
{Hoffleit} D.,  1987, Journal of the American Association of Variable Star
  Observers (JAAVSO), \href {http://adsabs.harvard.edu/abs/1987JAVSO..16...29H}
  {16, 29}

\bibitem[\protect\citeauthoryear{{Huang} et~al.,}{{Huang}
  et~al.}{2018}]{Huang-2018}
{Huang} C.~D.,  et~al., 2018, \mn@doi [\apj] {10.3847/1538-4357/aab6b3}, \href
  {https://ui.adsabs.harvard.edu/abs/2018ApJ...857...67H} {857, 67}

\bibitem[\protect\citeauthoryear{{Irwin} et~al.,}{{Irwin}
  et~al.}{2004}]{Irwin-2004}
{Irwin} M.~J.,  et~al., 2004, in {Quinn} P.~J.,  {Bridger} A.,  eds,  Society
  of Photo-Optical Instrumentation Engineers (SPIE) Conference Series Vol.
  5493, Optimizing Scientific Return for Astronomy through Information
  Technologies. pp 411--422, \mn@doi{10.1117/12.551449}

\bibitem[\protect\citeauthoryear{{Ita} et~al.,}{{Ita} et~al.}{2018}]{Ita-2018}
{Ita} Y.,  et~al., 2018, \mn@doi [\mnras] {10.1093/mnras/sty2539}, \href
  {http://adsabs.harvard.edu/abs/2018MNRAS.tmp.2428I} {}

\bibitem[\protect\citeauthoryear{{Ivezic} et~al.,}{{Ivezic}
  et~al.}{2008}]{Ivezic-2008}
{Ivezic} Z.,  et~al., 2008, \mn@doi [Serbian Astronomical Journal]
  {10.2298/SAJ0876001I}, \href
  {http://adsabs.harvard.edu/abs/2008SerAJ.176....1I} {176, 1}

\bibitem[\protect\citeauthoryear{{Ivezi{\'c}} et~al.,}{{Ivezi{\'c}}
  et~al.}{2019}]{Ivezic-2019}
{Ivezi{\'c}} {\v{Z}}.,  et~al., 2019, \mn@doi [\apj]
  {10.3847/1538-4357/ab042c}, \href
  {https://ui.adsabs.harvard.edu/abs/2019ApJ...873..111I} {873, 111}

\bibitem[\protect\citeauthoryear{{Kaiser} et~al.,}{{Kaiser}
  et~al.}{2002}]{Kaiser-2002}
{Kaiser} N.,  et~al., 2002, in {Tyson} J.~A.,  {Wolff} S.,  eds,  Society of
  Photo-Optical Instrumentation Engineers (SPIE) Conference Series Vol. 4836,
  Survey and Other Telescope Technologies and Discoveries. pp 154--164,
  \mn@doi{10.1117/12.457365}

\bibitem[\protect\citeauthoryear{{Kim}, {Protopapas}, {Byun}, {Alcock},
  {Khardon}  \& {Trichas}}{{Kim} et~al.}{2011}]{Kim-2011}
{Kim} D.-W.,  {Protopapas} P.,  {Byun} Y.-I.,  {Alcock} C.,  {Khardon} R.,
  {Trichas} M.,  2011, \mn@doi [\apj] {10.1088/0004-637X/735/2/68}, \href
  {http://adsabs.harvard.edu/abs/2011ApJ...735...68K} {735, 68}

\bibitem[\protect\citeauthoryear{Kim, Protopapas, Bailer-Jones, Byun, Chang,
  Marquette  \& Shin}{Kim et~al.}{2014}]{kim2014epoch}
Kim D.-W.,  Protopapas P.,  Bailer-Jones C.~A.,  Byun Y.-I.,  Chang S.-W.,
  Marquette J.-B.,   Shin M.-S.,  2014, Astronomy \& Astrophysics, 566, A43

\bibitem[\protect\citeauthoryear{{Lomb}}{{Lomb}}{1976}]{Lomb-1976}
{Lomb} N.~R.,  1976, \mn@doi [\apss] {10.1007/BF00648343}, \href
  {http://adsabs.harvard.edu/abs/1976Ap\%26SS..39..447L} {39, 447}

\bibitem[\protect\citeauthoryear{Long, El~Karoui, Rice, Richards  \&
  Bloom}{Long et~al.}{2012}]{long2012optimizing}
Long J.~P.,  El~Karoui N.,  Rice J.~A.,  Richards J.~W.,   Bloom J.~S.,  2012,
  Publications of the Astronomical Society of the Pacific, 124, 280

\bibitem[\protect\citeauthoryear{{Lucas} et~al.,}{{Lucas}
  et~al.}{2017}]{LucasSM-2017}
{Lucas} P.~W.,  et~al., 2017, \mn@doi [\mnras] {10.1093/mnras/stx2058}, \href
  {https://ui.adsabs.harvard.edu/abs/2017MNRAS.472.2990L} {472, 2990}

\bibitem[\protect\citeauthoryear{Mackenzie, Pichara  \& Protopapas}{Mackenzie
  et~al.}{2016}]{mackenzie2016clustering}
Mackenzie C.,  Pichara K.,   Protopapas P.,  2016, The Astrophysical Journal,
  820, 138

\bibitem[\protect\citeauthoryear{Mahabal, Sheth, Gieseke, Pai, Djorgovski,
  Drake, Graham  et~al.}{Mahabal et~al.}{2017}]{mahabal2017deep}
Mahabal A.,  Sheth K.,  Gieseke F.,  Pai A.,  Djorgovski S.~G.,  Drake A.,
  Graham M.,   et~al., 2017, arXiv preprint arXiv:1709.06257

\bibitem[\protect\citeauthoryear{{Mainzer} et~al.,}{{Mainzer}
  et~al.}{2011}]{NEOWISE}
{Mainzer} A.,  et~al., 2011, \mn@doi [\apj] {10.1088/0004-637X/731/1/53}, \href
  {https://ui.adsabs.harvard.edu/abs/2011ApJ...731...53M} {731, 53}

\bibitem[\protect\citeauthoryear{{McQuillan}, {Aigrain}  \&
  {Mazeh}}{{McQuillan} et~al.}{2013}]{McQuillan-2013}
{McQuillan} A.,  {Aigrain} S.,   {Mazeh} T.,  2013, \mn@doi [\mnras]
  {10.1093/mnras/stt536}, \href
  {http://adsabs.harvard.edu/abs/2013MNRAS.432.1203M} {432, 1203}

\bibitem[\protect\citeauthoryear{{McQuillan}, {Mazeh}  \&
  {Aigrain}}{{McQuillan} et~al.}{2014}]{McQuillan-2014}
{McQuillan} A.,  {Mazeh} T.,   {Aigrain} S.,  2014, \mn@doi [\apjs]
  {10.1088/0067-0049/211/2/24}, \href
  {https://ui.adsabs.harvard.edu/abs/2014ApJS..211...24M} {211, 24}

\bibitem[\protect\citeauthoryear{{Medina} et~al.,}{{Medina}
  et~al.}{2018}]{Medina-2018}
{Medina} N.,  et~al., 2018, \mn@doi [\apj] {10.3847/1538-4357/aacc65}, \href
  {https://ui.adsabs.harvard.edu/abs/2018ApJ...864...11M} {864, 11}

\bibitem[\protect\citeauthoryear{{Minniti} et~al.,}{{Minniti}
  et~al.}{2007}]{Minniti-2007exo}
{Minniti} D.,  et~al., 2007, \mn@doi [\apj] {10.1086/512722}, \href
  {https://ui.adsabs.harvard.edu/abs/2007ApJ...660..858M} {660, 858}

\bibitem[\protect\citeauthoryear{{Minniti} et~al.,}{{Minniti}
  et~al.}{2010}]{Minniti-2010}
{Minniti} D.,  et~al., 2010, \mn@doi [\na] {10.1016/j.newast.2009.12.002},
  \href {http://adsabs.harvard.edu/abs/2010NewA...15..433M} {15, 433}

\bibitem[\protect\citeauthoryear{{Minniti} et~al.,}{{Minniti}
  et~al.}{2015}]{MinnitiCR-2015}
{Minniti} D.,  et~al., 2015, \mn@doi [\apjl] {10.1088/2041-8205/810/2/L20},
  \href {https://ui.adsabs.harvard.edu/abs/2015ApJ...810L..20M} {810, L20}

\bibitem[\protect\citeauthoryear{{Minniti} et~al.,}{{Minniti}
  et~al.}{2017}]{Minniti-2017}
{Minniti} D.,  et~al., 2017, \mn@doi [\aj] {10.3847/1538-3881/aa5be4}, \href
  {http://adsabs.harvard.edu/abs/2017AJ....153..179M} {153, 179}

\bibitem[\protect\citeauthoryear{{Minniti} et~al.,}{{Minniti}
  et~al.}{2018}]{Minniti-2018ebv}
{Minniti} D.,  et~al., 2018, \mn@doi [\aap] {10.1051/0004-6361/201732099},
  \href {https://ui.adsabs.harvard.edu/abs/2018A&A...616A..26M} {616, A26}

\bibitem[\protect\citeauthoryear{{Mowlavi} et~al.,}{{Mowlavi}
  et~al.}{2018}]{Mowlavi-2018}
{Mowlavi} N.,  et~al., 2018, \mn@doi [\aap] {10.1051/0004-6361/201833366},
  \href {https://ui.adsabs.harvard.edu/abs/2018A&A...618A..58M} {618, A58}

\bibitem[\protect\citeauthoryear{{Navarro}, {Minniti}  \& {Contreras
  Ramos}}{{Navarro} et~al.}{2017}]{Navarro-2017}
{Navarro} M.~G.,  {Minniti} D.,   {Contreras Ramos} R.,  2017, \mn@doi [\apjl]
  {10.3847/2041-8213/aa9b29}, \href
  {https://ui.adsabs.harvard.edu/abs/2017ApJ...851L..13N} {851, L13}

\bibitem[\protect\citeauthoryear{{Navarro}, {Minniti}  \&
  {Contreras-Ramos}}{{Navarro} et~al.}{2018}]{Navarro-2018}
{Navarro} M.~G.,  {Minniti} D.,   {Contreras-Ramos} R.,  2018, \mn@doi [\apjl]
  {10.3847/2041-8213/aae08a}, \href
  {https://ui.adsabs.harvard.edu/abs/2018ApJ...865L...5N} {865, L5}

\bibitem[\protect\citeauthoryear{{Navarro}, {Minniti}, {Pullen}  \& {Contreras
  Ramos}}{{Navarro} et~al.}{2019}]{Navarro-2019}
{Navarro} M.~G.,  {Minniti} D.,  {Pullen} J.,   {Contreras Ramos} R.,  2019,
  arXiv e-prints, \href {https://ui.adsabs.harvard.edu/abs/2019arXiv191112897N}
  {p. arXiv:1911.12897}

\bibitem[\protect\citeauthoryear{{Nun}, {Pichara}, {Protopapas}  \&
  {Kim}}{{Nun} et~al.}{2014}]{Nun-2014}
{Nun} I.,  {Pichara} K.,  {Protopapas} P.,   {Kim} D.-W.,  2014, \mn@doi [\apj]
  {10.1088/0004-637X/793/1/23}, \href
  {http://adsabs.harvard.edu/abs/2014ApJ...793...23N} {793, 23}

\bibitem[\protect\citeauthoryear{Nun, Protopapas, Sim, Zhu, Dave, Castro  \&
  Pichara}{Nun et~al.}{2015}]{nun2015fats}
Nun I.,  Protopapas P.,  Sim B.,  Zhu M.,  Dave R.,  Castro N.,   Pichara K.,
  2015, arXiv preprint arXiv:1506.00010

\bibitem[\protect\citeauthoryear{{Paz-Chinch{\'o}n} et~al.,}{{Paz-Chinch{\'o}n}
  et~al.}{2015}]{Paz-Chinchon-2015}
{Paz-Chinch{\'o}n} F.,  et~al., 2015, preprint, \href
  {http://adsabs.harvard.edu/abs/2015arXiv150205051P} {} (\mn@eprint {arXiv}
  {1502.05051})

\bibitem[\protect\citeauthoryear{{Perryman}}{{Perryman}}{2005}]{Perryman-2005}
{Perryman} M.~A.~C.,  2005, in {Seidelmann} P.~K.,  {Monet} A.~K.~B.,  eds,
  Astronomical Society of the Pacific Conference Series Vol. 338, Astrometry in
  the Age of the Next Generation of Large Telescopes. p.~3

\bibitem[\protect\citeauthoryear{{Pichara} \& {Protopapas}}{{Pichara} \&
  {Protopapas}}{2013}]{Pichara-2013}
{Pichara} K.,  {Protopapas} P.,  2013, \mn@doi [\apj]
  {10.1088/0004-637X/777/2/83}, \href
  {http://adsabs.harvard.edu/abs/2013ApJ...777...83P} {777, 83}

\bibitem[\protect\citeauthoryear{{Pichara}, {Protopapas}  \&
  {Le{\'o}n}}{{Pichara} et~al.}{2016}]{Pichara-2016}
{Pichara} K.,  {Protopapas} P.,   {Le{\'o}n} D.,  2016, \mn@doi [\apj]
  {10.3847/0004-637X/819/1/18}, \href
  {http://adsabs.harvard.edu/abs/2016ApJ...819...18P} {819, 18}

\bibitem[\protect\citeauthoryear{{Pietrukowicz} et~al.,}{{Pietrukowicz}
  et~al.}{2010}]{Pietrukowicz-2010}
{Pietrukowicz} P.,  et~al., 2010, \mn@doi [\aap] {10.1051/0004-6361/200912141},
  \href {https://ui.adsabs.harvard.edu/abs/2010A&A...509A...4P} {509, A4}

\bibitem[\protect\citeauthoryear{{Rauer} et~al.,}{{Rauer}
  et~al.}{2014}]{Rauer-2014}
{Rauer} H.,  et~al., 2014, \mn@doi [Experimental Astronomy]
  {10.1007/s10686-014-9383-4}, \href
  {http://adsabs.harvard.edu/abs/2014ExA....38..249R} {38, 249}

\bibitem[\protect\citeauthoryear{{Rice}, {Reipurth}, {Wolk}, {Vaz}  \&
  {Cross}}{{Rice} et~al.}{2015}]{Rice-2015}
{Rice} T.~S.,  {Reipurth} B.,  {Wolk} S.~J.,  {Vaz} L.~P.,   {Cross} N.~J.~G.,
  2015, \mn@doi [\aj] {10.1088/0004-6256/150/4/132}, \href
  {http://adsabs.harvard.edu/abs/2015AJ....150..132R} {150, 132}

\bibitem[\protect\citeauthoryear{{Richards} et~al.,}{{Richards}
  et~al.}{2011}]{Richards-2011}
{Richards} J.~W.,  et~al., 2011, \mn@doi [\apj] {10.1088/0004-637X/733/1/10},
  \href {http://adsabs.harvard.edu/abs/2011ApJ...733...10R} {733, 10}

\bibitem[\protect\citeauthoryear{{Ricker} et~al.,}{{Ricker}
  et~al.}{2015}]{Ricker-2015}
{Ricker} G.~R.,  et~al., 2015, \mn@doi [Journal of Astronomical Telescopes,
  Instruments, and Systems] {10.1117/1.JATIS.1.1.014003}, \href
  {http://adsabs.harvard.edu/abs/2015JATIS...1a4003R} {1, 014003}

\bibitem[\protect\citeauthoryear{{Saito}, {Minniti}, {Angeloni}  \&
  {Catelan}}{{Saito} et~al.}{2012}]{SaitoTel-2012ATel}
{Saito} R.~K.,  {Minniti} D.,  {Angeloni} R.,   {Catelan} M.,  2012, The
  Astronomer's Telegram, \href
  {https://ui.adsabs.harvard.edu/abs/2012ATel.4426....1S} {4426, 1}

\bibitem[\protect\citeauthoryear{{Saito} et~al.,}{{Saito}
  et~al.}{2013}]{Saito-2013}
{Saito} R.~K.,  et~al., 2013, \mn@doi [\aap] {10.1051/0004-6361/201321260},
  \href {https://ui.adsabs.harvard.edu/abs/2013A&A...554A.123S} {554, A123}

\bibitem[\protect\citeauthoryear{{Scargle}}{{Scargle}}{1982}]{Scargle-1982}
{Scargle} J.~D.,  1982, \mn@doi [\apj] {10.1086/160554}, \href
  {http://adsabs.harvard.edu/abs/1982ApJ...263..835S} {263, 835}

\bibitem[\protect\citeauthoryear{{Shappee} \& {Stanek}}{{Shappee} \&
  {Stanek}}{2011}]{Shappee-2011}
{Shappee} B.~J.,  {Stanek} K.~Z.,  2011, \mn@doi [\apj]
  {10.1088/0004-637X/733/2/124}, \href
  {http://adsabs.harvard.edu/abs/2011ApJ...733..124S} {733, 124}

\bibitem[\protect\citeauthoryear{{Smith} et~al.,}{{Smith} et~al.}{2018}]{VIRAC}
{Smith} L.~C.,  et~al., 2018, \mn@doi [\mnras] {10.1093/mnras/stx2789}, \href
  {https://ui.adsabs.harvard.edu/abs/2018MNRAS.474.1826S} {474, 1826}

\bibitem[\protect\citeauthoryear{{Sokolovsky} et~al.,}{{Sokolovsky}
  et~al.}{2017}]{Sokolovsky-2017}
{Sokolovsky} K.~V.,  et~al., 2017, \mn@doi [\mnras] {10.1093/mnras/stw2262},
  \href {http://adsabs.harvard.edu/abs/2017MNRAS.464..274S} {464, 274}

\bibitem[\protect\citeauthoryear{{Soszy{\'n}ski} et~al.,}{{Soszy{\'n}ski}
  et~al.}{2009}]{Soszynski-2009}
{Soszy{\'n}ski} I.,  et~al., 2009, \actaa, \href
  {http://adsabs.harvard.edu/abs/2009AcA....59....1S} {59, 1}

\bibitem[\protect\citeauthoryear{{Stellingwerf}}{{Stellingwerf}}{1978}]{Stellingwerf-1978}
{Stellingwerf} R.~F.,  1978, \mn@doi [\apj] {10.1086/156444}, \href
  {http://adsabs.harvard.edu/abs/1978ApJ...224..953S} {224, 953}

\bibitem[\protect\citeauthoryear{{Su{\'a}rez Mascare{\~n}o}, {Rebolo}  \&
  {Gonz{\'a}lez Hern{\'a}ndez}}{{Su{\'a}rez Mascare{\~n}o}
  et~al.}{2016}]{Suarez-2016}
{Su{\'a}rez Mascare{\~n}o} A.,  {Rebolo} R.,   {Gonz{\'a}lez Hern{\'a}ndez}
  J.~I.,  2016, \mn@doi [\aap] {10.1051/0004-6361/201628586}, \href
  {https://ui.adsabs.harvard.edu/abs/2016A&A...595A..12S} {595, A12}

\bibitem[\protect\citeauthoryear{{Surot} et~al.,}{{Surot}
  et~al.}{2019}]{Surot-2019}
{Surot} F.,  et~al., 2019, \mn@doi [\aap] {10.1051/0004-6361/201833550}, \href
  {https://ui.adsabs.harvard.edu/abs/2019A&A...623A.168S} {623, A168}

\bibitem[\protect\citeauthoryear{{Taylor}}{{Taylor}}{2005}]{Taylor-2005}
{Taylor} M.~B.,  2005, in {Shopbell} P.,  {Britton} M.,   {Ebert} R.,  eds,
  Astronomical Society of the Pacific Conference Series Vol. 347, Astronomical
  Data Analysis Software and Systems XIV. p.~29

\bibitem[\protect\citeauthoryear{{Tonry} et~al.,}{{Tonry}
  et~al.}{2018}]{Tonry-2018}
{Tonry} J.~L.,  et~al., 2018, \mn@doi [\pasp] {10.1088/1538-3873/aabadf}, \href
  {http://adsabs.harvard.edu/abs/2018PASP..130f4505T} {130, 064505}

\bibitem[\protect\citeauthoryear{{Torres}, {Andersen}  \&
  {Gim{\'e}nez}}{{Torres} et~al.}{2010}]{Torres-2010}
{Torres} G.,  {Andersen} J.,   {Gim{\'e}nez} A.,  2010, \mn@doi [\aapr]
  {10.1007/s00159-009-0025-1}, \href
  {https://ui.adsabs.harvard.edu/abs/2010A&ARv..18...67T} {18, 67}

\bibitem[\protect\citeauthoryear{{Valenzuela} \& {Pichara}}{{Valenzuela} \&
  {Pichara}}{2018}]{Valenzuela-2018}
{Valenzuela} L.,  {Pichara} K.,  2018, \mn@doi [\mnras]
  {10.1093/mnras/stx2913}, \href
  {http://adsabs.harvard.edu/abs/2018MNRAS.474.3259V} {474, 3259}

\bibitem[\protect\citeauthoryear{{Wang} et~al.,}{{Wang}
  et~al.}{2017}]{Wang-2017}
{Wang} L.,  et~al., 2017, \mn@doi [\aj] {10.3847/1538-3881/153/3/104}, \href
  {http://adsabs.harvard.edu/abs/2017AJ....153..104W} {153, 104}

\bibitem[\protect\citeauthoryear{{Watson}, {Henden}  \& {Price}}{{Watson}
  et~al.}{2014}]{Watson-2014}
{Watson} C.,  {Henden} A.~A.,   {Price} A.,  2014, VizieR Online Data Catalog,
  \href {http://adsabs.harvard.edu/abs/2014yCat....102027W} {1, 2027}

\bibitem[\protect\citeauthoryear{{Wyrzykowski} et~al.,}{{Wyrzykowski}
  et~al.}{2015}]{Wyrzykowski-2015}
{Wyrzykowski} {\L}.,  et~al., 2015, \mn@doi [\apjs]
  {10.1088/0067-0049/216/1/12}, \href
  {https://ui.adsabs.harvard.edu/abs/2015ApJS..216...12W} {216, 12}

\bibitem[\protect\citeauthoryear{{Zechmeister} \& {K{\"u}rster}}{{Zechmeister}
  \& {K{\"u}rster}}{2009}]{Zechmeister-2009}
{Zechmeister} M.,  {K{\"u}rster} M.,  2009, \mn@doi [\aap]
  {10.1051/0004-6361:200811296}, \href
  {http://adsabs.harvard.edu/abs/2009A%26A...496..577Z} {496, 577}

\makeatother
\end{thebibliography}

\vspace{1cm}
\vspace{0.5cm}
\hspace{-0.6cm}$^{1}$National Institute For Space Research (INPE/MCTI), Av. dos Astronautas, 1758 – São José dos Campos – SP, 12227-010, Brazil  \\
$^{2}$SUPA (Scottish Universities Physics Alliance) Wide-Field Astronomy Unit, Institute for Astronomy, School of Physics and Astronomy, University of Edinburgh, Royal Observatory, Blackford Hill, Edinburgh EH9 3HJ, UK\\
$^{3}$Instituto de Astrof\'isica, Pontificia Universidad Cat\'olica de Chile, Av. Vicu\~{n}a Mackenna 4860, 7820436 Macul, Santiago, Chile\\
$^{4}$Millennium Institute of Astrophysics, Santiago, Chile\\
$^{5}$Departamento de Ciencias Fisicas, Facultad de Ciencias Exactas, Universidad Andres Bello, Av. Fernandez Concha 700, Las Condes, Santiago, Chile\\
$^{6}$Vatican Observatory, V00120 Vatican City State, Italy
$^{7}$Centre for Astrophysics Research, School of Physics, Astronomy and Mathematics, University of Hertfordshire, College Lane, Hatfield AL10 9AB, UK\\
$^{8}$Instituto de Investigación Multidisciplinar en Ciencia y Tecnolog\'ia, Universidad de La Serena, Av. R. Bitr\'an 1305, La Serena, Chile\\
$^{9}$Departamento de Astronom\'ia, Universidad de La Serena, Av. J. Cisternas 1200, La Serena, Chile\\
$^{10}$Departamento de F\'isica Te \'orica e Experimental, Universidade Federal do Rio Grande do Norte, Natal RN, Brazil\\
$^{11}$Centro de Astronom\'ia (CITEVA), Universidad de Antofagasta, Av. Angamos 601, Antofagasta, Chile\\
$^{12}$Computer Science Department, Pontificia Universidad Cat\'olica de Chile, Santiago, Chile\\
$^{13}$Institute for Applied Computational Science, Harvard University, Cambridge, MA USA\\
$^{14}$Departamento de F\'isica, Universidade Federal de Santa Catarina, Trindade 88040-900, Florian\'opolis, SC, Brazil\\
$^{15}$Universidade de São Paulo, IAG, Rua do Matão 1226, Cidade Universitária, São Paulo 05508-900, Brazil\\
$^{16}$N\'ucleo de Astroqu\'imica y Astrof\'isica, Instituto de Ciencias Qu\'imicas Aplicadas, Facultad de Ingenier\'ia, Universidad Aut\'onoma de Chile, Av. Pedro de Valdivia 425, 7500912 Santiago, Chile\\ 
$^{17}$Departamento de F\'isica, Facultad de Ciencias Básicas, Universidad Metropolitana de la Educación,Av. Jos\'e Pedro Alessandri 774,7760197 Nu\~noa, Santiago, Chile\\

\appendix
\section{Column description}\label{sec_columndescription}

All variability information found in this work is being released in order to facilitate forthcoming studies using the VVV database. Indeed, parameters like identifiers, coordinates, and  $ZYJHK_s$ default magnitudes were obtained from the VISTA Science Archive\footnote{\url{http://surveys.roe.ac.uk/vsa/index.html}} while the other ones were computed in the present work. The acronyms \textit{cflvsc} were added in the column description in order to identify the parameters that come from this work. Indeed, the \textit{vivaID} is unique and is equivalent to the \textit{sourceID} in VSA VVVDR4 data-release and hence it can be used to merge the current information with that provided in VSA tables. We have created two new tables in the VSA VVVDR4 release: \verb+vvvVivaCatalogue+ and \verb+vvvVivaXMatchCatalogue+ for the VIVA variable-star candidates (VVV-CVSC) and their cross-matched counterparts (VVV-CVSC-CROS) respectively. The two tables can be linked via the \textit{vivaID}. These tables can also be found in VVVDR5 and later releases, but in these cases \textit{vivaID} will not equal \textit{sourceID} so a joining neighbour table will be used. Examples of how to use the VIVA data with the rest of the VVV and external data are given in the VVV Guide\footnote{\url{http://horus.roe.ac.uk/vsa/vvvGuide.html\#VIVACatalogue}}.
The released parameters and their data types are listed below for the VVV-CVSC (\verb+vvvVivaCatalogue+);  

\begin{itemize}
\item \textbf{vivaID:}	UID in the VIVA catalogue, equivalent to the  merged band-pass detection (sourceID) in the VSA vvvSource (VVVDR4) table as assigned by merge algorithm (type: bigint, 8 bytes);
\item \textbf{raJ2000:} celestial right ascension in degrees, from VVVDR4 vvvSource (type: float, 8 bytes);
\item \textbf{decJ2000:} celestial declination in degrees, from VVVDR4 vvvSource (type: float, 8 bytes);
\item \textbf{glJ2000:} Galactic longitude in degrees, from VVVDR4 vvvSource (type: float, 8 bytes);
\item \textbf{gbJ2000:} Galactic latitude in degrees, from VVVDR4 vvvSource (type: float, 8 bytes);
\item \textbf{WAperMag3:} $W = [Z,Y,J,H,K_s]$ magnitudes using aperture corrected mag (2.0 arcsec aperture diameter, from VVVDR4 vvvVariability - type: float, 4 bytes);
\item \textbf{WAperMag3Err:} error in default point source $mag = [Z,Y,J,H,K_s]$ mag, from VVVDR4 vvvVariability (2.0 arcsec aperture diameter - type: float, 4 bytes)
\item \textbf{KsAperMagPawprint3:} $K_s$ mean magnitude using pawprint data (2.0 arcsec aperture diameter - type: float, 4 bytes);
\item \textbf{ED:} even dispersion parameter of $K_s$ pawprint data (type: float, 8 bytes);
\item \textbf{$ExpRMS_Noise$:} expected noise value for even dispersion parameter of $K_s$ pawprint data (type: float, 8 bytes);
\item \textbf{NgoodMeasurements:} number of good measurements found in the pawprint data (type: integer, 2 bytes);
\item \textbf{Xindex:} X variability index (type: float, 8 bytes);
\item \textbf{Kfi2:} even dispersion parameter of $K_s$ pawprint data (type: float, 8 bytes);
\item \textbf{L2:} expected noise value for even dispersion parameter of $K_s$ pawprint data (type: float, 8 bytes);
\item \textbf{Ncorrelation2:} number of correlated measurements (type: integer, 2 bytes);
\item \textbf{FAPcorrelation2:} false alarm probability to $K_{(fi)}$ variability index (type: float, 8 bytes);
\item \textbf{FlagDataType:} flag about data type, i.e correlated data (CCD) or non-correlated data (NCD) (type: string, 3 bytes);
\item \textbf{EJKs:} extinction computed from \citet[][]{Gonzalez-2012} (Galactic bulge) and \cite[][]{Minniti-2018ebv} (Galactic disk) (type: float, 4 bytes);
\item \textbf{EJKsErr:} rms related with the three nearest EJKs estimations (Galactic disk) (type: float, 4 bytes);
\item \textbf{FreqPKfi2:} main variability frequency using flux independent period method (type: float, 8 bytes);
\item \textbf{$H_{Kfi2}$:} PPSH of FreqPKfi2 considering PK method (type: float, 8 bytes);
\item \textbf{FreqPLfi2:} main variability frequency using panchromatic period method (type: float, 8 bytes);
\item \textbf{$H_{PL2}$:} height of FreqPLfi2 considering PL method (type: float, 8 bytes);
\item \textbf{FreqLSG:} main variability frequency using Lomb-Scargle generalized method (type: float, 8 bytes);
\item \textbf{$H_{LSG}$:} PPSH of FreqLSG considering LSG method (type: float, 8 bytes);
\item \textbf{FreqPDM:} main variability frequency using Phase Dispersion Minimization method (type: float, 8 bytes);
\item \textbf{$H_{PDM}$:} PPSH of FreqPDM considering PDM method (type: float, 8 bytes);
\item \textbf{FreqSTR:} main variability frequency using String Length Method method (type: float, 8 bytes);
\item \textbf{$H_{STR}$:} PPSH of FreqSTR considering STR method (type: float, 8 bytes);
\item \textbf{BestPeriod:} \textbf{the best period estimation, among the five methods, based in the signal to noise value (type: float, 8 bytes);}
\item \textbf{SNRfit:} \textbf{signal to noise value related with the best frequency estimation (type: float, 8 bytes);}
\item \textbf{Avar:} the difference between 5th and 95th percentile of magnitude  in order to provide a rough estimation of variability amplitude (type: float, 8 bytes);
\item \textbf{FlagNfreq:} number of frequencies in agreement with FreqLSG or its harmonic or subharmonic. It assumes values from 1 to 5 (type: integer, 2 bytes);
\item \textbf{FlagFbias6:} counts of periods within $10^{-6}$ periods (for more details see Sect. \ref{sec_mainperiod}) related with FreqLSG (type: integer, 2 bytes);
\item \textbf{FlagFbias7:} counts of periods within $10^{-7}$ periods (for more details see Sect. \ref{sec_mainperiod}) related with FreqLSG (type: integer, 2 bytes);

\end{itemize}

All this information can be used to perform a comprehensive variability search of any type of variable star. In particular, the variability frequencies and amplitudes help the users to select particular types of variable star. Indeed, the crossmatched sources can be used to set the limits on all parameters available. The crossmatched sample (see Sect. \ref{sec_cross}) is included in the \textit{VVV-CVSC} table. However, a new table is performed in order to facilitate the identification of crossmatched sample. All parameters found in the \textit{VVV-CVSC}  table plus the following information are available;

\begin{itemize}
\item \textbf{vivaID:}	UID in the VIVA catalogue, equivalent to the  merged band-pass detection (sourceID) in the VSA vvvSource (VVVDR4) table as assigned by merge algorithm (type: bigint, 8 bytes);
\item \textbf{LiteratureID:} the identifier found in the literature or "NONE" when the name is not available (type: string);

\item \textbf{CrossPeriod:} variability period found in the literature or $-99999999$ when the period is not available (type: float, 8 bytes); 
\item \textbf{MainVarType:} the single variability type adopted by us to group the crossmatched sources (type: string, length: irregular); 
\item \textbf{LiteratureVarType:} the variability types found in the literature (type: string, length: irregular);
\end{itemize}

Indeed, the column \textit{MainVarType} was introduced to summarize the variability types since some objects have multiple identifications  according to AAVSO\footnote{\url{https://www.aavso.org/vsx/index.php?view=about.vartypes}} and SIMBAD\footnote{\url{http://simbad.u-strasbg.fr/simbad/sim-display?data=otypes}} designations and number of cross-matched sources as following: 

\begin{itemize}
    \item \textbf{E:} AR, D, DM, ECL, SD, SB$*$
    \item \textbf{EA:} EA-BLEND, ED, EB$*$Algol, Al$*$
    \item \textbf{EB:} ESD, EB$*$WUMa, EB$*$betLyr, EB$*$, EB$*$Planet, bL$*$, Candidate\_EB$*$
    \item \textbf{EW:} EC, DW, K, KE, WU$*$, KW
    \item \textbf{I:} IA, IB, $*$iA
    \item \textbf{IN:} IT, INA, INB, IN(YY), INAT, INBT, INT, INT(YY)
    \item \textbf{INS:} INSB(YY), INST(YY), INSA, INSB, INST, Rapid\_Irreg\_V$*$
    \item \textbf{IS:} ISA, ISB, UXOR, Irregular\_V$*$
    \item \textbf{FU:} FUOR, FUOr
    \item \textbf{BE:} GCAS, Be$*$, Ae$*$, Candidate\_Ae$*$, Ae?
    \item \textbf{UV:} UVN, UVN(YY), Flare$*$
    \item \textbf{RCB:} DYPer, Erupt$*$RCrB, FF, DPV, DIP, Eruptive$*$
    \item \textbf{WR:} WR$*$, Candidate\_WR$*$
    \item \textbf{AHB:} AHB0, AHB1
    \item \textbf{BCEP:} BCEPS, PulsV$*$bCep
    \item \textbf{CEP:} CEP(B), Cepheid, Ce$*$, Candidate\_Cepheid
    \item \textbf{CW:} CWA, CWB, CW-FU, CW-FO
    \item \textbf{DCEP:} DCEP(B), DCEPS(B), DCEPS, DCEP-FU, DCEP-FO, PulsV$*$delSct, deltaCep
    \item \textbf{DSCT:} DSCTC, DSCTr, dS$*$, DS
    \item \textbf{RR:} RR(B), RRD, RRAB, RRC, RRLyr, RR$*$
    \item \textbf{SR:} SRA, SRB, SRC, SRD, SRS, semi-regV$*$, sr$*$
    \item \textbf{PVTEL:} PVTELI, PVTELII, PVTELIII
    \item \textbf{ZZ:} ZZA, ZZB, ZZLep, ZZO
    \item \textbf{HADS:} HADS(B), SXPHE, SXPHE(B)
    \item \textbf{L:} LB, LC, L:
    \item \textbf{RV:} RVA, RVB, PulsV$*$RVTau
    \item \textbf{GDOR:} gammaDor
    \item \textbf{LPV:} LP$*$, LP?, LPV$*$, Candidate\_LP$*$
    \item \textbf{M:} Mira, Mi?, Mi$*$, Candidate\_Mi$*$
    \item \textbf{roAm:} roAp
    \item \textbf{DWLYN:} V1093HER, V1093Her, V361HYA
    \item \textbf{PUL:} PULS, PulsV$*$, Pu$*$, Psr, Pulsar
    \item \textbf{TTau:} TTau$*$, TT$*$, Candidate\_TTau$*$
    \item \textbf{WVir:} PulsV$*$WVir, WV$*$
    \item \textbf{ACV:} ACVO, $*$alf2CVn, RotV$*$alf2CVn
    \item \textbf{ROT:} R, RotV$*$, RotV, CTTS
    \item \textbf{BY:} BY$*$
    \item \textbf{FKCOM:} RS, RSCVnRedSG$*$, RSCVn, SXARI
    \item \textbf{NSIN:} EllipVar, ELL
    \item \textbf{N:} NA, NB, NC, NL, NR, Nova, Nova-like, Symbiotic$*$, Sy1, No$*$, Candidate\_Nova
    \item \textbf{SN:} SNI, SNIa, SNIa-pec, SNIb, SNIb-pec, SNIc, SNIc-pec, SNIa-BL, SNIb-BL, SNIc-BL, SNIb|c, SNIax, SNIIn-pec, SNII, SNIIn, SNII-P, SNIIb, SNII-pec, SNII-L, SNIIP
    \item \textbf{CV:} CataclyV$*$, IBWD, V838MON, CBSS, Candidate\_CV$*$, C$*$, Candidate\_C$*$
    \item \textbf{X:} XB, XB$*$, XF, XI, XJ, XND, XNG, XP, XBPR, XR, XBP, XB?, Candidate\_XB$*$
    \item \textbf{HMXB:}Candidate\_HMXB, HXB, HX?
    \item \textbf{LMXB:}LXB
    \item \textbf{XPR:}XPRM
    \item \textbf{AGN:} AGN\_Candidate
    \item \textbf{GRB:} gamma, gammaBurst, gam, gB, SNR, SNR?
    \item \textbf{IR:} IR<10um, IR>30um, OH/IR, NIR
    \item \textbf{Radio:} Radio(cm), Radio(mm), Radio(sub-mm), radioBurst, mm, cm, smm, Maser, rB, FIR, RB?, Rad, Mas
    \item \textbf{YSO:} Y$*$O, Candidate\_YSO, Y$*$, Y$*$?
    \item \textbf{V$*$:} V$*$?
    \item \textbf{RGB:} RGB$*$, Candidate\_RGB$*$, RG$*$
    \item \textbf{Planet:} PN?, PN, Planet?, Pl, Pl?, Minorplanet
    \item \textbf{Microlens:} LensingEv, Lev
    \item \textbf{iC:} $*$iC, $*$iN, $*$inAssoc, $*$inCl, AGB$*$, Candidate\_AGB$*$, Candidate\_post-AGB$*$, post-AGB$*$
    \item \textbf{ISM:} PartofCloud, PoC, ComGlob, CGb, Bubble, bub, EmObj, EmO, Em$*$, EmG, Cloud, Cld, GalNeb, GNe, Cl$*$, Cl$*$?, BrNeb, BNe, DkNeb, DNe, RfNeb, RNe, MolCld, MoC, glb, OpCl, denseCore, cor, SFregion, SFR, HVCld, HVC, HII, $*$inNeb, sh, HI, Circumstellar, cir, outflow?, of?, Outflow, out, HH
    \item \textbf{Others:} $*$, $**$, Assoc$*$, BLLac, BLLac\_Candidate, Blazar, BlueSG$*$, Candidate\_BSG$*$, Candidate\_Hsd, Candidate\_brownD$*$, Candidate\_pMS$*$, DwarfNova, EP, Galaxy, GinGroup, GlCl, GlCl?, GroupG, HB$*$, HotSubdwarf, MISC, NON-CV, OH, Orion\_V$*$, PM$*$, Pec$*$, QSO, RedSG$*$, Region, S, S$*$, SIN, Seyfert\_1, Star, Transient, Unknown, VAR, WD$*$, brownD$*$, multiple\_object
\end{itemize}

Indeed, different surveys can assume different notation but the same meaning. For instance, RR is common used as RRAB or RRLyr. On the other side, many sources have a few objects or have a single notation and hence their notations were maintained: cPNB$[$e$]$, EXOR, SDOR, FSCMa, TTS, BYDra, ACEP, ACYG, BLAP, BXCIR, SPB, PPN, PSR, HB, UG, UGSS, UGSU, UGZ, UGWZ, UGER, ZAND, DQ, AM, XM, APER, PER, CST.

\section{SQL Queries}
\label{app:sql}
We have done all the selection from the VVVDR4 release via the VISTA Science Archive\footnote{http://surveys.roe.ac.uk/vsa}. The followinq query is designed to select light-curves using pawprint detections for sources in the range 515396075613 to 515396077613. Below we will step the curious reader through the design of this selection. The SQL Cookbook in the VSA\footnote{http://horus.roe.ac.uk/vsa/sqlcookbook.html} and the VVV Guide \footnote{http://horus.roe.ac.uk/vsa/vvvGuide.html} are helpful to build up complex queries.

\begin{verbatim}
    SELECT v.sourceID, v.frameSetID, v.ksMeanMag, 
    v.ksMagRms, v.variableClass, b.multiframeID, 
    b.seqNum, b.flag, m.filterID, m.mjdObs, o1SeqNum, 
    o2SeqNum, o3SeqNum, o4SeqNum, o5SeqNum, o6SeqNum, 
    do1.aperMag3 as o1AperMag3, do1.aperMag3Err as 
    o1AperMag3Err, do1.ppErrBits as o1ppErrBits, 
    do2.aperMag3 as o2AperMag3, do2.aperMag3Err as 
    o2AperMag3Err, do2.ppErrBits as o2ppErrBits,
    do3.aperMag3 as o3AperMag3, do3.aperMag3Err as 
    o3AperMag3Err, do3.ppErrBits as o3ppErrBits, 
    do4.aperMag3 as o4AperMag3, do4.aperMag3Err as 
    o4AperMag3Err, do4.ppErrBits as o4ppErrBits, 
    do5.aperMag3 as o5AperMag3, do5.aperMag3Err as 
    o5AperMag3Err, do5.ppErrBits as o5ppErrBits,
    do6.aperMag3 as o6AperMag3, do6.aperMag3Err as 
    o6AperMag3Err, do6.ppErrBits as o6ppErrBits 
    FROM vvvVariability as v, 
    vvvSourceXDetectionBestMatch AS b, vvvTileSet AS t, 
    vvvTilePawPrints AS p, Multiframe as m,
    (SELECT d.multiframeID,d.extNum,d.seqNum,
    d.aperMag3,d.aperMag3Err,d.ppErrBits FROM 
    vvvDetection as d,Multiframe as m where 
    m.multiframeID=d.multiframeID and m.offSetID=1) 
    AS do1,
    (SELECT d.multiframeID,d.extNum,d.seqNum,
    d.aperMag3,d.aperMag3Err,d.ppErrBits FROM 
    vvvDetection as d,Multiframe as m where 
    m.multiframeID=d.multiframeID and m.offSetID=2) 
    AS do2,
    (SELECT d.multiframeID,d.extNum,d.seqNum,
    d.aperMag3,d.aperMag3Err, d.ppErrBits FROM 
    vvvDetection as d,Multiframe as m where 
    m.multiframeID=d.multiframeID and m.offSetID=3) 
    AS do3,
    (SELECT d.multiframeID,d.extNum,d.seqNum,
    d.aperMag3,d.aperMag3Err,d.ppErrBits FROM 
    vvvDetection as d,Multiframe as m where 
    m.multiframeID=d.multiframeID and m.offSetID=4) 
    AS do4,
    (SELECT d.multiframeID,d.extNum,d.seqNum,
    d.aperMag3,d.aperMag3Err,d.ppErrBits FROM 
    vvvDetection as d,Multiframe as m where 
    m.multiframeID=d.multiframeID and m.offSetID=5) 
    AS do5,
    (SELECT d.multiframeID,d.extNum,d.seqNum,
    d.aperMag3,d.aperMag3Err,d.ppErrBits FROM 
    vvvDetection as d,Multiframe as m where 
    m.multiframeID=d.multiframeID and m.offSetID=6) 
    AS do6 
    WHERE v.sourceID=b.sourceID AND 
    (v.ksnGoodObs+v.ksnFlaggedObs) > 10 AND 
    v.sourceID BETWEEN 515396075613 AND 515396077613 
    AND v.frameSetID BETWEEN 515396075521 AND 
    515396075522 AND b.multiframeID=t.tlmfID AND 
    b.extNum=p.tlExtNum AND b.seqNum=p.tlSeqNum AND
    t.tileSetID=p.tileSetID AND
    (p.tlSeqNum>0 OR p.tileSetSeqNum<0) AND 
    m.multiframeID=t.tlmfID and m.filterID=5 AND
    do1.multiframeID=t.o1mfID and 
    do1.extNum=p.o1ExtNum and do1.seqNum=p.o1SeqNum 
    and do2.multiframeID=t.o2mfID and
    do2.extNum=p.o2ExtNum and do2.seqNum=p.o2SeqNum 
    and do3.multiframeID=t.o3mfID and 
    do3.extNum=p.o3ExtNum and do3.seqNum=p.o3SeqNum 
    and do4.multiframeID=t.o4mfID and 
    do4.extNum=p.o4ExtNum and do4.seqNum=p.o4SeqNum 
    and do5.multiframeID=t.o5mfID and 
    do5.extNum=p.o5ExtNum and do5.seqNum=p.o5SeqNum 
    and do6.multiframeID=t.o6mfID and
    do6.extNum=p.o6ExtNum and do6.seqNum=p.o6SeqNum
\end{verbatim}

This query can be broken into several parts:
\begin{itemize}
    \item Selection of sources with correct attributes from \verb+vvvVariability+ table
    \item Linking each source to an epoch via \verb+vvvSourceXDetectionBestMatch+ table
    \item Getting the individual pawprint detection photometry and flags for each epoch.
\end{itemize}

The main selection is on the \verb+vvvVariability+ catalogue where we select sources with at least 10 good or flagged $K_s$ band epochs (tile epochs) and {\bf sourceID} and {\bf framesetID} ranges. 

\begin{verbatim}
    (v.ksnGoodObs+v.ksnFlaggedObs) > 10 AND 
    v.sourceID BETWEEN 515396075613 AND 515396077613 
    AND v.frameSetID BETWEEN 515396075521 AND 
    515396075522 
\end{verbatim}

Joining to the \verb+vvvSourceXDetectionBestMatch+ and joining by sourceID links to all tiles that contain the source \verb+v.sourceID=b.sourceID+. The \verb+vvvSourceXDetectionBestMatch+ is in turn joined to \verb+vvvTilePawprints+ (and its companion table \verb+vvvTileSet+) via   
\verb+b.multiframeID=t.tlmfID AND b.extNum=p.tlExtNum AND+ \verb+b.seqNum=p.tlSeqNum AND t.tileSetID=p.tileSetID+. We also link to the \verb+Multiframe+ to select $K_s$ only epochs 
\verb+m.multiframeID=t.tlmfID and m.filterID=5+
\verb+vvvTilePawprints+ tells you which pawprint detections are linked to which tile detections, but does not include the photometric measurements, so joins to \verb+vvvDetection+ is necessary. Infact, we require 6 joins to \verb+vvvDetection+, one for each pawprint offset. However, \verb+vvvDetection+ is an extremely large table, 50 billion rows, with more than 100 attributes, so we do subqueries to select just pawprint data for the specific offset and with the minimal number of attributes:

\begin{verbatim}
    (SELECT d.multiframeID,d.extNum,d.seqNum,
    d.aperMag3,d.aperMag3Err,d.ppErrBits FROM 
    vvvDetection as d,Multiframe as m where 
    m.multiframeID=d.multiframeID and m.offSetID=6) 
    AS do6
\end{verbatim}

This selection returns a thin table of aperture photometry flags and the detection table primary key for all measurements that have an offsetID equal to 6 as table \verb+do6+, which is linked to a particular epoch through the \verb+vvvTilePawprints+, via \verb+do6.multiframeID=t.o6mfID+ \verb+ and do6.extNum=p.o6ExtNum and+ \verb+do6.seqNum=p.o6SeqNum+. 

\section{Acronyms list}\label{sec_acronyms}

The current section was introduced in order to facilitate the identification of the acronyms found along the paper. A wider definition, of the main acronyms used along the paper, are presented below;

\begin{itemize}

\item \textbf{A3:}  default aperture of 1 arcsec. This has a radius of 3 pixels and contains $\sim75\%$ of the total flux in stellar images;

\item \textbf{CD:} CD means data where correlated indices can be used properly. On the other hand CD-CVSC are the variable stars candidates that where selected using correlated indices;

\item \textbf{$K_{(fi)}^{(s)}$:} it means the flux independent indices that was used to select the variable stars in the CD data.

\item \textbf{FAP:} The false alarm probability for $K_{(fi)}^{(s)}$ to be performed by white noise. The ratio of $K_{(fi)}^{(s)}$ by the $FAP$ sets the noise data about 1 like X index.

\item \textbf{GraMi:} the catalogue of RRLyr stars found by \citet[][]{Gran-2015} and \citet[][]{Minniti-2017} selected from the VVV Survey. The GraMi and WFSC1 are used as comparison stars in   some plots of this paper.

\item \textbf{$H_{method}$:} means the
period power spectrum heights (PPSH) that was summarise as $H_{method}$;

\item \textbf{NITSA:} means the New Insight into Time Series Analysis project where one can found new tools and remarks about how analysis photometric data-sets.

\item \textbf{NCD:} the NCD means data where only statistical parameters (non-correlated indices) can be used. The correlated indice applied in NCD data can be over- or under- estimated. The NCD-CVSC are the variable stars candidates that where selected using statistical parameters;

\item \textbf{X:} means the ratio of a statistical parameter ($\sigma$) by its expected noise value ($\eta$). Such consideration imply that the noise data will be about 1;

\item \textbf{WFSC1:} it means the WFCAM variable star catalogue where comparison stars were used to test our approach. Indeed, the acronyms WFSC1- plus ZYZHKs also means the results considering a single waveband.

\item \textbf{VVVDR4:} it means the fourth data release of VVV data.

\end{itemize}

\bsp
\label{lastpage}
\end{document}